\documentclass[pdflatex,sn-mathphys-num]{sn-jnl}

\usepackage{graphicx}%
\usepackage{multirow}%
\usepackage{amsmath,amssymb,amsfonts}%
\usepackage{amsthm}%
\usepackage{mathrsfs}%
\usepackage[title]{appendix}%
\usepackage{xcolor}%
\usepackage{textcomp}%
\usepackage{manyfoot}%
\usepackage{booktabs}%
\usepackage{algorithm}%
\usepackage{algorithmicx}%
\usepackage{algpseudocode}%
\usepackage{listings}%

\theoremstyle{thmstyleone}%
\theoremstyle{thmstyletwo}%

\theoremstyle{thmstylethree}%

\begin{document}

\title[Device-to-protocol simulation of spectrally-resolved quantum photonic circuits]{Device-to-protocol simulation of spectrally-resolved quantum photonic circuits}


\author[1]{\fnm{Seongjin} \sur{Jeon}}\email{jsj8632@kaist.ac.kr}
\equalcont{These authors contributed equally to this work.}

\author[1]{\fnm{Youngbin} \sur{Kim}}\email{bin1091004@kaist.ac.kr}
\equalcont{These authors contributed equally to this work.}

\author[2]{\fnm{Joaquin} \sur{Matres}}\email{jmatres@gdsfactory.com}

\author[3]{\fnm{Dong-Yeop} \sur{Na}}\email{dyna22@postech.ac.kr}

\author*[1]{\fnm{Young-Ik} \sur{Sohn}}\email{youngik.sohn@kaist.ac.kr}

\affil*[1]{\orgdiv{School of Electrical Engineering}, \orgname{Korea Advanced Institute of Science and Technology (KAIST)}, \orgaddress{\street{291 Daehak-ro, Yuseong-gu}, \city{Daejeon}, \postcode{34141}, \country{Republic of Korea}}}

\affil[2]{\orgname{GDSFactory}, \orgaddress{\street{650 Castro St, Ste 120, PMB 98035}, \city{Mountain View}, \postcode{94041}, \state{CA}, \country{USA}}}

\affil[3]{\orgdiv{Department of Electrical Engineering}, \orgname{Pohang University of Science and Technology (POSTECH)}, \orgaddress{\street{77 Cheongam-ro, Nam-gu}, \city{Pohang}, \postcode{37673}, \state{Gyeongsangbuk-do}, \country{Republic of Korea}}}

\abstract{Predicting quantum-photonic circuit performance requires retaining source and device physics without making photon-counting calculations prohibitively expensive. Here we present PhoQuS, a framework that combines frequency-resolved source models, electromagnetic compact models and spatially correlated fabrication variation to simulate quantum photonic circuits. Covariance-Gram reduction after full-resolution Gaussian propagation lowers readout cost while retaining dominant spectral correlations. For continuous-variable state preparation, spectral reduction accelerates photon counting by $60.4\times$ in an approximate Gottesman–Kitaev–Preskill circuit. Frequency-resolved modeling guides the use of cascaded squeezing with temporal walk-off compensation, yielding $0.99$ fidelity to the single-mode circuit reference and $0.98$ conditional purity in lossless simulations. For discrete-variable state preparation, simulations of three-photon Greenberger–Horne–Zeilinger circuits across a virtual wafer show that component robustness alone does not predict circuit fidelity, which depends jointly on source spectra, circuit architecture and control. PhoQuS connects device physics to protocol-level performance through a unified workflow that retains spectral detail while reducing readout cost.}


\maketitle

Photonic quantum computing is entering a regime in which circuit scale and individual device fidelity can no longer be treated as separate computational problems. Large integrated processors coordinate sources, interferometric components, detectors and controls~\cite{Llewellyn2020Teleportation,Wang2018,Chen2024Heralded}, while remaining sensitive to source spectra, component dispersion and loss. In fusion-based architectures~\cite{Bartolucci2023Fusion}, these effects change resource-state quality, success probabilities and conditional output states. Hardware-oriented simulation must therefore represent large circuits while retaining the device physics that determines protocol performance.

Existing tools such as \textit{Strawberry Fields}~\cite{Killoran2019StrawberryFields} and \text{The Walrus}~\cite{Gupt2019Walrus} support quantum-circuit simulation and photon-counting calculations from specified input states and optical transformations. Predicting a physical implementation additionally requires models of source spectral correlations, device dispersion, loss and fabrication dependence~\cite{Weiss2026QuantumNonlinear}. Neglecting this structure can miss changes in multi-photon interference caused by source–device mode mismatch and dispersion. Resolving it requires additional internal modes, increasing computational costs as circuits grow. And even when state propagation remains tractable, evaluating photon-counting probabilities and heralded conditional states can become a bottleneck~\cite{Gupt2019Walrus,Bulmer2026}.

Classical photonic design automation (PDA)~\cite{Bogaerts2018CircuitDesign} offers a foundation for addressing this challenge. It supports scalable circuit design through reusable compact models, netlists that separate connectivity from device implementation~\cite{Ye2022SPICE}, and layout-aware workflows that link fabrication variation to circuit behavior~\cite{Lu2017, Bogaerts2019}. Building on these principles, we propose a framework for quantum photonic design automation (QPDA) that extends existing PDA workflows and tools with capabilities for modeling quantum-state generation, propagation and measurement.

QPDA requires physical dependencies to be carried through quantum-state generation, frequency-dependent propagation and measurement to protocol observables. To meet these requirements, we developed PhoQuS, which combines frequency-resolved source models, wavelength- and fabrication-dependent device compact models, and layout-aware parameter assignment. A common netlist specifies connectivity, while the layout determines local fabrication parameters. Source correlations, dispersion and loss are retained throughout propagation. Only at readout is the state reduced to spectral supermodes to lower photon-counting cost.

We apply PhoQuS to an approximate Gottesman–Kitaev–Preskill (GKP) state-preparation circuit~\cite{Larsen2025IntegratedGKP} that combines squeezed vacuum sources with linear and nonlinear interference, demonstrating its continuous-variable (CV) simulation capability and the importance of frequency-resolved modeling. Propagation retains all 408 internal bosonic modes. At readout, a rail-local unitary transformation reduces each heralding rail from 51 frequency-bin modes to 5 leading supermodes, accelerating photon counting by $60.4\times$ while closely reproducing the full-resolution state fidelity. Frequency-resolved simulation reveals gain-dependent mode mismatch that limits GKP-state fidelity. We then use PhoQuS to evaluate cascaded squeezing with temporal walk-off compensation, showing that it suppresses this mismatch and improves fidelity.

PhoQuS also simulates discrete-variable (DV) state preparation using heralded single photons. We demonstrate this capability with three-photon Greenberger–Horne–Zeilinger (GHZ) circuits~\cite{Chen2024Heralded} evaluated across a virtual wafer with spatially correlated fabrication variation. By comparing different sources, architectures and component designs, we examine how their combined effects shape quantum-state fidelity. The results show that component-level robustness does not necessarily translate into better circuit performance: the preferred device depends on the circuit architecture and available control. The study also identifies a practical limitation of circuit calibration: residual timing mismatch can limit fidelity, with source bandwidth influencing sensitivity to this mismatch. These findings highlight the need to evaluate light sources, optical components and control together when designing quantum photonic integrated circuits.

Together, these benchmarks demonstrate how PhoQuS connects source and device physics to protocol-level performance while keeping detailed circuit simulation tractable. By extending established PDA workflows with quantum-state propagation and measurement, PhoQuS provides a practical foundation for designing and evaluating quantum photonic integrated circuits and supports the development of a broader QPDA ecosystem.

\section{Results}

\subsection{PhoQuS carries physical information to protocol observables}

\begin{figure}[h]
\centering
\includegraphics[width=1.0\textwidth]{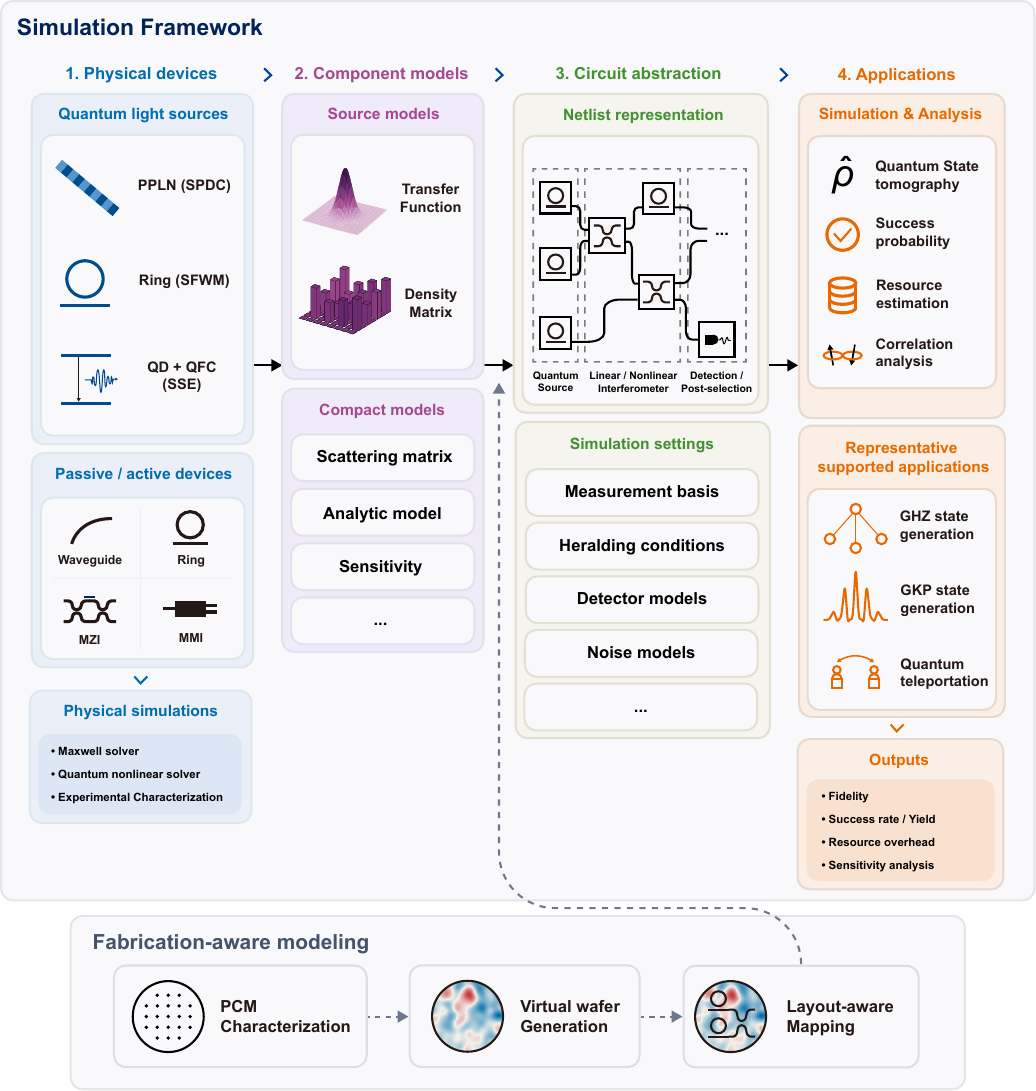}
\caption{\textbf{Multiscale simulation from physical models to protocol observables.} Physical simulations and measurements characterize quantum-light sources and passive or active photonic devices. Source solvers return frequency-resolved quantum states or input--output maps, whereas electromagnetic calculations are converted into complex wavelength- and fabrication-dependent compact models. A placed netlist connects these models, assigns local process parameters, and propagates the quantum state under the selected control settings. Measurement-basis transformations, heralding conditions and detector models are then applied to obtain photon-counting probabilities, conditional states and protocol-specific performance metrics. Fabrication-aware modeling supplies local component parameters through process characterization, virtual-wafer generation and layout mapping.}\label{fig1}
\end{figure}

PhoQuS begins from a physical model and ends with the observable used to evaluate the circuit (Fig.~\ref{fig1}). Under the quadratic undepleted-pump approximation, nonlinear-source solvers return frequency-resolved Bogoliubov transformations, converted into covariance matrices retaining source spectra, spectral phase, multi-mode correlations and loss. Non-Gaussian sources such as solid-state emitters (SSEs) can also be represented using auxiliary Gaussian states, linear interferometers and conditional measurements.

Passive and controlled components are represented by compact models obtained from electromagnetic simulations or experimental characterization. Their complex scattering matrices depend on wavelength and fabrication parameters and are evaluated on the spectral grid of the incident state. A netlist specifies how the components are connected, while their physical designs are defined separately. Sources, couplers and control settings can therefore be exchanged without changing the logical circuit. In fabrication-aware simulation, each placed component samples the virtual-wafer field at its coordinates and evaluates its model using the local parameters.

The wavelength-dependent transformations are applied sequentially to the frequency-resolved covariance matrix, preserving source correlations, component dispersion, loss and path-dependent spectral phase. Expensive source and electromagnetic calculations are performed before the circuit sweep and reused across configurations.

The photon-number-resolving (PNR) detectors complete the calculation. From the final covariance matrix, PhoQuS applies the measurement-basis transformations, heralding conditions and detector models, then evaluates photon-pattern probabilities. Accepted patterns determine the success probability and conditional signal statistics, from which state tomography, correlations or protocol observables are obtained.

\subsection{Physics-resolved modules as composable circuit primitives}

\begin{figure}[h]
\centering
\includegraphics[width=1.0\textwidth]{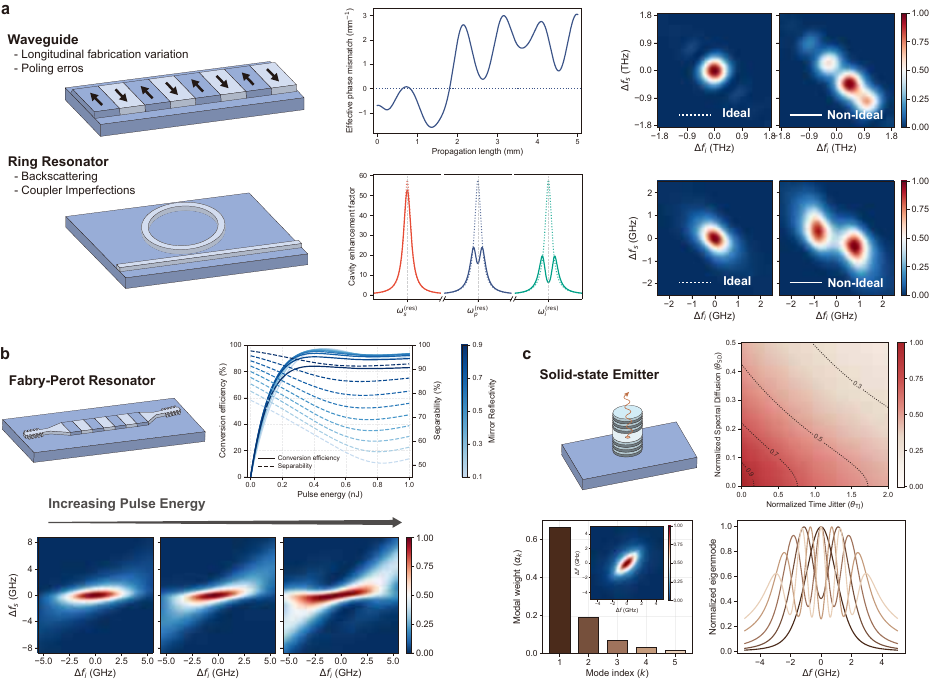}
\caption{
\textbf{Physics-resolved models for quantum-light generation and nonlinear interfacing.}
\textbf{a}, Nonlinear waveguide and resonator pair sources. Geometry and poling errors perturb waveguide phase matching, whereas resonator backscattering and coupler imperfections split the signal, pump or idler resonances and reshape the joint spectrum. Dotted and solid responses denote ideal and perturbed devices.
\textbf{b}, Fabry--Pérot QFC efficiency and modal separability versus pump energy and mirror reflectivity, with the corresponding converted spectra below.
\textbf{c}, Solid-state-emitter model showing purity loss from emission-time jitter and spectral diffusion, together with the modal weights and eigenmodes of the mixed state.
}
\label{fig:source}
\end{figure}

PhoQuS models three physical routes for generating or interfacing quantum light (Fig.~\ref{fig:source}): squeezing through spontaneous parametric down-conversion (SPDC) or spontaneous four-wave mixing (SFWM), nonlinear quantum frequency conversion (QFC), and solid-state emission under the single-emission assumption. Supplementary Notes 1 and 2 detail their implementation.

For pair sources (Fig.~\ref{fig:source}a), PhoQuS solves the full frequency-resolved Bogoliubov input--output map for poled waveguides~\cite{Chen2024AdaptedPoling}, ring resonators~\cite{Helt2010MicroringSFWM,Vernon2015LossyMicroring} and Fabry--Pérot resonators~\cite{Hwang2025HighQ}. Within the quadratic
undepleted-pump model, the calculation is non-perturbative in interaction strength and retains multi-mode squeezing, dispersion, resonant enhancement and propagation loss~\cite{Quesada2020HighGainTwinBeam,Kim2025SimulatingQuantumLight,Christ2013HighGain}. The solver also captures how geometry and poling variations modify phase matching and spectral correlations in SPDC waveguides~\cite{Santandrea2019FabricationLimits,Yang2016NonidealQPM}, and how backscattering hybridizes counter-propagating modes and splits their resonances in SFWM cavities~\cite{McCutcheon2021Backscattering}.

QFC acts on an incident state as a number-preserving channel (Fig. 2b). Its frequency-resolved kernel specifies conversion probability and spectro-temporal mode mapping~\cite{Reddy2013TemporalModeQFC, Brecht2015PhotonTemporalModes, Quesada2016ModeSelectiveQFC}. In the Fabry–Pérot example, mirror reflectivity and pump energy jointly determine resonant enhancement, conversion efficiency and modal separability; stronger driving also reshapes the kernel.

For a solid-state emitter (Fig.~\ref{fig:source}c), PhoQuS constructs a mixed single-photon density matrix incorporating radiative decay, pure dephasing, spectral diffusion and emission-time jitter~\cite{Kambs2018Indistinguishability}. The latter three processes reduce coherence and populate multiple spectro-temporal modes. Eigendecomposition identifies their profiles and weights, which determine modal purity~\cite{Gao2019OptimalFiltering}.

These outputs become distinct circuit primitives. Applied to vacuum, SPDC and SFWM Bogoliubov maps define Gaussian state preparation, whereas QFC defines a number-preserving Gaussian channel~\cite{Weedbrook2012GaussianQI}. The non-Gaussian emitter state is embedded into Gaussian propagation by pairing each occupied eigenmode with an auxiliary herald mode in a two-mode squeezed state whose strength is set by its modal weight. With lossless herald paths, mode-unresolved post-selection on exactly one photon across the herald modes recovers the target mixed single-photon state.

\begin{figure}[h]
\centering
\includegraphics[width=1.0\textwidth]{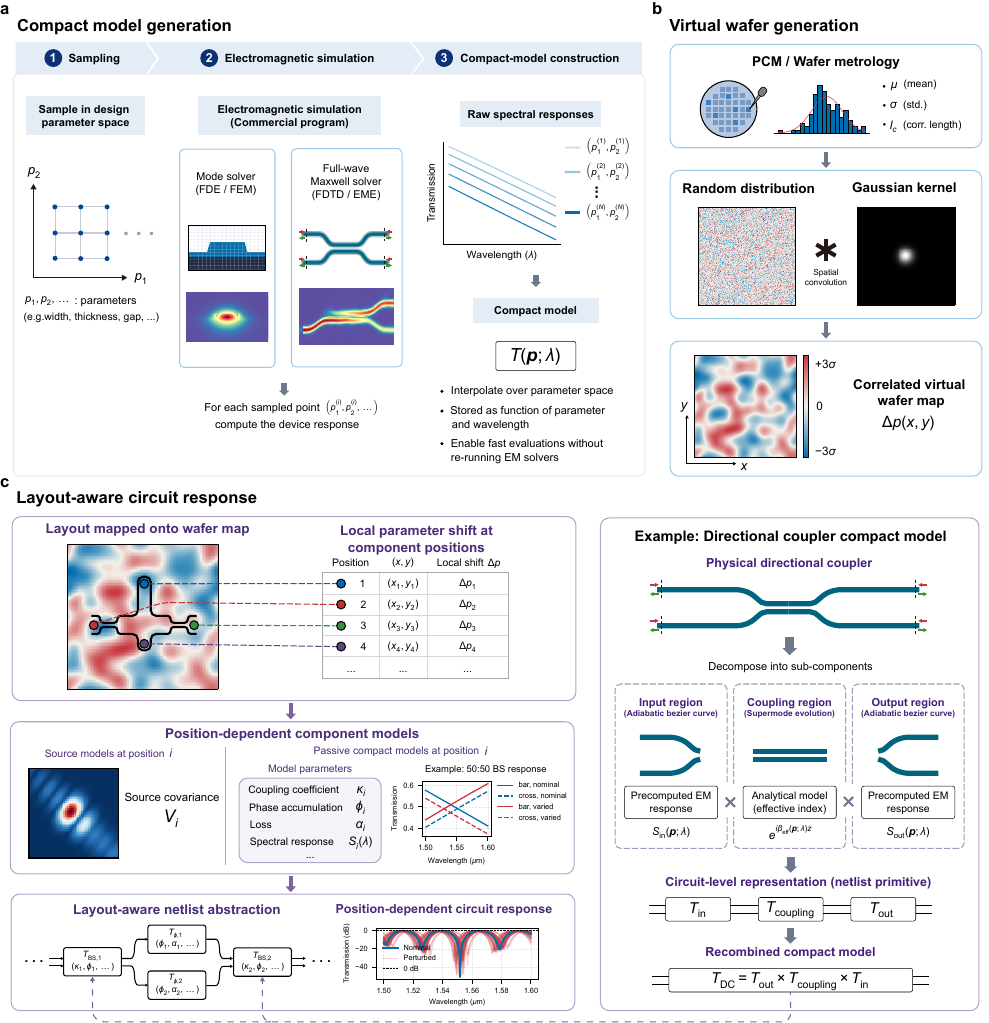}
\caption{
\textbf{Compact models and spatial process fields instantiate layout-aware photonic circuits.}
\textbf{a}, Mode and full-wave electromagnetic solvers sample a device over wavelength and its design and fabrication parameter space. The resulting complex responses are interpolated or parameterized as a reusable compact model, $T(\mathbf{p};\lambda)$.
\textbf{b}, Fabrication characterization or wafer metrology supplies the statistics of each process parameter. Spatial filtering generates a correlated virtual-wafer realization, $\Delta p(x,y)$; measured wafer maps can instead be used directly.
\textbf{c}, A placed layout samples the wafer fields at the positions of its components. The resulting local shifts update source or passive-device models, which are assembled according to the logical netlist to obtain position-dependent circuit responses. The directional-coupler example expands one passive compact model into precomputed input and output transition models and an analytically evaluated coupling region, which are recombined into the complete circuit-level transformation.
}
\label{fig:compact_layout}
\end{figure}

These state-preparation and nonlinear-channel primitives are combined with the complex transformations implemented by passive and tunable photonic devices. PhoQuS constructs these transformations from electromagnetic simulations sampled over wavelength and the relevant design and fabrication parameters, then interpolates or parameterizes the resulting complex responses as reusable compact models (Fig.~\ref{fig:compact_layout}a). Retaining the complex response is essential: its amplitude determines splitting and loss, whereas its spectral phase determines propagation delay and interference.

PhoQuS represents spatially correlated variation as fields over the wafer rather than independent errors assigned to schematic components (Fig.~\ref{fig:compact_layout}b). For each process parameter, a synthetic virtual wafer specifies a mean, standard deviation and correlation length. Gaussian filtering of an initially uncorrelated random field produces the prescribed correlated realization, 
$\Delta p(x,y)$. Linewidth and film thickness can use separate fields; spatially resolved metrology can instead supply measured or interpolated fields directly.

Each placed component samples these fields at 
$(x_i,y_i)$, updating its nominal parameter vector to 
$\mathbf{p}_i=\mathbf{p}_0+\Delta\mathbf{p}_i$ (Fig.~\ref{fig:compact_layout}c). The resulting local parameters modify its source covariance or passive-device response. For example, the DC compact model combines precomputed electromagnetic input and output transitions with analytical supermode propagation through the uniform coupling region, giving 
$T_{\mathrm{DC}}=T_{\mathrm{out}}T_{\mathrm{coupling}}T_{\mathrm{in}}$. This preserves wavelength-dependent splitting, loss and phase without rerunning the electromagnetic solver for every instance.

The netlist assembles these position-dependent primitives under the selected control settings. Separating topology from physical implementation allows sources, nonlinear interfaces, passive devices, layouts and controls to be exchanged while comparing their combined effects at the protocol level.

\subsection{Frequency-resolved simulation reveals spectro-temporal mode mismatch}
\label{sec:frequency-resolved}

We first examine how spectro-temporal mode mismatch reduces GKP-state fidelity relative to the output of the abstracted circuit model in Refs.~\cite{WalrusGKP, Tzitrin2020GKP}. We use a non-resonant periodically poled lithium tantalate (PPLT) SPDC source, with device geometry and poling optimized for high spectral purity. Interfering its signal and idler on a balanced beam splitter converts TMSV into two single-mode squeezed vacuum (SMSV) states when their spectro-temporal profiles match. We adopt this approach because high-purity TMSV is more readily engineered on this platform, and the conversion provides a direct setting for examining mode mismatch. We optimize the pump pulse duration to obtain similar signal and idler temporal widths and apply a separate delay to each field to compensate group-velocity walk-off.

\begin{figure}[h]
\centering
\includegraphics[width=1.0\textwidth]{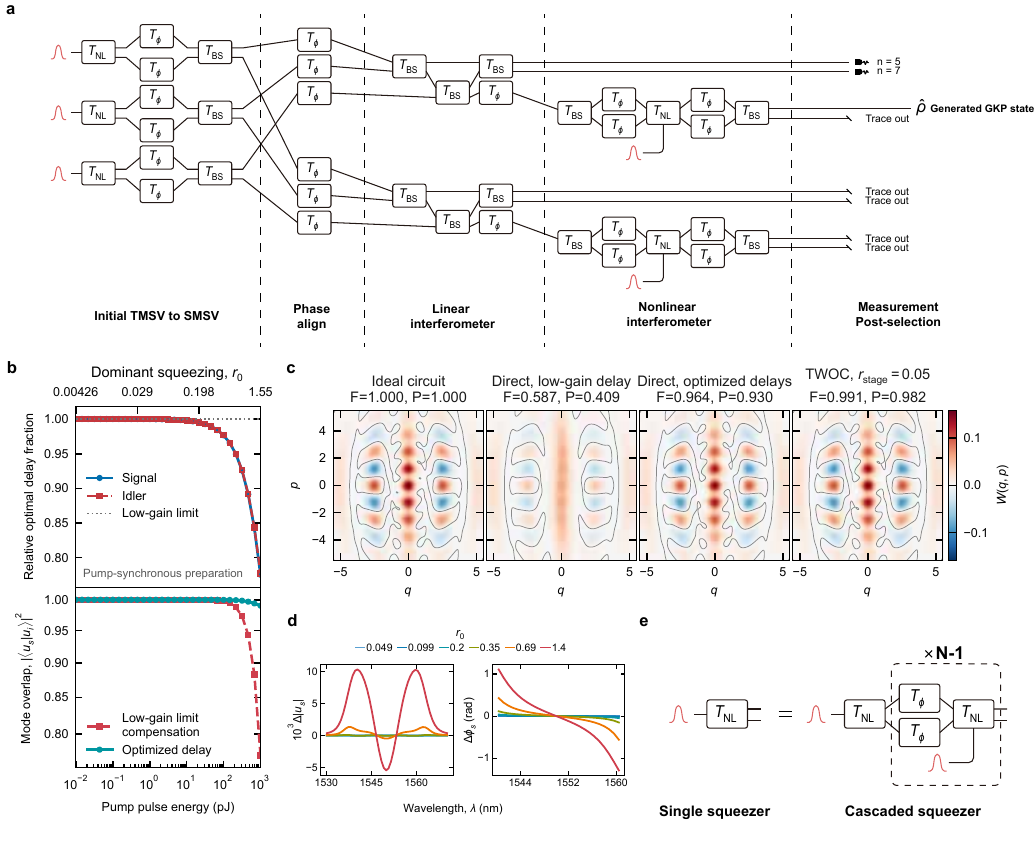}
\caption{
\textbf{Frequency-resolved simulation of Gottesman-Kitaev-Preskill (GKP) state generation circuit.}
\textbf{a}, Netlist description of GKP state generation circuit with TMSV-SMSV scheme. Each initial TMSV are converted into two SMSVs and align the phase to align desired squeezing axes. Then each SMSVs interferes in linear and nonlinear interferometer. Here, nonlinear interferometer is another single-mode squeezer which consists of SMSV-TMSV-SMSV conversion proceses. Here, $T_\mathrm{NL}$ denotes the nonlinear gate; $T_\phi$, the phase gates for squeezing-axes alignment and walk-off compensation and $T_\mathrm{BS}$, the beam splitter.
\textbf{b}, Optimal required delay relative to low-gain limit delay and mode overlap between signal and idler after low-gain limit delay is compensated.
\textbf{c}, Wigner-function comparison of the ideal circuit reference, single stage squeezers with low-gain limit delay compensation, single stage squeezers with optimized delays, and cascaded squeezers with TWOC where $r_{\mathrm{stage}}=0.05$ (from left to right). The normalized conditional state is $\rho=\tilde{\rho}/\operatorname{Tr}\tilde{\rho}$, evaluated within the retained Fock space. Its squared Uhlmann fidelity to the ideal reference $\rho_{\mathrm{ref}}$ and conditional purity are
$F=\left[\operatorname{Tr}\sqrt{\sqrt{\rho_{\mathrm{ref}}}\,\rho\,\sqrt{\rho_{\mathrm{ref}}}}\right]^2$ and $P=\operatorname{Tr}(\rho^2).$
\textbf{d}, Distortion of signal mode amplitude and phase error after low-gain limit delay compensation as a function of the squeezing level. Here $\Delta|u_s|=|u_s|-|u_{s,\mathrm{LG}}|$ and $\Delta\phi_s=\arg(u_su_{s,\mathrm{LG}}^*)$ where $u_{s,\mathrm{LG}}$ is signal mode of $r_0=0.05$
\textbf{e}, Netlist description of cascaded squeezer with interleaved TWOC. 
}
\label{fig:GKP_circuit}
\end{figure}

Figure~\ref{fig:GKP_circuit}a illustrates the TMSV-based GKP state generation circuit. High-fidelity state preparation requires accurate TMSV-to-SMSV conversion, together with phase alignment between successive squeezing operations. These conversions rely on the indistinguishability of the modes entering balanced beam splitter, requiring their spectral amplitudes to be matched through the squeezer geometry and pump configuration, and their relative phases to be corrected using phase shifters and walk-off compensators. Meanwhile, the circuit requires four distinct compression parameters $r=\{-1.38155,-1.21670,0.77988,0.19580\}$. Although phase alignment can be easily adjusted, the delay required for walk-off compensation varies with the squeezing strength (Fig.~\ref{fig:GKP_circuit}b), primarily due to gain-induced group delay arising from time-ordering effects~\cite{Thekkadath2024GD}. Figure~\ref{fig:GKP_circuit}c compares the Wigner functions of the ideal reference state and the states obtained without and with gain-dependent optimized temporal compensation.  The recovered Wigner function indicates that temporal mismatch is a major contributor to the fidelity loss.

Aside from the gain-induced delay, the time-ordering effect changes the spectro-temporal mode (Fig.~\ref{fig:GKP_circuit}d), causing residual infidelity that cannot be eliminated by delay compensation alone. These distortions arise even without parasitic $\chi^{(3)}$ nonlinearities, such as self-phase modulation (SPM) and cross-phase modulation (XPM), which could further degrade mode matching. To mitigate gain-dependent temporal mode variations, we use cascaded low-gain squeezers with interleaved temporal walk-off compensation (TWOC)~\cite{Kim2024TWOC, Reddy2015TemporalModeInterferometry}. This cascaded squeezer lowers required pump energy, mitigating the aforementioned undesired effects. The cascaded squeezing strength can also be controlled by tuning the relative phase between the squeezers. This separates the desired increase in squeezing from the stronger mode variation associated with driving a single stage at higher gain.

Figure~\ref{fig:GKP_circuit}e illustrates the replacement of a strong single stage by multiple compensated weak stages, restoring simulated GKP fidelity to 99.1~\% relative to the reference (Fig.~\ref{fig:GKP_circuit}c). This example highlights the importance of frequency-resolved simulation for capturing spectro-temporal effects and show how frequency-resolved simulation guides circuit design beyond an abstracted squeezing model.

\subsection{A reduced spectral supermode representation accelerates photon-counting readout}

While full frequency resolution supports accurate quantum light generation and propagation, retaining all spectral modes during readout introduces substantial computational overhead. The $\Lambda$-blocked loop-hafnian formulation~\cite{Bulmer2026} evaluates frequency-unresolved photon-counting probabilities without enumerating compatible frequency-resolved patterns; its memory and computational costs nevertheless grow with the retained internal mode dimension.

PhoQuS therefore reduces the final covariance in four steps. A rail-local covariance-Gram matrix is constructed from the vacuum-subtracted normal and anomalous covariance blocks. Its eigendecomposition identifies spectral correlation supermodes and ranks their correlation weights. The full covariance is rotated into this basis, after which the leading modes on each rail are retained, the remaining modes are traced out and the reduced covariance is passed to the photon-counting backend.

\begin{figure}[h]
\centering
\includegraphics[width=1.0\textwidth]{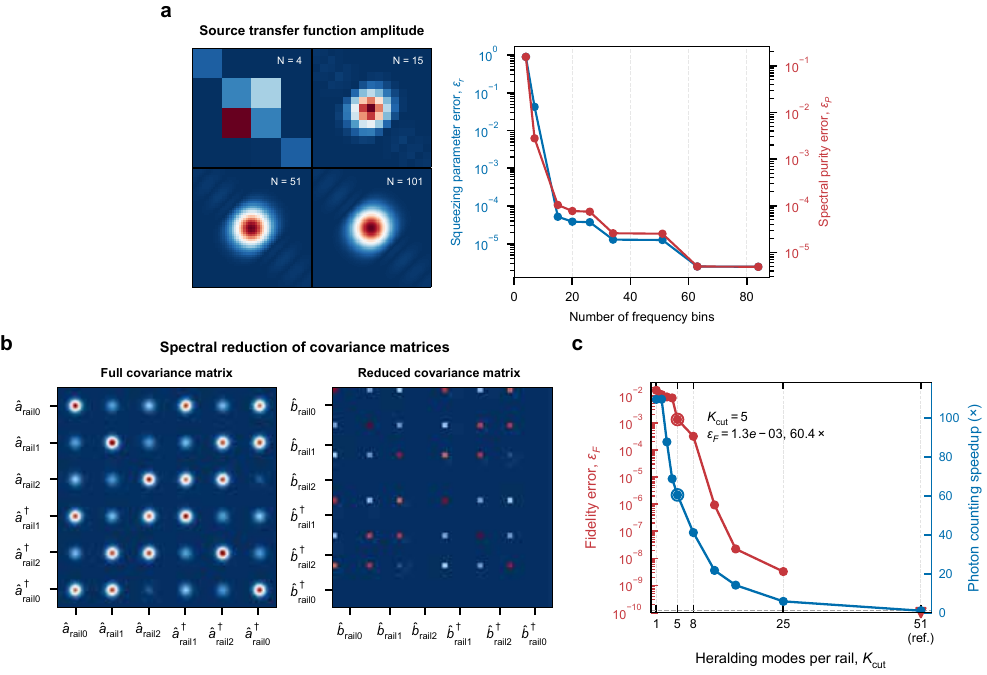}
\caption{
Spectral reduction enables scalable frequency-resolved quantum-circuit simulation.
\textbf{a}, Representative source transfer functions and convergence of the dominant squeezing parameter and spectral purity with increasing frequency-grid resolution. Relative errors are evaluated with respect to the 101-bin reference.
\textbf{b}, Illustration of spectral reduction for a final covariance matrix for compensation re-optimized GKP circuit in Ref.~\ref{sec:frequency-resolved}. Its full signal--idler covariance matrix in the frequency-bin basis is transformed into a reduced spectral-supermode representation by retaining the leading correlation modes.
\textbf{c}, Relative state fidelity error and photon counting speedup evaluated using \texttt{complex128} arithmetic as functions of the retained spectral dimension $K_{\mathrm{cut}}$. Retaining 5 of 51 spectral modes per rail gives a relative witness error of $1.3\times10^{-3}$ and a $60.4\times$ photon-counting speedup. The zero-error full reference at $K_{\mathrm{cut}}=51$ is indicated at the lower plotting boundary.
}
\label{fig:spectral_reduction}
\end{figure}

We benchmark this procedure using the approximate GKP circuit of Sec.~\ref{sec:frequency-resolved}. Four non-resonant PPLT SPDC sources populate eight spatial rails with 51 frequency modes each, giving 408 internal bosonic modes. Source-level convergence against a 101-bin reference confirms that the 51-bin grid resolves the dominant squeezing parameter and spectral purity (Fig.~\ref{fig:spectral_reduction}a), distinguishing readout truncation error from frequency discretization error.

Figure~\ref{fig:spectral_reduction}b shows the covariance transformation after full-resolution propagation. Keeping one supermode on the probing rail, we vary the cutoff on the heralding rails. Retaining 5 of 51 modes per heralding rail gives a relative fidelity error of $1.3\times10^{-3}$; 12 modes reduce it to $9.3\times10^{-7}$. At $K_{\mathrm{cut}}=5$, photon-counting time falls from 18261.38~s to 302.28~s, a $60.41\times$ speedup using \texttt{complex128} arithmetic on the GPU. These timings cover readout, excluding source generation and full-resolution propagation. Together, the mode-matching and reduction benchmarks demonstrate tractable readout while retaining the frequency-resolved physics needed to predict circuit performance precisely.

\subsection{Circuit-level simulation exposes the limits of component-level robustness}
\begin{figure}[h]
\centering
\includegraphics[width=1.0\textwidth]{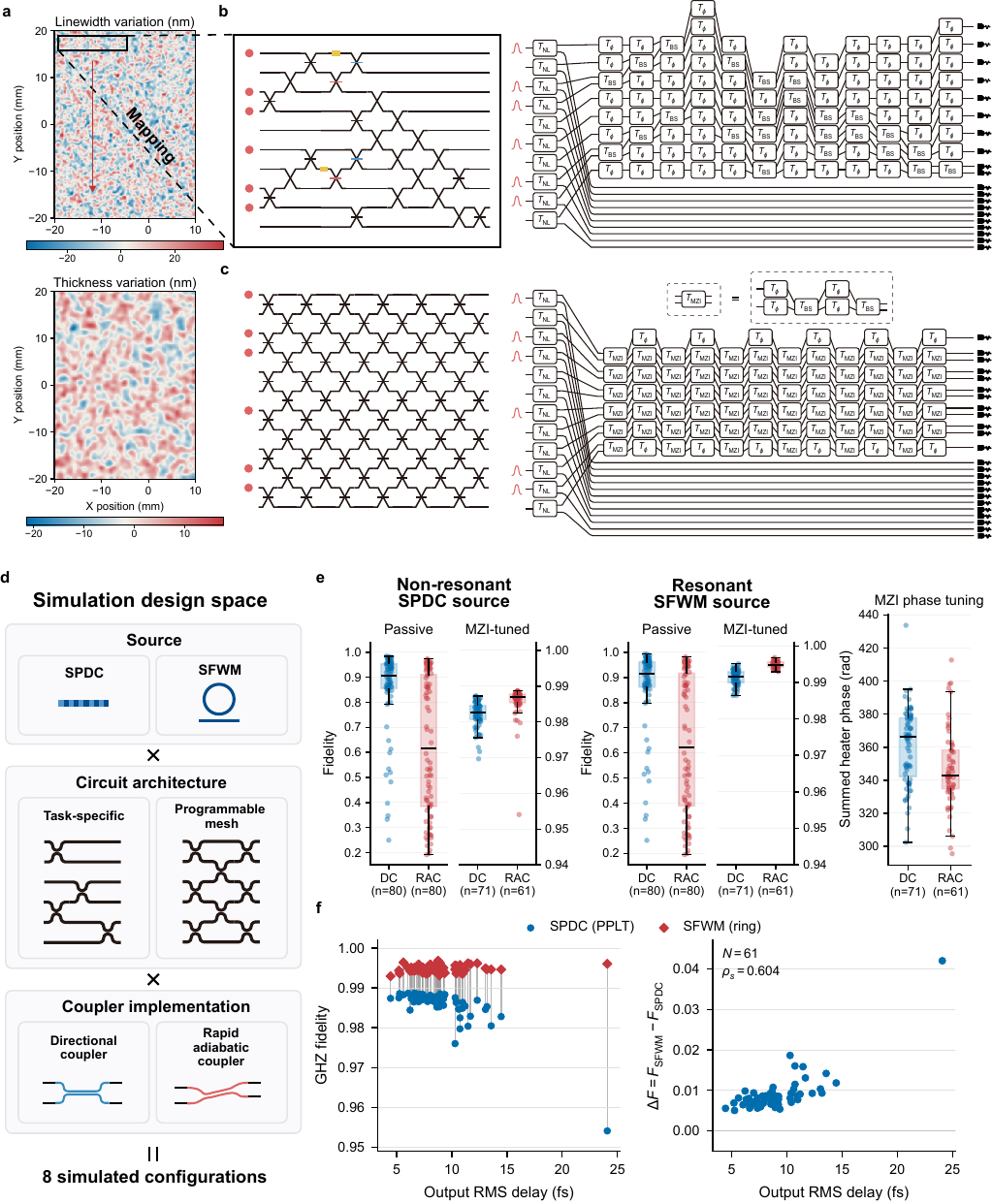}
\caption{\textbf{Fabrication-aware GHZ simulations reveal architecture-dependent coupler rankings and a source-dependent fidelity--delay association.}
\textbf{a}, Spatially correlated linewidth and thickness variations and translation of the circuit layout across one virtual-wafer realization.
\textbf{b}, Task-specific three-photon GHZ circuit. Black, red and blue coupler symbols indicate nominal bar-to-cross power ratios of $50{:}50$, $33{:}66$ and $75{:}25$, respectively. Yellow rectangles denote phase shifters providing a $\pi$ shift at $\lambda_0$.
\textbf{c}, Programmable Clements mesh; $T_{\mathrm{NL}}$, $T_{\phi}$, $T_{\mathrm{BS}}$ and $T_{\mathrm{MZI}}$ denote nonlinear-source, phase-shifter, beam-splitter and tunable-MZI transformations, respectively. All BS are $50{:}50$ splitter and phase shifters are not drawn.
\textbf{d}, Eight configurations spanning two fixed source models, two circuit architectures and two coupler implementations.
\textbf{e}, Reconstructed GHZ fidelities and programmed phase excursions across virtual-wafer positions. The passive-circuit distributions include all $n=80$ positions. For the programmable meshes, only solutions with mesh fidelity $\mathcal{F}_{U}\geq 0.99$ [Eq.~\eqref{eq:methods-mesh-objective}] are included, giving $n=71$ for DC--MZI and $n=61$ for RAC--MZI. Points denote individual positions; boxes span the 25th--75th percentiles, center lines denote medians and whiskers extend to the most extreme values within $1.5$ times the interquartile range. The aggregate programmed phase excursion $C_{\phi}$ is summed across all MZIs; no actuator-specific phase-to-power conversion is assumed.
\textbf{f}, Source swaps in the programmable RAC mesh. At each position, control solution obtained using the same source-independent target-unitary objective at $1550~\mathrm{nm}$, are each played with SPDC and SFWM without further retuning, giving $n=61$ paired source comparisons. Connected circles and diamonds denote matched SPDC and SFWM plays, respectively. The right panel shows $\Delta F_{\mathrm{swap}}=F_{\mathrm{SFWM}}-F_{\mathrm{SPDC}}$ against output-relative RMS delay.}
\label{fig:sim_layout_aware}
\end{figure}

As a second benchmark, we simulate a circuit adapted from the experimentally demonstrated three-photon GHZ-state generator in Ref.~\cite{Chen2024Heralded} (Fig.~\ref{fig:sim_layout_aware}a--d). Heralded three-photon GHZ states are building blocks for fusion-based photonic quantum computing, making their preparation a relevant test of circuit performance. The output fidelity depends on both the source spectrum and multiphoton interference throughout the circuit. Using this circuit, we test whether the performance of individual devices translates into high-fidelity quantum-state preparation. For example, based on splitting ratios alone, rapid adiabatic couplers (RACs) would appear preferable to directional couplers (DCs), because they maintain their target ratios over a wider range of fabrication variations. Full-circuit simulations, however, show that this advantage does not necessarily translate into higher fidelity. The combined effects of wavelength-dependent splitting, phase, loss and delay mean that the preferred coupler depends on the circuit architecture and the ability to retune its control settings.

The full-circuit benchmark compares all eight combinations of source type (non-resonant SPDC or resonant SFWM), circuit architecture (a task-specific passive circuit or a programmable Clements mesh) and coupler design (DC or RAC). The selected sources contrast broadband SPDC emission with narrowband resonant SFWM emission. This bandwidth difference reflects the chosen implementations rather than a fundamental distinction between SPDC and SFWM. The two circuit architectures compare fixed operation with programmable phase control, where they represent passive and active mesh, respectively. The DC–RAC comparison tests whether greater robustness of component power-splitting ratios to fabrication variation translates into more stable quantum-state fidelity at the circuit level.

Each layout is translated to 80 positions on the same spatially correlated virtual-wafer realization. At each position, local linewidth and film-thickness variations update only the passive-device models; the selected source model remains unchanged across positions. The task-specific circuit has no internal tuners, and its fidelity is evaluated by optimizing only the relative phase of the target GHZ state. In the programmable mesh, both phase settings of every universal Mach-Zehnder interferometer (MZI) are retuned at each wafer position.

Together, these tests comprise 640 simulations of GHZ-state generation across eight circuit configurations and 80 matched positions on the same spatially correlated virtual wafer. PhoQuS makes this Monte Carlo analysis of fabrication variation tractable by reusing precomputed source models, evaluating passive-device responses through compact models and reducing the spectral basis used for photon counting. Propagation retains the full spectral resolution needed to capture source correlations and device dispersion, while the subsequent readout approximation keeps the supermodes carrying the dominant output correlations.

The task-specific circuit exhibits a pronounced fidelity gap between the two coupler implementations (Fig.~\ref{fig:sim_layout_aware}e). The DC circuit reaches median fidelities of 0.908 with SPDC and 0.915 with SFWM, whereas the corresponding RAC medians are 0.617 and 0.622. At matched circuit positions, the median paired differences $F_{\mathrm{RAC}}-F_{\mathrm{DC}}$ are $-0.184$ for SPDC and $-0.187$ for SFWM. DC outperforms RAC at 59 of the 80 positions (73.8\%) for both sources. This comparison does not isolate the dominant cause of the fidelity gap; rather, it shows that component-level splitting-ratio robustness alone is insufficient to predict circuit fidelity, which also depends on layout footprint, phase, delay and loss.

For the programmable Clements meshes, we retain control solutions with reference-wavelength matrix fidelity $\mathcal{F}_U\geq 0.99$ (Eq.~\ref{eq:methods-mesh-objective}), leaving 71 DC and 61 RAC meshes. With the optimized control settings held fixed at each position, replacing the non-resonant SPDC source with the resonant SFWM source produces median paired fidelity advantages $F_{\mathrm{SFWM}}-F_{\mathrm{SPDC}}$ of 0.00847 for DC and 0.00801 for RAC. It also reduces the fidelity spread, measured by the interquartile range, by 21.9\% and 43.4\%, respectively.

These improvements are consistent with the narrower bandwidth of the resonant SFWM source. Compared with broadband SPDC photons, narrowband SFWM photons have longer temporal wave packets, so a given relative delay causes less loss of temporal overlap. This reduced sensitivity helps explain both the higher fidelities and their smaller variation across wafer positions, highlighting source bandwidth as an important circuit-design parameter.

Our control optimization targets the Clements mesh unitary only at $1550~\mathrm{nm}$ and is independent of the source model. Different tuning solutions can reproduce this transformation while yielding different spectral phases and group delays. Both fabrication-induced geometry changes and effective-index tuning modify component group indices, introducing small delay differences that can accumulate along paths through successive MZIs. The resulting timing mismatch can therefore limit circuit fidelity even after calibration at the central wavelength.

We compare SPDC and SFWM under identical circuit conditions to test whether larger estimated group-delay mismatches remain associated with a greater fidelity advantage for SFWM (Fig.~\ref{fig:sim_layout_aware}f). At each position, the layout and phase settings remain unchanged within each source comparison. For each fixed-control configuration, we calculate the effective port-averaged RMS delay spread using group delays derived from the scattering matrix. Within each output port, we evaluate the power-weighted RMS deviation of these delays from their weighted mean. This metric captures timing mismatch among input channels contributing to the same output, while excluding differences in mean delay between output ports. Among meshes with $\mathcal{F}_U\geq 0.99$, a larger output RMS delay is associated with a greater SFWM fidelity advantage, with a Spearman rank coefficient~\cite{Spearman1904Association} of $\rho_s=0.604$. This positive association persists after excluding the largest delay sample ($\rho_s=0.583$). At the largest sampled delay of $24.08~\mathrm{fs}$, the SFWM
advantage reaches 0.042, with SPDC and SFWM fidelities of
0.954 and 0.996, respectively. The larger advantage primarily reflects lower SPDC fidelity, whereas the SFWM fidelity remains comparatively stable. This behavior is consistent with the narrower-band resonant source being less sensitive to residual temporal mismatch.

In relatively shallow meshes, these femtosecond-scale delays may be masked by larger splitting, phase and loss errors. They become more consequential for broadband photons generated by ultrafast-pumped sources or in larger meshes, where small geometry- and tuning-induced delays accumulate over more components. Residual delay can then reduce temporal overlap even after retuning has corrected the center frequency unitary. By propagating source spectra and component spectral phases through the full circuit, PhoQuS makes this otherwise hidden scaling mechanism computationally accessible, directly connecting small device- and control-level perturbations to their accumulated impact on protocol fidelity.

\section{Discussion}


As quantum photonic computing approaches chip-scale processors, circuit performance must be evaluated from the joint effects of sources, dispersive devices, fabrication variation and control on multi-photon interference. PhoQuS connects these effects to protocol observables while reducing the computational bottleneck of frequency-unresolved readout through basis transformation.


The benchmarks demonstrate complementary aspects of this workflow. Frequency-resolved GKP simulation identifies gain-dependent mode mismatch and guides compensated cascaded squeezing; output-stage reduction then retains state fidelity to a relative error of 
$1.3\times10^{-3}$ while accelerating photon counting by $60.4\times$. The GHZ study shows that the circuit-level benefit of coupler robustness depends on architecture and control, while fixed-control source swaps connect residual group delay to source-dependent fidelity. Together, these results establish the central principle of PhoQuS: frequency-resolved physics is retained throughout circuit propagation, after which lower-weight spectral correlation modes are discarded at readout to reduce computational cost while preserving the accuracy of the target observable.


The current implementation has several limits. Nonlinear solvers address specific interaction classes and photonic structures; support for multi-comb squeezing or a structure-independent quantum feedback network formulation~\cite{GoughJames2009Series, GoughJames2009Feedback, Combes2017SLH} would broaden this layer. Gaussian covariance matrices efficiently represent state preparation and propagation, but arbitrary non-Gaussian evolution requires an additional backend. At the detection layer, the present supermode basis is constructed from second-order covariance and does not consider displacement. A displacement-aware extension would therefore be required for strongly displaced Gaussian states. Moreover, although this reduction accelerates photon counting by lowering the spectral dimension, it does not modify the underlying algorithm used to evaluate detection probabilities and therefore does not overcome the fundamental \#P-hard complexity of exact photon counting, which remains intractable at large photon numbers.


Within this scope, PhoQuS enables protocol-aware source–interface–circuit co-design. Separating netlist topology from physical implementation allows sources, QFC interfaces, devices and controls to be compared through common protocol observables, exposing trade-offs among fidelity, success probability, loss, footprint and tuning cost. These comparisons carry component changes through to state quality and conditional measurements, allowing design choices to be assessed at the level of their intended quantum task.

The framework also provides a route to layout-aware yield prediction and tolerance analysis. Mapping candidate layouts onto ensembles of calibrated, spatially correlated virtual wafers could identify the process parameters that most strongly affect performance and compare alternative layouts, devices and control ranges before tape-out. After fabrication, measured wafer maps, spectra and scattering matrices could update the same models to create circuit-specific digital twins for calibration and diagnosis.

Scaling further calls for hierarchical compilation. A source, QFC stage or entanglement gate could be exported as an effective state or channel with its success probability, loss and fabrication dependence, characterized within validated accuracy bounds. Such modules could replace detailed internal networks in simulations of multiplexed sources, fusion architectures, quantum networks and fault-tolerant protocols. Although PhoQuS is not yet a complete stack, it establishes a computational foundation for QPDA through reusable physics-derived primitives, layout-aware parameter assignment, end-to-end state-to-measurement simulation and controlled reduction of physical resolution that preserves protocol-level accuracy.

\section{Methods}

\subsection{Gaussian representation and circuit propagation}

Each optical mode is identified by its spatial rail, transverse mode and frequency bin. Frequency quadrature weights are absorbed into the sampled operators so that the discrete modes obey canonical bosonic commutation relations. For the zero-displacement states considered here, the ordering $(\hat{\boldsymbol a},\hat{\boldsymbol a}^{\dagger})^T$ gives the complex covariance
\begin{equation}
V=\begin{pmatrix}N+I/2&M\\M^*&N^*+I/2\end{pmatrix},\qquad N_{ij}=\langle\hat a_j^\dagger\hat a_i\rangle, \quad M_{ij}=\langle\hat a_i\hat a_j\rangle.
\label{eq:methods-covariance}
\end{equation}
The vacuum covariance is $I/2$. Independent sources and ancillary vacuum
modes are combined by direct sum and reordered into the circuit rail basis.

Source and device maps are represented as Gaussian channels, $V_{\mathrm{out}}=XV_{\mathrm{in}}X^\dagger+Y$, where $X$ describes the coherent transformation and $Y$ the noise from unobserved inputs. For a passive transmission matrix $R$ with vacuum loss channels, $X=R\oplus R^*$ and $Y=(I-XX^\dagger)/2$, provided $RR^\dagger\preceq I$. Linear maps act only on sample frequency bin modes, whereas nonlinear maps can couple frequency bin modes. The netlist specifies the order and connectivity of these operations, while component models and layout coordinates specify their physical responses. Source generation, circuit propagation and measurement-basis rotations retain the full frequency representation. The common state and channel conventions, including synchronization of different spectral grids, are detailed in
Supplementary Note~3.

\subsection{Nonlinear sources and single-photon interfaces}

SPDC, SFWM and quantum frequency conversion are calculated within a quadratic quantum-field model driven by undepleted classical pumps. Frequency integrals are discretized on uniform angular-frequency grids, and local propagation maps are composed in longitudinal order. This retains the non-commuting evolution responsible for time-ordering corrections. Ring and Fabry--P\'erot responses are obtained by imposing their device boundary conditions; inaccessible loss inputs contribute vacuum noise. Applying a source map to vacuum gives $V_{\mathrm{src}}=XX^\dagger/2+Y$, whereas a frequency-conversion map acts on the incident state. The interaction kernels, pump solvers and boundary closures are specified in Supplementary Note~1

The solid-state-emitter model instead constructs a single-photon density operator incorporating radiative decay, pure dephasing, spectral diffusion and emission-time jitter (Supplementary Note~2). For its decomposition $\rho_1=\sum_k p_k|1_{u_k}\rangle\langle1_{u_k}|$, an auxiliary herald mode is paired with each physical spectral mode $u_k$ in a two-mode squeezed vacuum satisfying $\tanh^2r_k=c p_k$, with $0<c<1/\max_k p_k$. Exactly one frequency-unresolved herald photon recovers $\rho_1$ when the herald paths are lossless.

\subsection{Electromagnetic compact models and waveguide phases}

Complex device transmissions are calculated using Ansys Lumerical eigenmode expansion (EME) solver. The built-in perturbative wavelength sweep assumes that the modal profiles vary weakly across the sampled band. Responses are sampled at 501 spectral points over $1500$--$1600~\mathrm{nm}$ and on a $3\times3$ fabrication grid with $\Delta w\in\{-50,0,50\}~\mathrm{nm}$ and $\Delta h\in\{-25,0,25\}~\mathrm{nm}$. Transmission amplitude, reference phase near $1550~\mathrm{nm}$ and reference phase subtracted unwrapped spectral phase are represented separately using quadratic Lagrange interpolation.

Reference-phase branches are initially selected by a smoothness criterion. Additional off-grid data at $\Delta w\in\{-25,25\}~\mathrm{nm}$ and $\Delta h\in\{-12.5,12.5\}~\mathrm{nm}$ constrain the branch calibration described in Supplementary Note~4. Residuals at those calibration points do not provide an independent validation of phase interpolation.

The directional coupler response is assembled as $S_{\mathrm{DC}}=T_{\mathrm{out}}T_{\mathrm{DC}}T_{\mathrm{in}}$. EME supplies the input waveguide-to-supermode and supermode-to-output waveguide transmission coefficients, while $T_{\mathrm{DC}}$ propagates the symmetric and antisymmetric supermodes through the uniform coupling region. This model neglects multiple reflections between transitions. RAC models use the complete EME forward transmission response.

Explicit waveguide paths are divided into segments. Effective and group indices are obtained from a finite-difference eigenmode (FDE) database using quadratic fits to the 64 nearest valid samples. A representative B\'ezier-path example compares the interpolated effective and group indices with direct FDE calculations at the sampled path coordinates (Supplementary Fig.~1). The phase is accumulated from
\begin{equation}
\phi(\omega)=\sum_s\Delta L_s \left[\frac{\omega_0}{c}n_{\mathrm{eff},s}+\frac{\omega-\omega_0}{c}n_{\mathrm g,s}+\frac{\omega}{c}\Delta n_s\right].
\label{eq:methods-path-phase}
\end{equation}
The index shift on phase shifter $\Delta n_s$ is treated as frequency independent but changes both of effective index and group index. Its phase contribution therefore also changes group delay. Routing waveguide loss is omitted in this paper. Each universal MZI combines two couplers with an input phase shifter to control the relative phase, and an internal phase shifter to adjust the split ratio. Model construction and interpolation details are given in Supplementary Note~4. Supplementary Figure~2 reports splitting ratio and relative phase errors at the off-grid points used for phase branch calibration.

\subsection{Virtual-wafer fields and layout mapping}

Linewidth and film-thickness fluctuations are represented by separate Gaussian-correlated fields with covariance
\begin{equation}
\mathbb E[\delta p_j(\boldsymbol r)\delta p_j(\boldsymbol r+\boldsymbol d)]=\sigma_j^2\exp[-\|\boldsymbol d\|^2/\ell_j^2].
\label{eq:methods-wafer-covariance}
\end{equation}
The linewidth field uses $\sigma_w=10~\mathrm{nm}$ and $\ell_w=0.5~\mathrm{mm}$. The thickness field uses $\sigma_h=5~\mathrm{nm}$ and $\ell_h=1.0~\mathrm{mm}$. The maps are generated by Gaussian filtering of independent white noise arrays and rescaled to the prescribed spatial means and standard deviations. The field generation and interpolated field maps are
illustrated in Supplementary Fig.~3. We assume no correlation between linewidth and thickness variations, and the mean of the variation field is set to nominal meaning no wafer-to-wafer offset.

All fabrication comparisons use one stored wafer realization and 80 identical layouts are positioned at $x=-20~\mathrm{mm}$, with the center of each circuit layout located at $y=19.75,19.25,\ldots,-19.75~\mathrm{mm}$. Each compact coupler samples one representative layout coordinate and is treated as locally uniform. Supplementary Figure~4 illustrates these two sampling procedures in a representative RAC--MZI, together with the resulting differential phases and changes in coupler power responses. Explicit routing paths sample the fields segment by segment. Sources are assumed to be ideal and unaffected by these variations, meaning that source covariances remain fixed during the whole simulations. Field generation and coordinate mapping are specified in Supplementary Note~5.

\subsection{Output-stage spectral reduction}

Before frequency-unresolved photon counting, the output covariance matrix is rotated into a rail-local spectral supermode basis. For covariance blocks connecting rails $r$ and $s$, define
\begin{equation}
C_r=[N_{r1}\ \cdots\ N_{rR}\ M_{r1}\ \cdots\ M_{rR}],\qquad G_r=C_rC_r^\dagger.
\label{eq:methods-gram}
\end{equation}
Eigenvectors of $G_r$ are ordered by decreasing eigenvalue. Retaining the first $K_r$ vectors minimizes the one-sided residual $\|(I-U_{r,K_r}U_{r,K_r}^\dagger)C_r\|_{\mathrm F}^2$ among passive rail-local subspaces of that dimension. This objective concerns covariance features, rather than a particular observable.

Writing $\mathcal U=\bigoplus_rU_r$, the rotated covariance is $V'=(\mathcal U^\dagger\oplus\mathcal U^T)V (\mathcal U^\dagger\oplus\mathcal U^T)^\dagger$. Tracing out the discarded supermodes gives the Gaussian marginal of the retained modes. So this readout approximation omits the correlation information contained in the discarded modes from the measured rails.

For truncation selection, let $\mathcal D$ denote the complete discarded subsystem and $V_{\mathcal D}$ its covariance. The two stopping quantities are
\begin{equation}
\epsilon_{\mathrm{PNR}}=1-\frac{1}{\sqrt{\det(V_{\mathcal D}+I/2)}},\qquad\epsilon_{\bar n}=\frac{\bar n_{\mathcal D}}{\bar n_{\mathrm{tot}}}.
\label{eq:methods-reduction-tails}
\end{equation}
The second ratio is set to zero for a numerically vacuum state. Starting from the user-specified minimum number of modes on each rail, one mode is added on the rail with the largest next Gram eigenvalue until both quantities fall below the tolerance. These criterions do not guarantees a relative error in a normalized heralded state directly, so the relevant conditional observable must also be checked for convergence. The construction and its limitations are detailed in
Supplementary Note~6. Supplementary Figure~5 illustrates the covariance-Gram basis, sequential mode allocation and stopping criteria for two source gain level.

\subsection{Photon counting and conditional-state reconstruction}

Detectors distinguish spatial rails and sum over their internal spectral and transverse modes. PNR probabilities are evaluated using the $\Lambda$-blocked loop-hafnian framework of Bulmer \textit{et al.}~\cite{Bulmer2026}, which sums compatible internal photon occupations without explicitly enumerating them. Threshold probabilities follow from inclusion--exclusion of vacuum probabilities of each rail. Details are given in Supplementary Note~3.

For conditional states described in full density matrices, the same unresolved detected occupations are summed incoherently in the bra and ket. This yields the unnormalized conditional density matrix $\widetilde\rho$. With a truncated Fock basis, $\rho=\widetilde\rho/\operatorname{Tr}\widetilde\rho$ is the conditional state restricted to that subspace. The computed trace contains only the probability mass represented within the cutoff. State comparisons use the squared Uhlmann fidelity
\begin{equation}
F(\rho,\rho_{\mathrm{ref}})=\left[\operatorname{Tr}\sqrt{\sqrt{\rho_{\mathrm{ref}}}\rho\sqrt{\rho_{\mathrm{ref}}}}\right]^2,\qquad P=\operatorname{Tr}(\rho^2).
\label{eq:methods-state-metrics}
\end{equation}
For a pure reference, this fidelity reduces to its expectation value in $\rho$.

\subsection{Frequency-resolved GKP state preparation}

he reference circuit is the three-mode approximate GKP preparation example in \textit{The Walrus}~\cite{WalrusGKP,Tzitrin2020GKP}. It contains three input squeezed vacuum, three beam splitters and one inline squeezer, with signed squeezing parameters $(r_1,r_2,r_3,r_4)=(-1.38155,-1.21670,0.77988,0.19580)$. The convention is $\hat S(r)=\exp[r(\hat a^2-\hat a^{\dagger2})/2]$. Detection of $n=(5,7)$ photons on 0, 1 rails heralds the state on rail 2.

Each frequency-resolved squeezers uses a $5~\mathrm{mm}$ PPLT SPDC waveguide driven at $775~\mathrm{nm}$ by an unchirped Gaussian pump with $\sigma_\omega=4.7\times10^{12}~\mathrm{rad\,s^{-1}}$. Signal and idler each contain 51 frequency bin modes over a nominal $40~\mathrm{nm}$ window near $1550~\mathrm{nm}$, with spacing $100~\mathrm{GHz}$. Fabrication variation, propagation loss, pump SPM and XPM are not considered. Ideal transverse mode separation and conversion allow signal and idler to interfere on a balanced beam splitter. The inline operation includes an additional vacuum rail and beam splitters before and after SPDC to ensure its SMSV-TMSV-SMSV conversion.

Pump energies of each squeezers are calibrated to the required dominant squeezing strengths. Walk-off compensation applies $\exp[-i(\omega-\bar\omega_\mu)\tau_\mu]$ with $\tau_\mu=L(v_{g,\mu}^{-1}-v_{g,p}^{-1})/2$. In the optimized-delay comparison, the relative delays are tuned to maximize the mode overlap between the dominant signal and the idler mode. At the same time, their common timing is aligned with the pump to synchronize photons across different squeezers. Inline input and output delays are optimized against the desired squeezing transformation in the incident spectral mode, with penalties for coupling to orthogonal modes and to the other rail. The final GKP fidelity is not targeted as an optimization objective throughout this process.

For cascaded squeezing, identical stages with pump energy $1.37924~\mathrm{pJ}$ and dominant squeezing $r_0=0.05$ are separated by walk-off compensators. Compensations are applied to the output and, for inline squeezers, also to the input. The four cascaded squeezers consist of 28, 25, 16, and 4 squeezer stages, respectively. The phase shift applied before the final stage fine-tunes the effective squeezing. Circuit parameters, delay objectives and calibrated settings are specified in Supplementary Note~7.

\subsection{Fabrication-aware GHZ benchmark and programmable control}

Eight configurations combine two sources, two circuit architectures and two coupler implementations. The non-resonant SPDC source uses a $775~\mathrm{nm}$, $0.1~\mathrm{pJ}$ pump and 51 frequency bins per field. The resonant SFWM source uses a $1.42~\mathrm{mm}$ ring, a $10~\mathrm{pJ}$, $0.3~\mathrm{ns}$ square pump pulse and 63 frequency bins per field. Both two-mode squeezer source models are lossless and remain unchanged across positions. To simulate a single photon input, the idler modes are heralded to produce the specific input pattern $[1,0,1,1,0,1,0,1,1,0]$.

The task-specific architecture uses a 10 rail passive network adapted from Ref.~\cite{Chen2024Heralded}. The wavelength- and fabrication-dependent splitting ratios, transmitted powers and relative spectral phases of its DC and RAC implementations are shown in Supplementary Fig.~6. The programmable architecture embeds its target in a 12-rail Clements mesh containing 66 MZIs, where waveguides without MZIs incorporate curved structures to achieve delay matching. Each MZI has two $500~\mu\mathrm m$ long tuning sections. A component-level differential evolution search, with Nelder--Mead refinement when needed, initializes a global L-BFGS-B optimization of the 132 index shifts. The global objective is
\begin{equation}
\mathcal{F}_U=\frac{|\operatorname{Tr}(U_{\mathrm{tar}}^\dagger
S_{\mathrm{mesh}}(\lambda_0))|^2}{12\operatorname{Tr}[S_{\mathrm{mesh}}^\dagger(\lambda_0)S_{\mathrm{mesh}}(\lambda_0)]},\qquad \lambda_0=1550~\mathrm{nm}.
\label{eq:methods-mesh-objective}
\end{equation}
This objective is independent of the source state. Index shifts are bounded by $|\Delta n|\leq5\times10^{-3}$. Three starts are allowed, with up to 1,000 global iterations per start and a stopping target $\mathcal F_U=0.9999$. The best encountered solution is retained when the target is not reached. However, because this approach does not guarantee the global minimum, the retained solution may not represent the maximum fidelity achievable by the circuit at that position.

Readout conditions on the heralding pattern $[1,1,1,0]$. For the heralded logical qubit rails, photon counting is performed within the subspace where only one photon exists per every two rails corresponding to the dual-rail encoding. The eight logical outcomes are normalized separately to evaluate the expectation values for each measurement basis. Reconstructions use 13 Pauli settings for the task-specific circuit and eight unique settings for the programmable mesh and source swaps. These measurement sets are informationally incomplete for general three-qubit states. Their use is motivated by the preparation of nearly pure GHZ states from sources with high spectral purity and by compressed-sensing approaches to approximately low-rank quantum state tomography~\cite{Flammia2012CompressedSensing}. We reconstruct a positive semidefinite, unit-trace density matrix by least-squares fitting of the simulated Pauli expectation values, using Clarabel solver. Before evaluating the photon counting probabilities for each measurement setting, the corresponding covariance matrix is reduced using the rail-local covariance-Gram procedure with tolerance $10^{-3}$.

For the task-specific circuit, we assumed that the target GHZ local phase is optimized:
\begin{equation}
F_{\mathrm{task}}=\frac{\widehat\rho_{000,000}+\widehat\rho_{111,111}}{2}+|\widehat\rho_{000,111}|.
\label{eq:methods-ghz-phase-fidelity}
\end{equation}
The programmable mesh instead use the fixed target $|\mathrm{GHZ}_+\rangle=(|000\rangle+|111\rangle)/\sqrt2$. Fidelities and purities refer to the reconstructed states. Representative reconstructed density matrices for SPDC inputs are shown in Supplementary Fig.~7a,b. Each example is selected from the position whose fidelity is closest to the median of the corresponding distribution.

The aggregate wrapped programmed phase excursion is
\begin{equation}
C_\phi=\sum_m\sum_{q\in\mathrm{input},\mathrm{internal}}\left[\frac{2\pi L_q}{\lambda_0}|\Delta n_{q,m}|\bmod 2\pi\right],\qquad L_q=500~\mu\mathrm m.
\label{eq:methods-phase-excursion}
\end{equation}
This is not an electrical power estimate. The physical simulations use the direct index shifts. Details are in Supplementary Note~8.

\subsection{Source swaps and delay statistics}

Each of the 80 stored RAC--MZI meshes is evaluated with both source covariances while its position, layout and phase settings remain fixed. The 160 broadband evaluations yield 80 paired fidelity differences, $\Delta F_{\mathrm{swap}}=F_{\mathrm{SFWM}}-F_{\mathrm{SPDC}}$. For the reported delay--fidelity analysis, we retain only meshes with $\mathcal{F}_{U}\geq 0.99$ [Eq.~\eqref{eq:methods-mesh-objective}], leaving 61 paired configurations.

The full $12\times12$ transfer matrix is assembled from the physical MZI and routing-waveguide responses, including the programmed phase settings. For every occupied input port $i$ and output port $o$, $\tau_{oi}$ is estimated by a linear fit of the unwrapped transmission phase against angular frequency over nine wavelength samples from $1549.6$ to $1550.4~\mathrm{nm}$. A channel is retained only if its power transmission $|S_{oi}(\omega_k)|^2$ is at least $10^{-4}$ at all nine sampled wavelengths. With $w_{oi}=|S_{oi}(\omega_0)|^2$ at
$\lambda_0=1550~\mathrm{nm}$ and $\bar{\tau}_o=\sum_i w_{oi}\tau_{oi}/\sum_i w_{oi}$, the output RMS delay is
\begin{equation}
\tau_{\mathrm{out,RMS}}=\left[\frac{\sum_o\sum_i w_{oi}(\tau_{oi}-\bar{\tau}_o)^2}{\sum_o\sum_i w_{oi}}\right]^{1/2}.
\label{eq:methods-output-relative-delay}
\end{equation}
This metric is shared by both sources within each pair and characterizes relative temporal mismatch among contributions reaching the same output; it is not a direct measure of multiphoton distinguishability.

Spearman coefficients is calculated over the 61 retained configurations for the paired fidelity differences. Sensitivity to the largest-delay sample is checked by repeating the correlation calculation after excluding that sample. The delay analysis is detailed in Supplementary Note~8. Supplementary Figure~7c,d shows the delay distribution and the source-specific fidelity and purity results for the retained configurations. The overlap of each source's dominant spectral mode with a delayed copy is shown separately in Supplementary Fig.~7e as a source-level diagnostic of sensitivity to temporal walk-off.

\subsection{Computational benchmarking}

Spectral-reduction accuracy and photon counting time were benchmarked for the GKP circuit using double-precision complex arithmetic (\texttt{complex128}). The benchmark used the frequency-resolved covariance matrix of the three logical rails after circuit propagation.

The two heralding rails retained $K_{\mathrm{cut}}$ spectral modes each, while the output rail retained one dominant spectral mode. The spectral basis and output-mode phase convention were held fixed throughout the sweep. Calculations used $K_{\mathrm{cut}}\in\{1,2,3,4,5,8,12,16,25,51\}$, the heralding outcome $(5,7)$, and an output density matrix of dimension 24, corresponding to Fock state with photon numbers $0$ through $23$. The convergence reference retained all 51 spectral modes on each heralding rail while keeping the same single output mode, giving retained dimensions $(51,51,1)$.

Let $F_K$ denote the fidelity of the normalized reconstructed GKP state to the ideal-circuit reference state. Convergence was quantified by
\begin{equation}
\epsilon_F(K)=\frac{|F_K-F_{51}|}{F_{51}},
\end{equation}
which measures convergence of the reported fidelity as the
heralding-mode cutoff is increased.

One timed evaluation was performed per retained dimension after a preliminary warm-up call to the photon counting routine. The density matrix building time, $t_\rho$, measures evaluation and assembly of the diagonal and off-diagonal matrix elements from the prepared adjacency matrix, including completion of CUDA operations.
The corresponding speedup is
\begin{equation}
S_\rho(K)=\frac{t_\rho^{(51,51,1)}}{t_\rho^{(K,K,1)}}.
\end{equation}
These timings characterize density matrix evaluation only rather than end-to-end GKP simulation.

Additional benchmarks evaluate frequency-unresolved PNR probabilities for collision-free output patterns in circuits comprising SPDC sources and independent Haar random interferometers for the signal and idler photons. The dependence of evaluation time on circuit size and total detected photon number is reported for single- and double-precision complex arithmetic in Supplementary Fig.~8. The protocol is described in Supplementary Note~9.

Benchmarks were performed on a workstation running Windows 11 with an AMD processor containing 8 physical and 16 logical CPU cores and an NVIDIA GeForce RTX 5060 Ti GPU. The software environment used Python 3.12.12, NumPy 2.4.2 and PyTorch 2.10.0 with CUDA 12.8.

\section{Data availability}

The numerical results shown in Figs.~2, 4--6 and the Supplementary Figures can be reproduced using the simulation scripts and supporting data available on Zenodo at https://doi.org/10.5281/zenodo.23251638~\cite{Jeon2026PhoQuS}.

\section{Code availability}

The PhoQuS source code and reproduction scripts are publicly available on Zenodo at https://doi.org/10.5281/zenodo.23251638~\cite{Jeon2026PhoQuS}.

\backmatter






\bibliography{sn-bibliography}

\end{document}


\begin{center}
{\Large\bfseries Supplementary Information for}\par
\vspace{0.8em}
{\LARGE\bfseries Device-to-protocol simulation of
spectrally-resolved quantum photonic circuits}\par
\vspace{1.8em}

{\large Seongjin Jeon$^{1}$, Youngbin Kim$^{1}$, Joaquin Matres$^{2}$, Dong-Yeop Na$^{3}$ and Young-Ik Sohn$^{1,*}$}\par
\vspace{1.2em}

$^{1}$School of Electrical Engineering, Korea Advanced Institute of Science and Technology (KAIST), 291 Daehak-ro, Yuseong-gu, Daejeon, 34141, Republic of Korea\par
$^{2}$GDSFactory, 650 Castro St, Ste 120, PMB 98035, Mountain View, 94041, CA, USA\par
$^{3}$Department of Electrical Engineering, Pohang University of Science and Technology (POSTECH), 77 Cheongam-ro, Nam-gu, Pohang, 37673, Gyeongsangbuk-do, Republic of Korea\par
\vspace{1.2em}

$^{*}$Correspondence: youngik.sohn@kaist.ac.kr
\end{center}

\vfill
\begin{center}
This PDF contains Supplementary Sections 1--9, Supplementary Figures,
Supplementary Tables and Supplementary Equations associated with the Article.
\end{center}
\vfill

\clearpage
\tableofcontents
\clearpage

\clearpage
\refstepcounter{section}
\section*{Supplementary Note \thesection.\\Nonlinear propagation and Bogoliubov maps}
\addcontentsline{toc}{section}{\thesection \quad Nonlinear propagation and Bogoliubov maps}
\label{sec:nonlinear-propagation}

This section presents the common mathematical framework used to describe nonlinear optical propagation in PhoQuS. We formulate nonlinear processes mediated by optical $\chi^{(2)}$ and $\chi^{(3)}$ interactions in a unified frequency-resolved framework that accommodates multiple spatial or spectral channels and both propagation directions. Strong pump fields are treated as undepleted semiclassical field, while the generated or converted fields retain their quantum operator description~\cite{Louisell1961ParametricNoise}. Under this approximation, the nonlinear evolution is linear in the quantum field operators and can therefore be represented by a frequency-resolved Bogoliubov transformation, which provides the common input--output description used throughout PhoQuS~\cite{Christ2013HighGain, Quesada2020HighGainTwinBeam}. Distributed optical loss and the associated quantum reservoirs are incorporated within the same propagation framework. Common resonant structures, including ring and Fabry--P\'erot resonators, are further constructed modularly by closing the propagation map with the corresponding boundary conditions.

\subsection{Model hierarchy and implemented nonlinear modules}
\label{subsec:nonlinear-modules}

The nonlinear modules currently implemented in PhoQuS describe two-mode squeezing processes, including spontaneous parametric down conversion (SPDC) and spontaneous four-wave mixing (SFWM), and quantum frequency-conversion processes, including sum-frequency generation (SFG), difference-frequency generation (DFG), and Bragg-scattering four-wave mixing (BSFWM). These processes are implemented for open waveguides, ring resonators, and Fabry--P\'erot resonators. Within the undepleted-pump approximation, the linearized quantum equations are solved nonperturbatively in the nonlinear gain, while pump self-phase modulation (SPM), cross-phase modulation (XPM) of the quantum fields, longitudinal time ordering, and optical loss are retained. Selected modules additionally support bidirectional propagation and linear backscattering. The principal modeling choices of the currently available modules are summarized in Table~\ref{tab:nonlin_phoqus_summary}.

The modules summarized in Table~\ref{tab:nonlin_phoqus_summary} differ primarily in their pump solver, quantum-mode basis, reservoir model, and boundary closure, while sharing the same underlying propagation framework. The open-waveguide modules provide single-pass $\chi^{(2)}$ solvers for SPDC and QFC. The Fabry--P\'erot modules extend these interactions to direction-resolved quantum fields and include all supported co- and counterpropagating processes. Their open-waveguide limit is recovered by setting the mirror reflectivities to zero. The ring-resonator modules implement several $\chi^{(3)}$ processes, including SFWM and BSFWM, together with specialized treatments of dual-pump interactions or forward--backward backscattering.

Conceptually, every nonlinear calculation follows the same four-stage workflow, although some single-pass modules combine the pump and quantum calculations within a single public method.

\begin{table}[!htbp]
    \centering
    \caption{
        Summary of the nonlinear modules currently implemented in
        PhoQuS.}
    \label{tab:nonlin_phoqus_summary}

    \begingroup
    \scriptsize
    \setlength{\tabcolsep}{3.2pt}
    \renewcommand{\arraystretch}{1.18}

    \renewcommand{\tabularxcolumn}[1]{m{#1}}

    \newcommand{\phoqpaircell}[1]{%
        \multirow[c]{2}{=}[-1.4\baselineskip]{#1}%
    }

    \begin{tabularx}{\textwidth}{
        @{}
        >{\hsize=0.95\hsize\linewidth=\hsize
          \raggedright\arraybackslash}X
        >{\hsize=1.00\hsize\linewidth=\hsize
          \raggedright\arraybackslash}X
        >{\hsize=0.86\hsize\linewidth=\hsize
          \raggedright\arraybackslash}X
        >{\hsize=1.25\hsize\linewidth=\hsize
          \raggedright\arraybackslash}X
        >{\hsize=0.96\hsize\linewidth=\hsize
          \raggedright\arraybackslash}X
        @{}
    }

    \toprule
    \textbf{Implementation}
    &
    \textbf{Main Nonlinear processes}
    &
    \textbf{Pump solver}
    &
    \textbf{Quantum solver}
    &
    \textbf{Notes}
    \\
    \midrule\midrule

    \multicolumn{5}{@{}l}{
        \textit{Open $\chi^{(2)}$ waveguides}
    }
    \\[0.4em]

    \addlinespace

    \texttt{TwoModeSPDC}
    &
    Unidirectional SPDC,
    $\omega_p=\omega_s+\omega_i$
    &
    \multirow[c]{2}{=}[-0.0\baselineskip]{
        Forward single-pass pump with propagation loss and SPM
    }
    &
    \multirow[c]{2}{=}[0.5\baselineskip]{
        Two-mode basis $(\hat{\bm \Psi}_s,\hat{\bm \Psi}_i^\dagger)^T$ or $(\hat{\bm \Psi}_s,\hat{\bm \Psi}_i)^T$;\newline
        Propagator interconnection;\newline
        Optical loss via distributed beam-splitter coupling to vacuum channel $\hat{\bm f}$
    }
    &
    \multirow[c]{2}{=}[-0.0\baselineskip]{
        Supports longitudinally varying dispersion, nonlinear coupling, and poling
    }
    \\
    \addlinespace

    \cmidrule(lr){1-2}

    \texttt{TwoModeQFC}
    &
    Unidirectional SFG and DFG,
    $\omega_p=\omega_i-\omega_s$;
    &
    &
    &
    \\

    \addlinespace
    \midrule

    \multicolumn{5}{@{}l}{
        \textit{Fabry--P\'erot $\chi^{(2)}$ devices}
    }
    \\[0.4em]

    \addlinespace\addlinespace
    \texttt{FabryPerotSPDC\_\allowbreak{}SinglePassPump}
    &
    Bidirectional SPDC
    &
    \multirow[c]{2}{=}[0.5\baselineskip]{
        Non-resonant pump; Forward single-pass pump with propagation loss
        and SPM
    }
    &
    \multirow[c]{2}{=}[2.8\baselineskip]{
        Direction-resolved two-mode basis $(\hat{\bm \Psi}_{s+},
      \hat{\bm \Psi}_{i+}^{\dagger},
      \hat{\bm \Psi}_{s-},
      \hat{\bm \Psi}_{i-}^{\dagger})^T$ or $(\hat{\bm \Psi}_{s+},
      \hat{\bm \Psi}_{i+},
      \hat{\bm \Psi}_{s-},
      \hat{\bm \Psi}_{i-})^T$;\newline
      Scattering matrix interconnection;\newline
      Direction-resolved distributed loss;\newline
      Two-mirror boundary closure 
    }
    &
    \multirow[c]{2}{=}[2.8\baselineskip]{
        Bidirectional simulation allows both co- and counter-propagation;\newline
        Supports longitudinally varying structures;\newline
        Open-device limit obtained at zero mirror reflectivity
    }
    \\

    \addlinespace\addlinespace
    \cmidrule(lr){1-2}
    \addlinespace\addlinespace

    \texttt{FabryPerotQFC\_\allowbreak{}SinglePassPump}
    &
    Bidirectional SFG and DFG
    &
    &
    &
    \\

    \addlinespace\addlinespace
    \midrule

    \multicolumn{5}{@{}l}{
        \textit{Single-bus $\chi^{(3)}$ ring resonators}
    }
    \\[0.4em]
    \addlinespace

    \texttt{SingleRingSPSFWM\_\allowbreak{}Homogeneous}
    &
    Unidirectional SFWM,
    $2\omega_p=\omega_s+\omega_i$
    &
    Nonlinear Ikeda map;\newline
    Loss and SPM
    &
    Two-mode basis $(\hat{\bm \Psi}_s,\hat{\bm \Psi}_i^\dagger)^T$;\newline
    Distributed optical loss;\newline
    Single-bus ring feedback closure
    &
    Homogeneous single-pump ring
    \\[0.45em]

    \addlinespace
    \texttt{SingleRingSPSFWM\_\allowbreak{}Backscattering}
    &
    Bidirectional SFWM with linear backscattering
    &
    Coupled Ikeda map between forward-, backward- pumps;\newline
    Loss, SPM, counterpropagating-pump XPM
    &
    Direction-resolved two-mode basis 
    $(\hat{\bm \Psi}_{s+},
      \hat{\bm \Psi}_{i+}^{\dagger},
      \hat{\bm \Psi}_{s-},
      \hat{\bm \Psi}_{i-}^{\dagger})^T$;\newline
    Lossless mirror for internal backscattering;\newline
    Four-port lossy coupler boundary closure
    &
    Supports classical, quantum backscattering
    \\

    \addlinespace
    \texttt{SingleRing\_\allowbreak{}FourSideBand\_\allowbreak{}Homogeneous}
    &
    Unidirectional, dual-pump;\newline
    Selectable BSFWM, single-pump SFWM, dual-pump SFWM
    &
    Dual-pump solver based on temporal coupled-mode theory;\newline
    Pump loss, SPM, and mutual pump XPM
    &
    Four-sideband Nambu basis
    $(\hat{\bm \Psi}_{s1},
      \hat{\bm \Psi}_{i1},
      \hat{\bm \Psi}_{s2}^{\dagger},
      \hat{\bm \Psi}_{i2}^{\dagger})^T$;\newline
    Ordered cell-propagator interconnection;\newline
    Distributed optical loss and single-bus ring feedback closure
    &
    Supports BSFWM and dual-pump SFWM
    \\[0.45em]

    \addlinespace
    \bottomrule
    \end{tabularx}

    \vspace{0.45em}
    \noindent
    \parbox{\textwidth}{
        \scriptsize
        \textit{Notation and common assumptions.}
        The mode labels appearing in the reduced bases denote their
        corresponding frequency-discretized field-operator vectors,
        and the subscripts $+$ and $-$ denote forward and backward
        propagation, respectively. All modules treat the pump fields
        semiclassically under the undepleted-pump approximation,
        include pump-induced XPM of the quantum fields, and solve the
        linearized quantum evolution nonperturbatively in the
        nonlinear gain.
    }

    \endgroup
\end{table}

\begin{enumerate}

\item \textit{Physical and numerical specification.}
The device is first constructed using \texttt{core.structure} or,
for reciprocal multiport devices with backscattering,
\texttt{core.structure\_port}. These objects describe the nonlinear
waveguide and, where required, its couplers and mirrors. They may
include longitudinally varying poling and
mode-dependent reflection and transmission coefficients. The
participating optical modes are defined using \texttt{core.mode},
which stores their dispersion, propagation loss, nonlinear
coefficients, and modal overlaps. The incident pump energy and
temporal envelope are supplied through \texttt{core.pulse}, while
\texttt{core.grid} specifies the frequency grids used to discretize
the quantum fields. These objects are then passed to the selected
nonlinear-module constructor.

\item \textit{Longitudinal discretization.}
The method \texttt{generate\_cell} divides the propagation region
into longitudinal cells and stores the position and length of each
cell. Where distributed optical loss is included, selected cells are
tagged as loss-coupling locations. These tags specify how much
distributed loss is represented at each location; the corresponding
reservoir transformations are subsequently applied during quantum
propagation. The backscattering
module instead uses direction-resolved lumped propagation-loss
channels.

\item \textit{Classical pump calculation.}
The strong pump field is calculated before propagating the quantum
fields. For ring-resonator modules, \texttt{run\_pump} explicitly
evaluates the intracavity pump dynamics using a nonlinear Ikeda map,
coupled forward--backward Ikeda equations, or temporal coupled-mode
equations for interacting pumps. For the open-waveguide and present
Fabry--P\'erot modules, the nonresonant single-pass pump is evaluated
internally from its propagation loss and SPM, and a separate call to
\texttt{run\_pump} is therefore unnecessary. The resulting pump
envelopes $\beta_m(z,t)$ determine the XPM, frequency-conversion, and
pair-generation kernels used in the quantum calculation.

\item \textit{Quantum propagation and boundary closure.}
The exact local propagator can be written formally as~\cite{Dyson1949}
\begin{equation}
    \mathbf U_c
    =\mathcal T_z
    \exp\left[
        i\int_{z_c}^{z_{c+1}}
        \mathbf Q(z)\,dz
    \right],
\end{equation}
where $\mathcal{T}_z$ denotes longitudinal ordering.
The method \texttt{run} evaluates the pump-dependent quantum
generator in each cell and constructs the corresponding local
propagator. In PhoQuS, this ordered exponential is evaluated using finite longitudinal cells
and the cell-propagator approximation is implemented by each module.
The propagators are interconnected in propagation order, thereby
retaining finite-gain and time-ordering effects. At tagged cells,
optical loss is introduced through beam-splitter coupling to vacuum
operators $\hat{\bm f}_c$. For bidirectional Fabry--P\'erot propagation, each
cell map is converted to scattering form before interconnection.
After completing the one-pass propagation, the appropriate mirror,
single-bus coupler, or four-port boundary condition is applied.
Open-waveguide modules return the one-pass map directly.

\end{enumerate}

The resulting transformation can be written in the common form
\begin{equation}
\begin{aligned}
    \mathrm{Propagator}: \hat{\bm \Psi}(z_{k+1}) &= \bm U_{aa}^{(k)} \hat{\bm\Psi}(z_k) + \bm U_{af}^{(k)}\hat{\bm f}_k
    \\ &\downarrow \\
    \mathrm{Scatterer}: \hat{\bm \Psi}_f &= \bm S_{aa}^{(k)} \hat{\bm\Psi}_i + \bm S_{af}^{(k)}\hat{\bm f}_k
    \\ &\downarrow \\
        \mathrm{Boundary}:
    \hat{\bm\Psi}_{\mathrm{out}}
    &=
    \bm M_{aa}\hat{\bm\Psi}_{\mathrm{in}}
    +
    \sum_r
    \bm M_{af}^{(r)}\hat{\bm f}_r.
    \label{eq:common_nonlinear_map}
\end{aligned}
\end{equation}

Equation~\eqref{eq:common_nonlinear_map} summarizes the three successive
representations used by the nonlinear modules. At the propagation level,
$\bm U^{(k)}$ is the transfer map across longitudinal cell $k$, and
$\bm U_{af}^{(k)}$ describes coupling from the vacuum reservoir introduced
at that cell. For a bidirectional module, the transfer relation is rearranged
into the scattering form $\bm S^{(k)}$, in which the fields incident on the
two ends of the cell are grouped into
$\hat{\bm\Psi}_{\mathrm{init}}$ and the fields leaving the cell are grouped
into $\hat{\bm\Psi}_{\mathrm{final}}$. The cell-scattering maps are then
interconnected in their physical order to obtain the one-pass waveguide
scattering map.

At the final level, the device-specific coupler or mirror boundary condition
is applied and the internal cavity-side fields are eliminated, yielding the
external device map $\bm M$. Accordingly, an open unidirectional waveguide
returns its accumulated transfer map directly, whereas an open bidirectional
waveguide returns its accumulated scattering map. For a ring or
Fabry--P\'erot resonator, $\bm M$ is obtained by closing the corresponding
one-pass transfer or scattering map with the resonator boundary condition.
The reservoir blocks undergo the same rearrangement, interconnection, and
boundary closure as the optical block. In the final expression, $r$ labels
all independent propagation-loss and boundary-loss reservoirs retained by
the device. Their inclusion ensures that the complete transformation
preserves the canonical commutation relations.

This modular hierarchy also provides a natural route for future development. The present device-specific boundary solvers may be reformulated as reusable canonical scattering components and interconnected through a general quantum feedback network framework, thereby separating the local nonlinear propagation model from the global photonic-circuit topology. The same nonlinear component could then be embedded in arbitrary feedforward or feedback topologies, including coupled resonators and multi-component photonic networks. Additional material effects, including two-photon absorption and photorefraction, may be incorporated into the pump and reservoir models.

\subsection{Nonlinear propagation dynamics from Maxwell's equations}

This subsection derives the longitudinal quantum propagation equations directly from Maxwell's equations. We first state the scopes underlying the derivation:\newline
1. The reference waveguide is linear, reciprocal, nonmagnetic, and lossless. At real frequencies, its relative-permittivity tensor therefore satisfies \(\bm\epsilon_b^{\mathsf T}=\bm\epsilon_b\) and \(\bm\epsilon_b^\dagger=\bm\epsilon_b\). Material absorption is introduced later through an explicit Langevin channel rather than being incorporated into a complex mode normalization.\newline
2. The field is expanded in a finite set of local guided modes. Radiation modes can be added as additional channels but are not written explicitly.\newline
3. The medium is stationary, so frequency is conserved at a linear static perturbation. 

We begin by expanding the quantized electromagnetic field over the guided optical modes supported by the structure. Each channel is indexed by $c=(\mu_c, \sigma_c)$, where $\mu$ labels a transverse mode or spectral window and $\sigma_c=\pm1$ denotes the propagation direction. Separating the rapidly accumulating longitudinal phase from the frequency-resolved field operator, the positive-frequency electric and magnetic fields are written as
\begin{subequations}
    \begin{align}
        \hat{\bm E}^{(+)}(\bm r, t) &= \sum_{c} \sqrt{\cfrac{\hbar\bar\omega_{\mu_c}}{2\pi}}\int_{\mathfrak B_{\mu_c}} d\omega \ \bm e_c(\bm r_\perp, z;\omega)\hat\psi_c(z,\omega)e^{i\left(\theta_c(z,\omega)-\omega t\right)}, 
        \\
        \hat{\bm H}^{(+)}(\bm r, t) &= \sum_{c} \sqrt{\cfrac{\hbar\bar\omega_{\mu_c}}{2\pi}}\int_{\mathfrak B_{\mu_c}} d\omega \ \bm h_c(\bm r_\perp, z;\omega)\hat\psi_c(z,\omega)e^{i\left(\theta_c(z,\omega)-\omega t\right)},
    \end{align}
    \label{eq:nonlin_em_fields}
\end{subequations}
where $\mathfrak B_{\mu_c}$ indicates the spectral window of mode $\mu_c$. The full Hermitian fields are $\hat{\bm E} = \hat{\bm E}^{(+)}+\hat{\bm E}^{(-)}$ with $\hat{\bm E}^{(-)}=\left(\hat{\bm E}^{(+)}\right)^\dagger$, and similarly for $\hat{\bm H}$. The frequency-resolved operators satisfy the bosonic commutation relation
\begin{equation}
    \left[\hat \psi_c(z,\omega), \hat \psi_d^\dagger(z,\omega')\right] = \delta_{cd}\delta(\omega-\omega'),
\end{equation}
separated from rapidly accumulated longitudinal phase
\begin{equation}
    \theta_c(z,\omega)=\int_{z_0}^z d\zeta \ k_c(\zeta,\omega), \quad k_c(z,\omega)=\sigma_ck_{\mu_c}(z,\omega), \quad k_{\mu_c}>0.
\end{equation}
The backward mode accumulates the phase in opposite sign when it is represented on the same global propagation coordinate $z$.

Here, $\bm e_c$ and $\bm h_c$ are local electric- and magnetic-field mode profiles. For two local modes at the same $(z,\omega)$, define the sesquilinear flux product
\begin{equation}
    \braket{\bm u_c, \bm u_d}_S = \int_A da\left(\bm e_d\times \bm h_c^* - \bm h_d\times \bm e_c^*\right)\cdot \hat {\bm z},
\end{equation}
where $\bm u_c = (\bm e_c, \bm h_c)$. The directed modes are normalized to satisfy
\begin{equation}
    \braket{\bm u_c, \bm u_d}_S = \sigma_c\delta_{cd}.
    \label{eq:nonlin_norm}
\end{equation}
Under the field normalization Eq.~\eqref{eq:nonlin_norm}, the local photon number operator in channel $c$ is
\begin{equation}
    \hat N_c(z) = \int_{\mathfrak B_{\mu_c}} d\omega \hat\psi_c^\dagger(z,\omega)\hat\psi_c(z,\omega).
\end{equation}

The longitudinal evolution equation is obtained by projecting Maxwell's equations onto the local guided modes. Using the Fourier convention
\begin{equation}
    \hat {\bm X}^{(+)}(\bm r, \omega) = \cfrac{1}{\sqrt{2\pi}}\int dt  \hat {\bm X}^{(+)}(\bm r, t) e^{i\omega t},
\end{equation}
the positive-frequency fields satisfy
\begin{subequations}
    \begin{align}
        \nabla \times \hat{\bm E}^{(+)}(\bm r, \omega) &= i\omega\mu_0\hat{\bm H}^{(+)}(\bm r, \omega), 
        \\
        \nabla \times \hat{\bm H}^{(+)}(\bm r, \omega) &= -i\omega\left[\epsilon_0\bm \epsilon_b(\bm r,\omega)\hat{\bm E}^{(+)}(\bm r, \omega)+\hat{\bm P}^{(+)}_\mathrm{NL}(\bm r, \omega)\right].
    \end{align}
    \label{eq:nonlin_maxwell}
\end{subequations}
Here, $\bm \epsilon_b$ defines the linear reference waveguide, whereas $\hat{\bm P}^{(+)}_\mathrm{NL}$ contains the nonlinear polarization contributions treated as sources terms relative to that reference structure. Applying the Fourier transform to Eq.~\eqref{eq:nonlin_em_fields} gives the frequency-domain fields
\begin{subequations}
    \begin{align}
       \hat{\bm E}^{(+)}(\bm r, \omega) &= \sum_d \sqrt{\hbar\bar\omega_{\mu_d}}\bm e_d(\bm r_\perp, z;\omega) \hat\psi_d(z,\omega) e^{i\theta_d(z,\omega)}, \\
       \hat{\bm H}^{(+)}(\bm r, \omega) &= \sum_d \sqrt{\hbar\bar\omega_{\mu_d}}\bm h_d(\bm r_\perp, z;\omega) \hat\psi_d(z,\omega) e^{i\theta_d(z,\omega)}.
    \end{align}
    \label{eq:nonlin_em_fields_freq}
\end{subequations}
Meanwhile, at each longitudinal position $z$, the phase-dressed modal fields, $\mathcal E_c=\bm e_c e^{i\theta_c}$ and $\mathcal H_c=\bm h_c e^{i\theta_c}$, satisfy
\begin{subequations}
    \begin{align}
        \nabla\times\mathcal E_c &= i\omega\mu_0\mathcal H_c + e^{i\theta_c}\hat{\bm z}\times \partial_z\bm e_c, \\
        \nabla\times\mathcal H_c &= -i\omega\epsilon_0\bm\epsilon_b\mathcal E_c + e^{i\theta_c}\hat{\bm z}\times \partial_z\bm h_c.
    \end{align}
    \label{eq:nonlin_modal_profile_relation}
\end{subequations}

We now project the total electromagnetic field onto channel $c$. The vector identity for the divergence of a cross product is
\begin{equation}
    \begin{aligned}
    \nabla\cdot\left(\hat{\bm E}^{(+)}\times \mathcal H_c^* - \hat{\bm H}^{(+)}\times \mathcal E_c^*\right) =& \mathcal H_c^*\cdot\left(\nabla\times \hat{\bm E}^{(+)}\right) - \hat{\bm E}^{(+)}\cdot\left(\nabla \times \mathcal H_c^*\right) \\
    & -\mathcal E_c^*\cdot\left(\nabla\times \hat{\bm H}^{(+)}\right) + \hat{\bm H}^{(+)}\cdot\left(\nabla \times \mathcal E_c^*\right).
    \end{aligned}
    \label{eq:nonlin_vector_identity}
\end{equation}
Substituting Maxwell's equations, Eq.~\eqref{eq:nonlin_maxwell}, and the conjugated modal-field relations obtained from Eq.~\eqref{eq:nonlin_modal_profile_relation} into Eq.~\eqref{eq:nonlin_vector_identity} yields the differential Lorentz reciprocity relation
\begin{equation}
    \nabla\cdot\left(\hat{\bm E}^{(+)}\times \mathcal H_c^* - \hat{\bm H}^{(+)}\times \mathcal E_c^*\right) = i\omega\mathcal E_c^*\cdot \hat{\bm P}_\mathrm{NL}^{(+)} + e^{-i\theta_c}\hat{\bm z}\cdot\left[\hat{\bm E}^{(+)}\times\partial_z\bm h_c^* - \hat {\bm H}^{(+)}\times\partial_z\bm e_c^*\right].
    \label{eq:nonlin_lorentz}
\end{equation}

We next apply Gauss's divergence theorem to an infinitesimal longitudinal slice. Taking the transverse integration domain sufficiently large that the guided fields vanish at its boundary, the transverse boundary-flux term vanishes, and hence
\begin{equation}
    \int_A da \ \nabla\cdot \bm F = \partial_z\int_A da \bm F\cdot \hat{\bm z}.
\end{equation}
Taking $\int_A da\ (\cdot)$ to the reciprocity relation, and expanding electromagnetic fields in Eq.~\eqref{eq:nonlin_em_fields_freq}, the left-hand side becomes
\begin{equation}
    \begin{aligned}
        \int_A da \nabla\cdot\left(\hat{\bm E}^{(+)}\times \mathcal H_c^* - \hat{\bm H}^{(+)}\times \mathcal E_c^*\right) &= \partial_z\int_A da\left(\hat{\bm E}^{(+)}\times \mathcal H_c^* - \hat{\bm H}^{(+)}\times \mathcal E_c^*\right)\cdot \hat{\bm z} \\
        &= \partial_z\sum_d \sqrt{\hbar\bar\omega_{\mu_d}}e^{i(\theta_d-\theta_c)}\hat\psi_d\braket{\bm u_c, \bm u_d}_S \\
        &= \sigma_c \sqrt{\hbar\bar\omega_{\mu_c}}\partial_z\hat\psi_c.
    \end{aligned}
\end{equation}
The second term of the right-hand side becomes
\begin{equation}
    \int_A da \ e^{-i\theta_c}\hat{\bm z}\cdot\left[\hat{\bm E}^{(+)}\times\partial_z\bm h_c^* - \hat {\bm H}^{(+)}\times\partial_z\bm e_c^*\right] = \sqrt{\hbar\bar\omega_{\mu_c}}\sum_d\mathcal C_{cd}(z,\omega)e^{i(\theta_d-\theta_c)}\hat\psi_d(z,\omega),
\end{equation}
where $\mathcal C_{cd} = \sqrt{\bar\omega_{\mu_d}/\bar\omega_{\mu_c}}\braket{\partial_z\bm u_c, \bm u_d}_S$ is the linear mode coupling between modes $c$ and $d$. Combining these gives the propagation equation for the field operator in a longitudinally varying waveguide~\cite{Lin2007Nonlinear, Kim2024TWOC, Laegsgaard2012Modeling}:
\begin{equation} 
    \begin{aligned} 
       \sigma_c\partial_z\hat\psi_c(z,\omega) 
       & = \sum_d \mathcal C_{cd}(z,\omega)e^{i(\theta_d(z,\omega)-\theta_c(z,\omega))}\hat\psi_d(z,\omega) 
       \\
       &+ i\sqrt{\cfrac{\bar\omega_{\mu_c}}{2\pi\hbar}}\int_A da\int dt \bm e_c^*(\bm r_\perp, z;\omega) \cdot \hat{\bm P}_\mathrm{NL}^{(+)}(\bm r, t) e^{-i(\theta_c(z,\omega)-\omega t)}.
    \end{aligned}
    \label{eq:nonlin_prop_equation}
\end{equation}
For current PhoQuS modules, longitudinal variations are assumed to be sufficiently smooth that nonadiabatic coupling between distinct local guided modes can be neglected; $\mathcal C_{cd}\simeq 0$.

\subsection{Semiclassical reduction and pump dynamics}
\label{subsec:nonlinear-pump}

Within the semiclassical approximation, the pump field is represented by its coherent mean-field amplitude,
\begin{equation}
    \beta_p(z,\omega) = \sqrt{\hbar\bar\omega_p} \braket{\hat\psi_p(z,\omega)}.
\end{equation}
All nonlinear modules in PhoQuS solve the pump dynamics in the time domain using the temporal envelope obtained by Fourier transforming $\beta_p(z,\omega)$,
\begin{equation}
    \tilde\beta_p(z,t) = \cfrac{1}{\sqrt{2\pi}}
    \int_{\mathfrak B_p} d\omega\,\beta_p(z,\omega) e^{-i(\omega-\bar\omega_p)t}.
\end{equation}
Over the bandwidth retained by the current nonlinear modules, the propagation constant is expanded to first order about the carrier frequency as $k(\omega)=\bar k_p+(\omega-\bar\omega_p)/v_p+\mathcal O\left((\omega-\bar\omega_p)^2\right)$, where $\bar k_p=k(\bar\omega_p)$ and $v_p^{-1}=\partial_\omega k(\bar\omega_p)$. This approximation relates the electric field to the temporal envelope in the retarded-time coordinate as $\mathbf E_p^{(+)}\propto\tilde\beta_p(z,\tau)e^{i(\bar k_pz-\bar\omega_p t)}$, where $\tau=t-z/v_p$. With this convention, the envelope is power normalized such that
\begin{equation}
    P_p(z,\tau)=|\tilde\beta_p(z,\tau)|^2.
\end{equation}

The pump normalization is common to all modules, whereas its propagation and boundary conditions are device specific. \texttt{TwoModeSPDC}, \texttt{TwoModeQFC}, and the two \texttt{FabryPerot\allowbreak{}\_\allowbreak{}Single\allowbreak{}Pass\allowbreak{}Pump} modules use a forward single-pass pump with loss and SPM. \texttt{SingleRingSPSFWM\allowbreak{}\_\allowbreak{}Homogeneous} uses a nonlinear Ikeda map, while \texttt{SingleRingSPSFWM\allowbreak{}\_\allowbreak{}Backscattering} uses a coupled forward--backward Ikeda map including linear backscattering and counterpropagating-pump XPM. \texttt{SingleRing\allowbreak{}\_\allowbreak{}FourSideBand\allowbreak{}\_\allowbreak{}Homogeneous} uses temporal coupled-mode equations for interacting dual pumps. These single-pass and resonant pump solvers are described below.

\subsubsection{Single-pass pump propagation dynamics}

For an instantaneous Kerr nonlinearity with the usual permutation symmetry, the component of the third-order nonlinear polarization oscillating at the pump carrier frequency is
\begin{equation}
    \left[\hat{\bm P}^{(+)}_\mathrm{SPM}\right]_i
    =
    3\epsilon_0
    \sum_{jkl}\chi_{ijkl}^{(3)}
    \hat E^{(-)}_{p,j}
    \hat E^{(+)}_{p,k}
    \hat E^{(+)}_{p,l}.
\end{equation}
Substituting this polarization into Eq.~\eqref{eq:nonlin_prop_equation}, projecting it onto the pump mode, and including linear propagation loss gives
\begin{equation}
    \partial_z\tilde\beta_p(z,\tau)
    =
    -\frac{\alpha_p}{2}\tilde\beta_p(z,\tau)
    +
    i\gamma_\mathrm{SPM}
    \left|\tilde\beta_p(z,\tau)\right|^2
    \tilde\beta_p(z,\tau),
    \label{eq:nonlin_single_pass_pump_eom}
\end{equation}
where $\alpha_p$ is the power-attenuation coefficient and
$\gamma_\mathrm{SPM}$ is the effective SPM coefficient obtained after the
transverse-mode overlap and susceptibility-tensor contraction.

The solution of Eq.~\eqref{eq:nonlin_single_pass_pump_eom} takes the analytic form
\begin{subequations}
\label{eq:nonlin_single_pass_pump}
    \begin{align}
        \tilde\beta_p(z,\tau)
        &=
        \tilde\beta_p(0,\tau)
        \exp\left[
            -\frac{\alpha_pz}{2}
            +i\Phi_\mathrm{SPM}(z,\tau)
        \right],\\
        \Phi_\mathrm{SPM}(z,\tau)
        &=
        \gamma_\mathrm{SPM}
        z_{\mathrm{eff},p}
        P_p(0,\tau),
    \end{align}
\end{subequations}
where
$z_{\mathrm{eff},p}=(1-e^{-\alpha_pz})/\alpha_p$
is the effective nonlinear length, with
$z_{\mathrm{eff},p}\rightarrow z$ as $\alpha_p\rightarrow0$.

Equation~\eqref{eq:nonlin_single_pass_pump} directly supplies the pump field
for the open-waveguide and Fabry--P\'erot modules. It also provides the
single-pass propagation block used in the ring-resonator modules, supplemented
by pump XPM and forward--backward coupling where applicable.

\subsubsection{Resonant pump dynamics}

For the ring-resonator modules, the single-pass propagation in
Eq.~\eqref{eq:nonlin_single_pass_pump} is closed by the bus--ring coupler
boundary condition. The resulting pump solver depends on the number of pump
fields and the presence of forward--backward coupling.

\paragraph{Single-pump Ikeda map.}

Let $\tau_p$ and $i\rho_p$ denote the self- and cross-coupling amplitudes of
the bus--ring coupler, respectively, satisfying the lossless condition
$|\tau_p|^2+|\rho_p|^2=1$. Combining the coupler boundary condition with
Eq.~\eqref{eq:nonlin_single_pass_pump}, the intracavity pump at $z=0$ is
iteratively updated according to
\begin{subequations}
\label{eq:nonlin_single_pump_ikeda}
    \begin{align}
        \tilde\beta_p(0,t)
        &=
        \tau_p
        \mathcal M_\mathrm{RT}
        [\tilde\beta_p(0,\cdot)](t-T_p)
        +
        i\rho_p\tilde\beta_p^\mathrm{in}(t),\\
        \mathcal M_\mathrm{RT}[\beta](t)
        &=
        \exp\left[
            -\frac{\alpha_pL}{2}
            +i\left(
                \Delta_pT_p
                +
                \gamma_\mathrm{SPM}L_\mathrm{eff}
                |\beta(t)|^2
            \right)
        \right]\beta(t).
    \end{align}
\end{subequations}
Here, $T_p$ is the pump round-trip time,
$\Delta_p=\bar\omega_p-\omega_p^0$ is the detuning from the pump resonance.
Equation~\eqref{eq:nonlin_single_pump_ikeda} is the Ikeda map for a
unidirectional single pump with higher-order pump dispersion neglected~\cite{Ikeda1979RingCavity}.

\paragraph{Coupled Ikeda map with counterpropagating modes.}

In the module \texttt{SingleRingSPSFWM\allowbreak{}\_\allowbreak{}Backscattering}, linear backscattering couples the forward- and backward-propagating intracavity pumps. Their propagation is governed by
\begin{equation}
    \left[\sigma\partial_z + \alpha_p/2 - i\gamma_\mathrm{SPM}\left(\left|\tilde\beta_{p,\sigma}(z,\tau)\right|^2 + 2\left|\tilde\beta_{p,-\sigma}\left(z, \tau+2\sigma z/v_p\right)\right|^2\right)\right]\tilde\beta_{p,\sigma}(z,\tau) = 0. 
    \label{eq:nonlin_pump_bidirectional}
\end{equation}
The factor of two multiplying the counterpropagating pump power arises from XPM between two directions.

Under the long-pulse approximation, $2T_p\left|\partial_t\ln\left|\tilde\beta_{p,\sigma}\right|^2\right|\ll1$, the pump power varies negligibly over the relative delay accumulated during one round trip. Equation~\eqref{eq:nonlin_pump_bidirectional} then has the approximate analytic solutions
\begin{subequations}
    \begin{align}
    \tilde\beta_{p,+}(L,t)&\simeq\tilde\beta_{p,+}(0,t)\times \exp\left[-\cfrac{\alpha_pL}{2}+i\gamma_\text{spm}L_\text{eff}\left(\left|\tilde\beta_{p,+}(0,t)\right|^2+2\left|\tilde\beta_{p,-}(L,t+T_p)\right|^2\right)\right],
    \\
    \tilde\beta_{p,-}(0,t)&\simeq\tilde\beta_{p,-}(L,t)\times \exp\left[-\cfrac{\alpha_pL}{2}+i\gamma_\text{spm}L_\text{eff}\left(\left|\tilde\beta_{p,-}(L,t)\right|^2+2\left|\tilde\beta_{p,+}(0,t-T_p)\right|^2\right)\right].
    \end{align}
\end{subequations}

The resonator boundary is represented by an effective four-port scattering matrix in the channel ordering
$(\mathrm{ring}+,\mathrm{bus}+,\mathrm{ring}-,\mathrm{bus}-)$. With this ordering, the boundary matrix is
\begin{equation}
    \bm S_p
    =
    \begin{pmatrix}
    t_1 & \kappa_{12} & r_{1,R} & \xi_R\\
    \kappa_{21} & t_2 & \xi_R & r_{2,R}\\
    r_{1,L} & \xi_L & t_1 & \kappa_{21}\\
    \xi_L & r_{2,L} & \kappa_{12} & t_2
    \end{pmatrix}.
    \label{eq:nonlin_pump_backscatter_boundary}
\end{equation}
Here, $\bm S_p$ is the effective scattering matrix obtained by interconnecting the bus--ring coupler with the internal reflector. Consequently, $r_{j,L/R}$ and $\xi_{L/R}$ describe direction-changing backscattering pathways, while $\kappa_{ij}$ describes co-propagating transfer between the bus and ring.

To express the resulting coupled Ikeda map, define
\begin{equation}
    \bm \beta_p^{(n)} = \begin{pmatrix}
        \tilde\beta_{p,+}(0, nT_p) \\ \tilde\beta_{p,-}(L, (n+1)T_p)e^{-i\Delta_p T_p}
    \end{pmatrix}, \quad s^{(n)} = \tilde\beta_{p,+}^\mathrm{in}(nT_p).
\end{equation}
The map implemented in PhoQuS is then
\begin{subequations}
    \begin{align}
        &\bm\beta_p^{(n)} = \bm{S}_{p,\mathrm{cav}}\bm{\mathcal M}_\mathrm{RT}\left[\bm{\beta}_p^{(n-1)}\right] + \bm d_ps^{(n)},\\
        &\bm S_{p,\mathrm{cav}}  = \begin{pmatrix}
            t_1 & r_{1,R} \\ r_{1,L} & t_1
        \end{pmatrix}, 
        \ \ 
        \bm d_p = \begin{pmatrix}
            \kappa_{12}\\ \xi_L
        \end{pmatrix}.
    \end{align}
\end{subequations}
Here, $\bm S_{p,\mathrm{cav}}$ is the intracavity subblock of $\bm S_p$, and $\bm d_p$ describes injection from the forward bus channel. The nonlinear round-trip map is
\begin{subequations}
\label{eq:nonlin_round_trip_backscatter}
    \begin{align}
    \bm{\mathcal M}_\mathrm{RT}
    \left[\bm\beta_p^{(n)}\right]
    &=
    \exp\left[-\frac{\alpha_pL}{2}+i\Delta_pT_p\right]
    \begin{pmatrix}
        \Phi_{p,+}^{(n)} & 0 \\
        0 & \Phi_{p,-}^{(n)}
    \end{pmatrix}
    \bm\beta_p^{(n)},
    \\
    \Phi_{p,\sigma}^{(n)}
    &=
    \exp\left[i\gamma_\mathrm{SPM}L_\mathrm{eff}
    \left(\bm\beta_p^{(n)}\right)^\dagger
    \left(\frac{3}{2}\bm I_2 -\frac{\sigma}{2}\bm Z\right)\bm\beta_p^{(n)}\right],
    \end{align}
\end{subequations}
where $\bm Z = \mathrm{diag}(+1, -1)$ is Pauli-$z$ matrix.

\paragraph{Temporal coupled-mode equations for dual pumps.}
Although a coupled Ikeda map is naturally defined for counterpropagating components of the same pump, two pumps with different group velocities generally have different round-trip times and therefore do not share a common round-trip index. The pump solver in \texttt{SingleRing\allowbreak{}\_\allowbreak{}FourSideBand\allowbreak{}\_\allowbreak{}Homogeneous} describes two co-propagating pump fields, denoted by $p_1$ and $p_2$, to support BSFWM and dual-pump SFWM processes. In the
retarded-time frame of pump $p_m$, their single-pass propagation is governed by
\begin{equation}
    \left[\partial_z + \alpha_{p_m}/2 - i\gamma_{\mathrm{SPM},m}\left|\tilde\beta_{p_m}(z,t)\right|^2 - 2i\gamma_{\mathrm{XPM},mn}\left|\tilde\beta_{p_n}(z, t-\delta_{nm}z)\right|^2\right]\tilde\beta_{p_m}(z,t) = 0
    \label{eq:nonlin_dual_pump_dynamcis}
\end{equation}
where $m,n\in{1,2}$ and $\delta_{nm}=v_{p_n}^{-1}-v_{p_m}^{-1}$ accounts for the group-velocity mismatch between the pumps.

When both pumps are resonant, the round-trip dynamics are described using the mean-field limit of the corresponding coupled Ikeda maps. The resulting temporal coupled-mode equations are
\begin{equation}
    \begin{aligned}
        \partial_t \tilde\beta_{p_m}(0,t) = - \left[
        \kappa_{\mathrm{tot}, p_m} - i\left(\Delta_{p_m} 
        +
        \cfrac{\gamma_{\mathrm{SPM},m}L}{T_{p_m}}\left|\tilde\beta_{p_m}\right|^2
        +
        \cfrac{2\gamma_{\mathrm{XPM},mn}L}{T_{p_m}}\left|\tilde\beta_{p_n}\right|^2
        \right)
        \right]\tilde\beta_{p_m}
        +
        i\cfrac{\rho_{p_m}}{T_{p_m}}\tilde\beta_{p_m}^\mathrm{in}(t).
    \end{aligned}
    \label{eq:nonlin_dual_pump_tcmt}
\end{equation}
Here, $\kappa_{\mathrm{tot},p_m}$ is the total amplitude-decay rate arising from intrinsic propagation loss and bus--ring coupling, written as
\begin{equation}
    \kappa_{\mathrm{tot},p_m}
    =
    \frac{
    \alpha_{p_m}L/2-\ln|\tau_{p_m}|
    }{T_{p_m}}.
\label{eq:nonlin_kappa_tot}
\end{equation}

\subsection{Frequency-resolved nonlinear kernels}
\label{subsec:nonlinear-kernels}

Once the classical pump fields have been obtained, the corresponding pump
operators in the nonlinear polarization are replaced by their coherent
amplitudes. Retaining only terms that are linear in the weak quantum fields
gives
\begin{equation}
    \sigma_j\partial_z\hat\psi_j(z,\omega)
    =
    i\sum_l\int_{\mathfrak B_l}d\omega'\,
    \left[
        \mathcal K_{jl}(z;\omega,\omega')
        \hat\psi_l(z,\omega')
        +
        \mathcal G_{jl}(z;\omega,\omega')
        \hat\psi_l^\dagger(z,\omega')
    \right].
    \label{eq:nonlin_continuous_kernel_action}
\end{equation}
Here, the normal kernel $\mathcal K_{jl}$ describes XPM or coherent
frequency conversion, whereas the anomalous kernel $\mathcal G_{jl}$
describes pair generation. Which of these kernels appears is determined by
the operator basis selected by each module.

The nonlinear kernels depend on the pump fields obtained using the pump solvers described in Subsection~\ref{subsec:nonlinear-pump}. The pump-dependent spectra entering the implemented kernels are
\begin{subequations}
\label{eq:nonlin_continuous_pump_spectra}
    \begin{align}
        \mathcal E_{m\sigma_m}(z;\Omega)
        &=
        \int dt\,
        \left|\tilde\beta_{p_m,\sigma_m}(z,t)\right|^2e^{i\Omega t},
        \\
        \mathcal B_{m\sigma_m}(z;\Omega)
        &=
        \frac{1}{\sqrt{2\pi}}
        \int dt\,
        \tilde\beta_{p_m,\sigma_m}(z,t)e^{i\Omega t}
        =\beta_{p_m,\sigma_m}
        (z,\bar\omega_{p_m}+\Omega),
        \\
        \mathcal P_{mn}(z;\Omega)
        &=
        \int dt\,
        \tilde\beta_{p_m}(z,t)
        \tilde\beta_{p_n}
        \!\left(z,t-\Delta T_{nm}(z)\right)e^{i\Omega t},
        \\
        \mathcal P_{mm,\sigma_m}(z;\Omega)
        &=
        \int dt\,
        \tilde\beta_{p_m,\sigma_m}^{\,2}(z,t)e^{i\Omega t},
        \\
        \mathcal C_{mn}(z;\Omega)
        &=
        \int dt\,
        \tilde\beta_{p_m}^{*}(z,t)
        \tilde\beta_{p_n}
        \!\left(z,t-\Delta T_{nm}(z)\right)e^{i\Omega t}.
    \end{align}
\end{subequations}
Thus, $\mathcal E$, $\mathcal B$, $\mathcal P$, and $\mathcal C$ supply the
pump factors for XPM, $\chi^{(2)}$ interactions, SFWM, and BSFWM,
respectively. For the dual-pump ring,
$\Delta T_{nm}(z)=(v_{p_n}^{-1}-v_{p_m}^{-1})z$, consistent with the pump
walk-off defined in the preceding subsection.

Table~\ref{tab:nonlin_quantum_kernel_summary} lists the resonant nonlinear
polarization retained for each process and the corresponding
continuous-frequency kernel. The arguments $(z;\omega,\omega')$ of the kernel and phase symbols are suppressed in the table. Tensor indices and transverse-mode profiles are suppressed in the polarization column; their contraction is included in the effective coefficient $\gamma_\rho$. We write $\Omega_j=\omega-\bar\omega_j$ and $\Omega_l'=\omega'-\bar\omega_l$, and define
\begin{equation}
    K_\rho(z)
    =
    \int_{z_0}^{z}d\zeta\,
    \Delta\bar k_\rho(\zeta).
\end{equation}
The carrier-wave mismatches entering $K_\rho(z)$ are
\begin{subequations}
\label{eq:nonlin_carrier_mismatches}
\begin{align}
    \Delta\bar k_{\mathrm{pd}}
    &=
    \bar k_p-\sigma_s\bar k_s-\sigma_i\bar k_i,
    \\
    \Delta\bar k_{\mathrm{fc}}
    &=
    \bar k_p+\sigma_s\bar k_s-\sigma_i\bar k_i,
    \\
    \Delta\bar k_{\mathrm{sp},m}
    &=
    2\sigma_m\bar k_{p_m}
    -\sigma_s\bar k_s-\sigma_i\bar k_i,
    \\
    \Delta\bar k_{\mathrm{dp}}
    &=
    \bar k_{p_1}+\bar k_{p_2}-\bar k_s-\bar k_i,
    \\
    \Delta\bar k_{\mathrm{bs}}
    &=
    \bar k_{p_2}-\bar k_{p_1}+\bar k_s-\bar k_i.
\end{align}
\end{subequations}

To state the relations between the two orderings of each kernel, define the
transpose and adjoint of a continuous-frequency kernel as
\begin{equation}
    A^{\mathsf T}(z;\omega,\omega')
    =
    A(z;\omega',\omega),
    \qquad
    A^{\ddagger}(z;\omega,\omega')
    =
    A^{*}(z;\omega',\omega).
    \label{eq:nonlin_integral_kernel_adjoint}
\end{equation}
The pair-generation and frequency-conversion kernels then satisfy
\begin{equation}
    \mathcal G_{is}^{(\rho)}
    =
    \left(\mathcal G_{si}^{(\rho)}\right)^{\mathsf T},
    \qquad
    \mathcal K_{si}^{(\rho)}
    =
    \left(\mathcal K_{is}^{(\rho)}\right)^{\ddagger}.
    \label{eq:nonlin_kernel_partner_relations}
\end{equation}
The first relation contains a transpose but no complex conjugation because
$\mathcal G_{si}$ and $\mathcal G_{is}$ describe the same pair-creation
amplitude with the two bosonic operators exchanged. By contrast, reversing
a frequency-conversion process conjugates the pump-assisted transition
amplitude, giving the adjoint relation between $\mathcal K_{is}$ and
$\mathcal K_{si}$. The same relations apply independently to every
forward--backward direction pair in the Fabry--P\'erot modules.

For a quantum channel $(j,\sigma_j)$, the total XPM kernel is obtained by
summing over all pump modes and propagation directions present, $\mathcal X_{j\sigma_j} = \sum_{m,\sigma_m} \mathcal K_{j\sigma_j}^{(\mathrm{x};m\sigma_m)}$. The signed poling profile of a $\chi^{(2)}$ device is included in
$\gamma_{\mathrm{pd}}(z)$ or $\gamma_{\mathrm{fc}}(z)$, while
$\Phi_\rho$ contains the accumulated carrier-wave mismatch and temporal
walk-off among the participating modes.

In addition to the pump-dependent XPM kernel, some modules retain a
mode-diagonal linear contribution. We write the complete diagonal block as
\begin{subequations}
    \begin{align}
    \mathcal H_{j\sigma_j}(z;\omega,\omega')
    &=
    \mathcal D_{j\sigma_j}(z;\omega,\omega')
    +
    \mathcal X_{j\sigma_j}(z;\omega,\omega'),
    \\
    \mathcal D_{j\sigma_j}(z;\omega,\omega')
    &=
    d_{j\sigma_j}(z,\omega)\delta(\omega-\omega').
    \end{align}
    \label{eq:nonlin_diagonal_kernel}
\end{subequations}
Here, $d_{j\sigma_j}(z,\omega)$ contains the residual linear detuning and
first-order group-velocity mismatch of channel $(j,\sigma_j)$ relative to the
reference pump frame. Linear phase contributions already included in
$\Phi_\rho$ are not repeated in $\mathcal D_{j\sigma_j}$.

The vector form of the continuous-frequency propagation equation is
\begin{equation}
    \partial_z\hat{\bm\psi}_{\nu}(z,\omega)
    =
    i\int d\omega'\,
    \bm Q_{\nu}(z;\omega,\omega')
    \hat{\bm\psi}_{\nu}(z,\omega'),
    \label{eq:nonlin_operator_evolution_general}
\end{equation}
where the integration is taken over the frequency window associated with
each column operator. The module-specific operator basis $\hat{\bm \psi}_\nu$ and generator
$\bm Q_\nu$ are summarized in
Table~\ref{tab:nonlin_continuous_operator_bases}.

The two-mode squeezing and frequency-conversion generators differ in their
operator structure. Pair generation couples annihilation and creation
operators and therefore produces the signed
$\mathcal G$--$\mathcal G^{\ddagger}$ blocks. Frequency conversion acts
within the annihilation sector and produces the
$\mathcal K$--$\mathcal K^{\ddagger}$ blocks.

For the direction-resolved generators, both propagation directions are
expressed using the same global coordinate $z$. The signs of the backward
rows in Table~\ref{tab:nonlin_continuous_operator_bases} therefore follow
from the factor $\sigma_j$ in Eq.~\eqref{eq:nonlin_continuous_kernel_action}. The Fabry--P\'erot SPDC module retains all four combinations $(\sigma_s,\sigma_i)\in\{+,-\}^2$. The ring-backscattering module uses the same operator ordering but sets $\mathcal G_{+-}=\mathcal G_{-+}=0$ during nonlinear propagation. Its forward--backward quantum mixing is introduced by the four-port boundary condition.

The four-sideband generator combines two BSFWM, two single-pump SFWM, and
two dual-pump SFWM pathways. A block is included only when both participating
quantum modes are enabled and its nonlinear coefficient is nonzero. These
continuous-frequency kernels are discretized and assembled into cell
propagators in Subsection~\ref{subsec:nonlinear-cell-propagation}.

\begin{table}[!htbp]
    \centering
    \caption{Nonlinear polarizations, pump-dependent kernels, and accumulated
    phases implemented in PhoQuS. For pair generation,
    $(\omega,\omega')=(\omega_s,\omega_i)$; for frequency conversion,
    $(\omega,\omega')=(\omega_i,\omega_s)$.}
    \label{tab:nonlin_quantum_kernel_summary}

    \begingroup
    \scriptsize
    \setlength{\tabcolsep}{2.8pt}
    \renewcommand{\arraystretch}{1.20}
    \renewcommand{\tabularxcolumn}[1]{m{#1}}

    \begin{tabularx}{\textwidth}{
        @{}
        >{\raggedright\arraybackslash}m{0.20\textwidth}
        >{\centering\arraybackslash}m{0.24\textwidth}
        >{\centering\arraybackslash}m{0.31\textwidth}
        >{\centering\arraybackslash}X
        @{}
    }
        \toprule
        \textbf{Process}
        &
        \textbf{Nonlinear polarization}
        &
        \textbf{Pump-dependent kernel}
        &
        \textbf{Accumulated phase}
        \\
        \midrule

        XPM
        \newline
        $\bar\omega_{p_m}+\bar\omega_j=\bar\omega_{p_m}+\bar\omega_j$
        &
        $\displaystyle
        \begin{aligned}
        \left.\hat P_{\mathrm{NL}}^{(+)}\right|_j
        &=6\epsilon_0\chi^{(3)}
        \\
        &\quad\times
        E_{p_m}^{(-)}E_{p_m}^{(+)}\hat E_j^{(+)}
        \end{aligned}$
        &
        $\displaystyle
        \begin{aligned}
        \mathcal K_{j\sigma_j}^{(\mathrm{x};m\sigma_m)}
        &=
        \frac{2\gamma_{\mathrm{XPM},jm}}{2\pi}
        \\
        &\quad\times
        \mathcal E_{m\sigma_m}
        (z;\Omega_j-\Omega_j')
        \\
        &\quad\times e^{i\Phi_{\mathrm{x},j}^{(m)}}
        \end{aligned}$
        &
        $\displaystyle
        \begin{aligned}
        \Phi_{\mathrm{x},j}^{(m)}
        &=
        (\Omega_j-\Omega_j')
        \\
        &\quad\times
        (\sigma_mT_{p_m}-\sigma_jT_j)
        \end{aligned}$
        \\

        \addlinespace[0.6em]

        SPDC
        \newline
        $\bar\omega_p=\bar\omega_s+\bar\omega_i$
        &
        $\displaystyle
        \begin{aligned}
        \left.\hat P_{\mathrm{NL}}^{(+)}\right|_s
        &=2\epsilon_0\chi^{(2)}
        \\
        &\quad\times E_p^{(+)}\hat E_i^{(-)}
        \end{aligned}$
        &
        $\displaystyle
        \begin{aligned}
        \mathcal G_{si}^{(\mathrm{pd})}
        &=
        \frac{\gamma_{\mathrm{pd}}(z)}{\sqrt{2\pi}}
        \\
        &\quad\times
        \mathcal B_p(z;\Omega_s+\Omega_i')
        \\
        &\quad\times e^{i\Phi_{\mathrm{pd}}}
        \end{aligned}$
        &
        $\displaystyle
        \begin{aligned}
        \Phi_{\mathrm{pd}}
        &=
        K_{\mathrm{pd}}(z)
        \\
        &\quad+\Omega_s(T_p-\sigma_sT_s)
        \\
        &\quad+\Omega_i'(T_p-\sigma_iT_i)
        \end{aligned}$
        \\

        \addlinespace[0.6em]

        SFG and DFG
        \newline
        $\bar\omega_i=\bar\omega_s+\bar\omega_p$
        &
        $\displaystyle
        \begin{aligned}
        \left.\hat P_{\mathrm{NL}}^{(+)}\right|_i
        &=2\epsilon_0\chi^{(2)}
        \\
        &\quad\times E_p^{(+)}\hat E_s^{(+)}
        \end{aligned}$
        &
        $\displaystyle
        \begin{aligned}
        \mathcal K_{is}^{(\mathrm{fc})}
        &=
        \frac{\gamma_{\mathrm{fc}}(z)}{\sqrt{2\pi}}
        \\
        &\quad\times
        \mathcal B_p(z;\Omega_i-\Omega_s')
        \\
        &\quad\times e^{i\Phi_{\mathrm{fc}}}
        \end{aligned}$
        &
        $\displaystyle
        \begin{aligned}
        \Phi_{\mathrm{fc}}
        &=
        K_{\mathrm{fc}}(z)
        \\
        &\quad+\Omega_i(T_p-\sigma_iT_i)
        \\
        &\quad-\Omega_s'(T_p-\sigma_sT_s)
        \end{aligned}$
        \\

        \addlinespace[0.6em]

        Single-pump SFWM
        \newline
        $2\bar\omega_{p_m}=\bar\omega_s+\bar\omega_i$
        &
        $\displaystyle
        \begin{aligned}
        \left.\hat P_{\mathrm{NL}}^{(+)}\right|_s
        &=3\epsilon_0\chi^{(3)}
        \\
        &\quad\times
        E_{p_m}^{(+)}E_{p_m}^{(+)}\hat E_i^{(-)}
        \end{aligned}$
        &
        $\displaystyle
        \begin{aligned}
        \mathcal G_{si}^{(\mathrm{sp},m)}
        &=
        \frac{\gamma_{\mathrm{sp},m}}{2\pi}
        \\
        &\quad\times
        \mathcal P_{mm,\sigma_m}(z;\Omega_s+\Omega_i')
        \\
        &\quad\times e^{i\Phi_{\mathrm{sp}}^{(m)}}
        \end{aligned}$
        &
        $\displaystyle
        \begin{aligned}
        \Phi_{\mathrm{sp}}^{(m)}
        &=
        K_{\mathrm{sp},m}(z)
        \\
        &\quad+\Omega_s
        (\sigma_mT_{p_m}-\sigma_sT_s)
        \\
        &\quad+\Omega_i'
        (\sigma_mT_{p_m}-\sigma_iT_i)
        \end{aligned}$
        \\

        \addlinespace[0.6em]

        Dual-pump SFWM
        \newline
        $\bar\omega_{p_1}+\bar\omega_{p_2}=\bar\omega_s+\bar\omega_i$
        &
        $\displaystyle
        \begin{aligned}
        \left.\hat P_{\mathrm{NL}}^{(+)}\right|_s
        &=6\epsilon_0\chi^{(3)}
        \\
        &\quad\times
        E_{p_1}^{(+)}E_{p_2}^{(+)}\hat E_i^{(-)}
        \end{aligned}$
        &
        $\displaystyle
        \begin{aligned}
        \mathcal G_{si}^{(\mathrm{dp})}
        &=
        \frac{2\gamma_{\mathrm{dp}}}{2\pi}
        \\
        &\quad\times
        \mathcal P_{12}(z;\Omega_s+\Omega_i')
        \\
        &\quad\times e^{i\Phi_{\mathrm{dp}}}
        \end{aligned}$
        &
        $\displaystyle
        \begin{aligned}
        \Phi_{\mathrm{dp}}
        &=
        K_{\mathrm{dp}}(z)
        \\
        &\quad+\Omega_s(T_{p_1}-T_s)
        \\
        &\quad+\Omega_i'(T_{p_1}-T_i)
        \end{aligned}$
        \\

        \addlinespace[0.6em]

        BSFWM
        \newline
        $\bar\omega_i-\bar\omega_s=\bar\omega_{p_2}-\bar\omega_{p_1}$
        &
        $\displaystyle
        \begin{aligned}
        \left.\hat P_{\mathrm{NL}}^{(+)}\right|_i
        &=6\epsilon_0\chi^{(3)}
        \\
        &\quad\times
        E_{p_1}^{(-)}E_{p_2}^{(+)}\hat E_s^{(+)}
        \end{aligned}$
        &
        $\displaystyle
        \begin{aligned}
        \mathcal K_{is}^{(\mathrm{bs})}
        &=
        \frac{2\gamma_{\mathrm{bs}}}{2\pi}
        \\
        &\quad\times
        \mathcal C_{12}(z;\Omega_i-\Omega_s')
        \\
        &\quad\times e^{i\Phi_{\mathrm{bs}}}
        \end{aligned}$
        &
        $\displaystyle
        \begin{aligned}
        \Phi_{\mathrm{bs}}
        &=
        K_{\mathrm{bs}}(z)
        \\
        &\quad+\Omega_i(T_{p_1}-T_i)
        \\
        &\quad-\Omega_s'(T_{p_1}-T_s)
        \end{aligned}$
        \\

        \bottomrule
    \end{tabularx}

    \vspace{0.45em}
    \noindent
    \parbox{\textwidth}{
        \scriptsize
        \textit{Notation.}
        The pump is forward propagating in the current $\chi^{(2)}$
        modules, so that $\mathcal B_p\equiv\mathcal B_{p,+}$.
        Fixed propagation-direction labels are omitted, while those
        required for bidirectional propagation are retained explicitly
        through $\sigma_j$. The table displays
        $\mathcal G_{si}$ for pair generation and $\mathcal K_{is}$ for
        frequency conversion. Their opposite-order kernels are not
        independent and are determined by Eq.~\eqref{eq:nonlin_kernel_partner_relations}. For forward-only modules,
        $\mathcal P_{mm} = \mathcal P_{mm,+}$.
    }
    
    \endgroup
\end{table}

\begin{table}[!htbp]
    \centering
    \caption{Continuous-frequency operator bases and generator structures
    used by the implemented nonlinear modules. Frequency arguments
    $(z;\omega,\omega')$ of the matrix elements are suppressed.}
    \label{tab:nonlin_continuous_operator_bases}

    \begingroup
    \scriptsize
    \setlength{\tabcolsep}{3.0pt}
    \setlength{\arraycolsep}{2.0pt}
    \renewcommand{\arraystretch}{1.15}
    \renewcommand{\tabularxcolumn}[1]{m{#1}}

    \begin{tabularx}{\textwidth}{
        @{}
        >{\raggedright\arraybackslash}m{0.27\textwidth}
        >{\centering\arraybackslash}m{0.20\textwidth}
        >{\centering\arraybackslash}X
        @{}
    }
        \toprule
        \textbf{Implementation}
        &
        \textbf{Continuous operator basis}
        &
        \textbf{Generator $\bm Q_\nu$}
        \\
        \midrule

        \texttt{TwoModeSPDC}
        \newline
        \texttt{SingleRingSPSFWM\_\allowbreak{}Homogeneous}
        &
        $\displaystyle
        \hat{\bm\psi}_{\mathrm{sq}}
        =
        \begin{pmatrix}
            \hat\psi_s\\
            \hat\psi_i^\dagger
        \end{pmatrix}$
        &
        $\displaystyle
        \bm Q_{\mathrm{sq}}
        =
        \begin{pmatrix}
            \mathcal H_s
            &
            \mathcal G_{si}
            \\
            -\mathcal G_{si}^{\ddagger}
            &
            -\mathcal H_i^{*}
        \end{pmatrix}$
        \\

        \addlinespace[0.6em]

        \texttt{TwoModeQFC}
        &
        $\displaystyle
        \hat{\bm\psi}_{\mathrm{fc}}
        =
        \begin{pmatrix}
            \hat\psi_s\\
            \hat\psi_i
        \end{pmatrix}$
        &
        $\displaystyle
        \bm Q_{\mathrm{fc}}
        =
        \begin{pmatrix}
            \mathcal H_s
            &
            \mathcal K_{is}^{\ddagger}
            \\
            \mathcal K_{is}
            &
            \mathcal H_i
        \end{pmatrix}$
        \\

        \addlinespace[0.6em]

        \texttt{FabryPerotSPDC\_\allowbreak{}SinglePassPump}
        \newline
        \texttt{SingleRingSPSFWM\_\allowbreak{}Backscattering}
        &
        $\displaystyle
        \hat{\bm\psi}_{\mathrm{dir,sq}}
        =
        \begin{pmatrix}
            \hat\psi_{s+}\\
            \hat\psi_{i+}^{\dagger}\\
            \hat\psi_{s-}\\
            \hat\psi_{i-}^{\dagger}
        \end{pmatrix}$
        &
        $\displaystyle
        \bm Q_{\mathrm{dir,sq}}
        =
        \begin{pmatrix}
            \mathcal H_{s+}
            &
            \mathcal G_{++}
            &
            0
            &
            \mathcal G_{+-}
            \\
            -\mathcal G_{++}^{\ddagger}
            &
            -\mathcal H_{i+}^{*}
            &
            -\mathcal G_{-+}^{\ddagger}
            &
            0
            \\
            0
            &
            -\mathcal G_{-+}
            &
            -\mathcal H_{s-}
            &
            -\mathcal G_{--}
            \\
            \mathcal G_{+-}^{\ddagger}
            &
            0
            &
            \mathcal G_{--}^{\ddagger}
            &
            \mathcal H_{i-}^{*}
        \end{pmatrix}$
        \\

        \addlinespace[0.6em]

        \texttt{FabryPerotQFC\_\allowbreak{}SinglePassPump}
        &
        $\displaystyle
        \hat{\bm\psi}_{\mathrm{dir,fc}}
        =
        \begin{pmatrix}
            \hat\psi_{s+}\\
            \hat\psi_{i+}\\
            \hat\psi_{s-}\\
            \hat\psi_{i-}
        \end{pmatrix}$
        &
        $\displaystyle
        \bm Q_{\mathrm{dir,fc}}
        =
        \begin{pmatrix}
            \mathcal H_{s+}
            &
            \mathcal K_{++}^{\ddagger}
            &
            0
            &
            \mathcal K_{-+}^{\ddagger}
            \\
            \mathcal K_{++}
            &
            \mathcal H_{i+}
            &
            \mathcal K_{+-}
            &
            0
            \\
            0
            &
            -\mathcal K_{+-}^{\ddagger}
            &
            -\mathcal H_{s-}
            &
            -\mathcal K_{--}^{\ddagger}
            \\
            -\mathcal K_{-+}
            &
            0
            &
            -\mathcal K_{--}
            &
            -\mathcal H_{i-}
        \end{pmatrix}$
        \\

        \addlinespace[0.6em]

        \texttt{SingleRing\_\allowbreak{}FourSideBand%
        \_\allowbreak{}Homogeneous}
        &
        $\displaystyle
        \hat{\bm\psi}_{4\mathrm{sb}}
        =
        \begin{pmatrix}
            \hat\psi_{s1}\\
            \hat\psi_{i1}\\
            \hat\psi_{s2}^{\dagger}\\
            \hat\psi_{i2}^{\dagger}
        \end{pmatrix}$
        &
        $\displaystyle
        \bm Q_{4\mathrm{sb}}
        =
        \begin{pmatrix}
            \mathcal H_{s1}
            &
            \mathcal K_{\mathrm{bs},1}^{\ddagger}
            &
            \mathcal G_{\mathrm{sp},1}
            &
            \mathcal G_{\mathrm{dp},1}
            \\
            \mathcal K_{\mathrm{bs},1}
            &
            \mathcal H_{i1}
            &
            \mathcal G_{\mathrm{dp},2}
            &
            \mathcal G_{\mathrm{sp},2}
            \\
            -\mathcal G_{\mathrm{sp},1}^{\ddagger}
            &
            -\mathcal G_{\mathrm{dp},2}^{\ddagger}
            &
            -\mathcal H_{s2}^{*}
            &
            -\mathcal K_{\mathrm{bs},2}^{\mathsf T}
            \\
            -\mathcal G_{\mathrm{dp},1}^{\ddagger}
            &
            -\mathcal G_{\mathrm{sp},2}^{\ddagger}
            &
            -\mathcal K_{\mathrm{bs},2}^{*}
            &
            -\mathcal H_{i2}^{*}
        \end{pmatrix}$
        \\

        \bottomrule
    \end{tabularx}

    \vspace{0.45em}
    \noindent
    \parbox{\textwidth}{
        \scriptsize
        \textit{Compact kernel notation.}
        For the direction-resolved squeezing generator,
        $\mathcal G_{\sigma_s\sigma_i}
        \equiv
        \mathcal G_{s\sigma_s,i\sigma_i}^{(\rho)}$, where
        $\rho=\mathrm{pd}$ for
        \texttt{FabryPerotSPDC\_\allowbreak{}SinglePassPump} and
        $\rho=\mathrm{sp}$ for
        \texttt{SingleRingSPSFWM\_\allowbreak{}Backscattering}.
        For direction-resolved frequency conversion,
        $\mathcal K_{\sigma_i\sigma_s}
        \equiv
        \mathcal K_{i\sigma_i,s\sigma_s}^{(\mathrm{fc})}$.
        In the four-sideband generator,
        $\mathcal K_{\mathrm{bs},a}
        \equiv\mathcal K_{i a,s a}^{(\mathrm{bs})}$ for
        $a\in\{1,2\}$,
        $\mathcal G_{\mathrm{sp},1}
        \equiv\mathcal G_{s1,s2}^{(\mathrm{sp},1)}$,
        $\mathcal G_{\mathrm{sp},2}
        \equiv\mathcal G_{i1,i2}^{(\mathrm{sp},2)}$,
        $\mathcal G_{\mathrm{dp},1}
        \equiv\mathcal G_{s1,i2}^{(\mathrm{dp})}$, and
        $\mathcal G_{\mathrm{dp},2}
        \equiv\mathcal G_{i1,s2}^{(\mathrm{dp})}$.
    }

    \endgroup
\end{table}

\subsection{Frequency discretization and cell propagation}
\label{subsec:nonlinear-cell-propagation}

The continuous-frequency kernels introduced in
Subsection~\ref{subsec:nonlinear-kernels} are converted into finite-dimensional
matrices before propagation. PhoQuS then divides the nonlinear medium into
longitudinal cells, constructs a local propagator for each cell, and
interconnects the cell maps in their physical propagation order. Once the
operator basis of a module has been selected, its basis label is suppressed
throughout this subsection.

\paragraph{Frequency discretization.}

All mutually coupled quantum bands are discretized on uniform
angular-frequency grids with a common spacing $\Delta\omega$. For
$\omega_{j,n}=\omega_{j,0}+n\Delta\omega$, where
$n=0,\ldots,N_j-1$, the discretized frequency-domain operator is defined by
\begin{equation}
    \hat\Psi_{j,n}(z)
    =
    \hat\psi_j(z,\omega_{j,n}).
    \label{eq:nonlin_discrete_psi_operators}
\end{equation}
The discrete vector $\hat{\bm\Psi}$ is assembled from these operators in the
same mode ordering and, where applicable, Nambu ordering as the corresponding
continuous vector in Table~\ref{tab:nonlin_continuous_operator_bases}.

For a nonsingular continuous kernel, the frequency integral is replaced by
the rectangular quadrature
\begin{equation}
\label{eq:nonlin_frequency_quadrature}
    \begin{aligned}
        &\left.
        \int_{\mathfrak B_l}d\omega'\,
        \mathcal Q_{jl}(z;\omega,\omega')
        \hat\psi_l(z,\omega')
        \right|_{\omega=\omega_{j,n}}
        \\
        &\longrightarrow
        \sum_{m=0}^{N_l-1}
        \Delta\omega\,
        \mathcal Q_{jl}
        (z;\omega_{j,n},\omega_{l,m})
        \hat\Psi_{l,m}(z)=
        \left[
            \bm Q_{jl}(z)\hat{\bm\Psi}_l(z)
        \right]_n.
    \end{aligned}
\end{equation}
Consequently, a regular kernel contributes
\begin{equation}
    \left[\bm Q_{jl}(z)\right]_{nm}
    =
    \Delta\omega\,
    \mathcal Q_{jl}
    (z;\omega_{j,n},\omega_{l,m})
    \label{eq:nonlin_discrete_kernel}
\end{equation}
to the discrete generator. A term proportional to
$d_j(z,\omega)\delta(\omega-\omega')$, such as the residual linear kernel in
Eq.~\eqref{eq:nonlin_diagonal_kernel}, instead becomes
$d_j(z,\omega_{j,n})\delta_{nm}$ and does not acquire an additional factor of
$\Delta\omega$.

\paragraph{Ordered cell propagation.}

Let the nonlinear region be divided as
$0=z_0<z_1<\cdots<z_{N_z}=L$, with
$C_k=[z_k,z_{k+1}]$ and
$\Delta z_k=z_{k+1}-z_k$. The cell midpoint is
$\bar z_k=(z_k+z_{k+1})/2$. Denoting the discretized generator by
$\bm Q(z)$, the propagator within cell $C_k$ is defined by
\begin{equation}
    \partial_z\bm U(z,z_k)
    =
    i\bm Q(z)\bm U(z,z_k).
    \label{eq:nonlin_cell_propagator}
\end{equation}
The propagator satisfies
$\bm U(z_k,z_k)=\bm I$. The local cell propagator is defined as
$\bm U_k=\bm U(z_{k+1},z_k)$. The one-pass propagation matrix is obtained by
interconnecting the cell propagators in their physical propagation order,
\begin{subequations}
\label{eq:nonlin_discrete_ordered_propagator}
    \begin{align}
        \hat{\bm\Psi}(L)
        &=
        \bm U(L,0)\hat{\bm\Psi}(0),
        \\
        \bm U(L,0)
        &=
        \bm U_{N_z-1}
        \cdots
        \bm U_1
        \bm U_0.
    \end{align}
\end{subequations}
This ordered product retains the noncommuting evolution between different
longitudinal cells. The construction of each local propagator $\bm U_k$ is
described below.

\paragraph{Cell-integrated propagation.}

When the discretized generator remains longitudinally dependent, PhoQuS
constructs the local propagator by integrating the generator separately
within each cell. Using the first-order Magnus approximation~\cite{Blanes2009Magnus}, the integrated
generator and cell propagator are
\begin{subequations}
\label{eq:nonlin_cell_magnus_propagator}
    \begin{align}
        \bm A_k
        &=
        \int_{z_k}^{z_{k+1}}dz\,
        \bm Q(z),
        \\
        \bm U_k
        &\simeq
        \exp\left(i\bm A_k\right).
    \end{align}
\end{subequations}
This approximation neglects longitudinal noncommutation within a single
cell. The matrix exponential itself, however, retains the nonlinear
interaction nonperturbatively in the integrated gain. Noncommutation between
different cells is retained through the ordered product in
Eq.~\eqref{eq:nonlin_discrete_ordered_propagator}.

The direct evaluation of $\bm A_k$ would require constructing the full
frequency-resolved generator at every longitudinal integration point.
PhoQuS avoids this operation by separating each nonlinear block into a
pump-dependent spectral factor and a rapidly varying longitudinal factor.
We use $\rho\in\{\mathrm{x},\mathrm{pd},\mathrm{fc},\mathrm{sp},\mathrm{dp}, \mathrm{bs}\}$ to label the nonlinear process. Mode and propagation-direction labels
associated with each block are suppressed unless required explicitly.

A nonsingular discrete block associated with process $\rho$ can be written
in the generic form
\begin{equation}
    \left[
        \bm Q^{(\rho)}(z)
    \right]_{nm}
    =
    \Delta\omega\,
    c_\rho\,
    \mathcal F_\rho
    \left(
        z;\Omega_\rho^{nm}
    \right)
    \gamma_\rho(z)
    \exp\left[
        i\Phi_\rho
        (z;\omega_{j,n},\omega_{l,m})
    \right].
    \label{eq:nonlin_kernel_block_factorization}
\end{equation}
Here, $c_\rho$ contains the numerical and permutation factors of the
interaction. The pump-dependent spectrum $\mathcal F_\rho$ is selected from
$\mathcal E$, $\mathcal B$, $\mathcal P$, and $\mathcal C$ in
Eq.~\eqref{eq:nonlin_continuous_pump_spectra}, and
$\Omega_\rho^{nm}$ denotes its corresponding sum- or difference-frequency
argument. The coefficient $\gamma_\rho(z)$ is the effective nonlinear
coupling and includes the signed poling profile where applicable.
The phase $\Phi_\rho$ contains the accumulated carrier mismatch and
frequency-dependent walk-off.

The corresponding block of the integrated generator is
\begin{equation}
    \left[
        \bm A_k^{(\rho)}
    \right]_{nm}
    =
    \Delta\omega\,
    c_\rho
    \int_{z_k}^{z_{k+1}}dz\,
    \mathcal F_\rho
    \left(
        z;\Omega_\rho^{nm}
    \right)
    \gamma_\rho(z)
    \exp\left[
        i\Phi_\rho
        (z;\omega_{j,n},\omega_{l,m})
    \right].
    \label{eq:nonlin_exact_integrated_block}
\end{equation}
If the pump-dependent spectrum varies slowly within cell $C_k$, it is
evaluated at a representative position
$z_k^{\mathrm{ref}}\in C_k$. Equation~\eqref{eq:nonlin_exact_integrated_block}
then reduces to
\begin{equation}
    \left[
        \bm A_k^{(\rho)}
    \right]_{nm}
    \simeq
    \Delta\omega\,
    c_\rho\,
    \mathcal F_\rho
    \left(
        z_k^{\mathrm{ref}};
        \Omega_\rho^{nm}
    \right)
    \mathcal I_{\rho,k}^{nm},
    \label{eq:nonlin_cell_integrated_block}
\end{equation}
where
\begin{equation}
    \mathcal I_{\rho,k}^{nm}
    =
    \int_{z_k}^{z_{k+1}}dz\,
    \gamma_\rho(z)
    \exp\left[
        i\Phi_\rho
        (z;\omega_{j,n},\omega_{l,m})
    \right].
    \label{eq:nonlin_scalar_kernel_integral}
\end{equation}
When the pump-dependent spectrum varies appreciably within the cell,
Eq.~\eqref{eq:nonlin_exact_integrated_block} is evaluated without extracting
$\mathcal F_\rho$ from the longitudinal integral.

A separate moment expansion is used when the signed nonlinear coefficient
or carrier mismatch varies rapidly. The accumulated phase is first separated
as
\begin{equation}
    \Phi_\rho
    (z;\omega_{j,n},\omega_{l,m})
    =
    K_\rho(z)
    +
    \varphi_\rho
    (z;\omega_{j,n},\omega_{l,m}),
    \label{eq:nonlin_scalar_phase_separation}
\end{equation}
where $K_\rho(z)$ is the accumulated carrier-wave mismatch and
$\varphi_\rho(z;\omega_{j,n},\omega_{l,m})$ contains the frequency-dependent
walk-off. The rapidly varying longitudinal factor is defined by
\begin{equation}
    h_\rho(z)
    =
    \gamma_\rho(z)
    e^{iK_\rho(z)}.
    \label{eq:nonlin_rapid_longitudinal_factor}
\end{equation}
Its zeroth- and first-order moments about the cell midpoint are
\begin{equation}
    M_{q,k}^{(\rho)}
    =
    \int_{z_k}^{z_{k+1}}dz\,
    (z-\bar z_k)^q
    h_\rho(z),
    \label{eq:nonlin_scalar_phase_moments}
\end{equation}
where $q=0,1$.

Expanding only the slowly varying walk-off phase to first order gives
\begin{equation}
    \mathcal I_{\rho,k}^{nm}
    \simeq
    e^{i\bar\varphi_k^{nm}}
    \left[
        M_{0,k}^{(\rho)}
        +
        i\varphi_k^{\prime\,nm}
        M_{1,k}^{(\rho)}
    \right].
    \label{eq:nonlin_scalar_moment_approximation}
\end{equation}
The cell-averaged walk-off phase and longitudinal slope are
\begin{subequations}
    \label{eq:nonlin_cell_average_walkoff_phase}
    \begin{align}
    \bar\varphi_k^{nm}
    &=
    \frac{
        \varphi_\rho
        (z_k;\omega_{j,n},\omega_{l,m})
        +
        \varphi_\rho
        (z_{k+1};\omega_{j,n},\omega_{l,m})
    }{2}, \\
    \varphi_k^{\prime\,nm}
    &=
    \frac{
        \varphi_\rho
        (z_{k+1};\omega_{j,n},\omega_{l,m})
        -
        \varphi_\rho
        (z_k;\omega_{j,n},\omega_{l,m})
    }{\Delta z_k}.
    \end{align}
\end{subequations}
The rapidly varying signed poling profile and carrier-wave mismatch are thus
evaluated only through the scalar moments
$M_{0,k}^{(\rho)}$ and $M_{1,k}^{(\rho)}$. These moments are independent of
the frequency-grid indices $n$ and $m$ and can therefore be reused throughout
the corresponding frequency-resolved block. The remaining frequency
dependence is restored through
$\bar\varphi_k^{nm}$ and $\varphi_k^{\prime\,nm}$. This construction avoids
repeatedly assembling and exponentiating the full frequency-resolved
generator on the fine longitudinal grid required to resolve rapid poling or
carrier-phase oscillations.

\paragraph{Optical loss.}

Optical attenuation cannot be described by a closed propagation matrix acting
only on the retained optical modes, because the resulting nonunitary
transformation does not preserve their canonical commutation relations.
Following Loudon's traveling-wave reservoir picture~\cite{Jeffers1993TravelingWaveLoss}, distributed optical loss
is represented by local beam-splitter coupling to independent vacuum modes.

For the optical operator vector used throughout the nonlinear modules, the
grid-weighted commutator signature is
\begin{equation}
    \bm J
    =
    \operatorname{diag}_{\mu}
    \left(s_\mu\bm I_\mu\right).
    \label{eq:nonlin_optical_commutator_signature}
\end{equation}
Here, $s_\mu=+1$ for an annihilation-operator sector and $s_\mu=-1$ for a
creation-operator sector. Thus, $\bm J=\bm I$ for an all-annihilation basis,
whereas each creation-operator sector contributes a negative block in a
Nambu basis.

In PhoQuS, selected longitudinal cells are tagged as \texttt{lossy} cells.
A loss tag at cell $k$ represents the propagation distance $d_{\ell,k}$
accumulated since the preceding loss tag. For an optical mode $\mu$ with
power-attenuation coefficient $\alpha_\mu$, the local transmission and
reservoir-coupling matrices are
\begin{subequations}
\label{eq:nonlin_local_loss_matrices}
    \begin{align}
        \bm T_{\ell,k}
        &=
        \operatorname{diag}_{\mu,n}
        \left(
            e^{-\alpha_\mu d_{\ell,k}/2}
        \right),
        \\
        \bm R_{\ell,k}
        &=
        i\bm J
        \left(
            \bm I-\bm T_{\ell,k}^{2}
        \right)^{1/2}.
    \end{align}
\end{subequations}
The corresponding local dilation is
\begin{equation}
    \hat{\bm\Psi}_k^{+}
    =
    \bm T_{\ell,k}\hat{\bm\Psi}_k^{-}
    +
    \bm R_{\ell,k}\hat{\bm f}_k,
    \label{eq:nonlin_local_reservoir_dilation}
\end{equation}
where $\hat{\bm f}_k$ is an independent vacuum input and the superscripts
$-$ and $+$ indicate the optical fields immediately before and after the
loss coupling. The factor $\bm J$ in $\bm R_{\ell,k}$ supplies the opposite
beam-splitter phase for a creation-operator sector. Since
$\bm J_{f,k}=\bm J$, the local loss map satisfies
\begin{equation}
    \bm T_{\ell,k}\bm J\bm T_{\ell,k}^{\dagger}
    +
    \bm R_{\ell,k}\bm J\bm R_{\ell,k}^{\dagger}
    =
    \bm J.
    \label{eq:nonlin_local_reservoir_canonicality}
\end{equation}
Accordingly, the complete cell blocks appearing in
Eq.~\eqref{eq:common_nonlinear_map} are
\begin{subequations}
\label{eq:nonlin_local_cell_blocks}
    \begin{align}
        \bm U_{aa}^{(k)}
        &=
        \bm T_{\ell,k}\bm U_k,
        \\
        \bm U_{af}^{(k)}
        &=
        \bm R_{\ell,k}
    \end{align}
\end{subequations}
for a lossy cell. For a lossless cell,
$\bm U_{aa}^{(k)}=\bm U_k$ and the reservoir block is absent.

When a loss map is inserted after the accumulated propagation map, its
optical and reservoir blocks are updated as
\begin{subequations}
\label{eq:nonlin_cumulative_loss_update}
    \begin{align}
        \bm U_{aa}
        &\leftarrow
        \bm T_{\ell,k}\bm U_{aa},
        \\
        \bm U_{af}
        &\leftarrow
        \left[
            \bm T_{\ell,k}\bm U_{af}
            \;\;
            \bm R_{\ell,k}
        \right].
    \end{align}
\end{subequations}
Here, square brackets denote horizontal matrix concatenation; if no loss
reservoir has previously been introduced, $\bm R_{\ell,k}$ initializes
$\bm U_{af}$. In general, the concatenated reservoir vector and its
cumulative coupling block are defined by
\begin{subequations}
\label{eq:nonlin_concatenated_reservoir_blocks}
    \begin{align}
        \hat{\bm f}
        &=
        \operatorname{col}_{k\in\mathcal L_f}
        \left(\hat{\bm f}_k\right),
        \\
        \bm U_{af}
        &=
        \begin{bmatrix}
            \bm U_{af}^{(k_1)}
            &
            \bm U_{af}^{(k_2)}
            &
            \cdots
        \end{bmatrix}.
    \end{align}
\end{subequations}
The same concatenation convention is used for $\bm S_{af}$ and
$\bm M_{af}$. After all coherent cell propagators and loss maps have been
interconnected in longitudinal order, the one-pass transformation is
\begin{equation}
    \hat{\bm\Psi}(L)
    =
    \bm U_{aa}\hat{\bm\Psi}(0)
    +
    \sum_{k\in\mathcal L_f}
    \bm U_{af}^{(k)}\hat{\bm f}_k,
    \label{eq:nonlin_one_pass_reservoir_map}
\end{equation}
where $\mathcal L_f$ denotes the set of loss-coupling cells and
$\bm U_{af}^{(k)}$ maps the reservoir introduced at cell $k$ to the output
optical fields. The complete one-pass map preserves the canonical
commutation relations when
\begin{equation}
    \bm U_{aa}\bm J\bm U_{aa}^{\dagger}
    +
    \sum_{k\in\mathcal L_f}
    \bm U_{af}^{(k)}
    \bm J_{f,k}
    \left(\bm U_{af}^{(k)}\right)^{\dagger}
    =
    \bm J.
    \label{eq:nonlin_one_pass_reservoir_canonicality}
\end{equation}
Increasing the number of loss tags refines the spatial distribution of the
reservoir coupling without changing the total one-pass attenuation. For
direction-resolved propagation, the same construction is applied in the
scattering representation described in
Subsection~\ref{subsec:nonlinear-boundary-closure}.

\subsection{Bidirectional scattering and device boundary closure}
\label{subsec:nonlinear-boundary-closure}

The notation in this subsection distinguishes three levels of the calculation.
A matrix \(\bm U\) denotes a transfer map through a nonlinear propagation
region, \(\bm S\) denotes the corresponding scattering map after the
counterpropagating fields have been rearranged into incident and outgoing
ports, and \(\bm M\) denotes the external device map after the resonator
boundary has been closed. The subscripts \(aa\) and \(af\) distinguish the
optical-to-optical and reservoir-to-optical blocks, respectively.

We use \(\hat{\bm\Psi}_{\mathrm{init}}\) and
\(\hat{\bm\Psi}_{\mathrm{final}}\) for the cavity-side fields immediately
before and after one-pass propagation. The labels
\(\hat{\bm\Psi}_{\mathrm{in}}\) and
\(\hat{\bm\Psi}_{\mathrm{out}}\) are reserved for the externally incident
and emitted fields. For a bidirectional nonlinear region, the internal
vectors are ordered as
\begin{equation}
    \hat{\bm\Psi}_{\mathrm{init}}
    =
    \begin{pmatrix}
        \hat{\bm\Psi}_{+}(0)\\
        \hat{\bm\Psi}_{-}(L)
    \end{pmatrix},
    \qquad
    \hat{\bm\Psi}_{\mathrm{final}}
    =
    \begin{pmatrix}
        \hat{\bm\Psi}_{+}(L)\\
        \hat{\bm\Psi}_{-}(0)
    \end{pmatrix}.
    \label{eq:nonlin_cavity_initial_final_vectors}
\end{equation}
For a unidirectional region, these reduce to
\(\hat{\bm\Psi}_{\mathrm{init}}=\hat{\bm\Psi}(0)\) and
\(\hat{\bm\Psi}_{\mathrm{final}}=\hat{\bm\Psi}(L)\). An open waveguide
requires no further boundary closure, whereas a resonator eliminates the
internal fields to obtain \(\bm M\).

\paragraph{Transfer-to-scattering conversion.}

The direction-resolved Fabry--P\'erot modules first construct the optical
transfer map of each cell,
\begin{equation}
    \begin{pmatrix}
        \hat{\bm\Psi}_{+}(z_{k+1})\\
        \hat{\bm\Psi}_{-}(z_{k+1})
    \end{pmatrix}
    =
    \begin{pmatrix}
        \bm U_{11}^{(k)} & \bm U_{12}^{(k)}\\
        \bm U_{21}^{(k)} & \bm U_{22}^{(k)}
    \end{pmatrix}
    \begin{pmatrix}
        \hat{\bm\Psi}_{+}(z_k)\\
        \hat{\bm\Psi}_{-}(z_k)
    \end{pmatrix}.
    \label{eq:nonlin_bidirectional_transfer_map}
\end{equation}
Because the backward field is incident on the cell at \(z_{k+1}\), the
transfer map is converted to a scattering map by eliminating
\(\hat{\bm\Psi}_{-}(z_k)\). The lossless cell-scattering matrix is
\begin{equation}
    \bm S_k
    =
    \begin{pmatrix}
        \bm U_{11}^{(k)}
        -
        \bm U_{12}^{(k)}\left(\bm U_{22}^{(k)}\right)^{-1}\bm U_{21}^{(k)}
        &
        \bm U_{12}^{(k)}\left(\bm U_{22}^{(k)}\right)^{-1}
        \\
        -\left(\bm U_{22}^{(k)}\right)^{-1}\bm U_{21}^{(k)}
        &
        \left(\bm U_{22}^{(k)}\right)^{-1}
    \end{pmatrix}.
    \label{eq:nonlin_transfer_to_scattering_optical_block}
\end{equation}
At a tagged loss cell, the distributed-loss dilation is applied in this
scattering representation,
\begin{subequations}
\label{eq:nonlin_bidirectional_loss_cell}
    \begin{align}
        \bm S_{aa}^{(k)}
        &=
        \bm T_{\ell,k}\bm S_k,
        \\
        \bm S_{af}^{(k)}
        &=
        \bm R_{\ell,k}.
    \end{align}
\end{subequations}
Here, $\bm S_k$ denotes the coherent scattering matrix obtained from
$\bm U_k$, whereas $(\bm S_{aa}^{(k)},\bm S_{af}^{(k)})$ denotes the
complete cell map after the local loss dilation has been applied.
The corresponding cell relation is~\cite{PovedaHospital2026TransferMatrix}
\begin{equation}
    \begin{pmatrix}
        \hat{\bm\Psi}_{+}(z_{k+1})\\
        \hat{\bm\Psi}_{-}(z_k)
    \end{pmatrix}
    =
    \bm S_{aa}^{(k)}
    \begin{pmatrix}
        \hat{\bm\Psi}_{+}(z_k)\\
        \hat{\bm\Psi}_{-}(z_{k+1})
    \end{pmatrix}
    +
    \bm S_{af}^{(k)}\hat{\bm f}_k.
    \label{eq:nonlin_bidirectional_scattering_map}
\end{equation}
For a lossless cell, \(\bm S_{aa}^{(k)}=\bm S_k\) and the reservoir
term is absent.

The cell-scattering maps are interconnected in longitudinal order through
the Redheffer star product~\cite{Redheffer1962StarProduct},
\begin{equation}
    \bm S_{aa}
    =
    \bm S_{aa}^{(N_z-1)}
    \star\cdots\star
    \bm S_{aa}^{(1)}
    \star
    \bm S_{aa}^{(0)}.
    \label{eq:nonlin_bidirectional_scattering_cascade}
\end{equation}
The reservoir block is updated by the same internal-port elimination. The
completed bidirectional one-pass map is therefore
\begin{equation}
    \hat{\bm\Psi}_{\mathrm{final}}
    =
    \bm S_{aa}
    \hat{\bm\Psi}_{\mathrm{init}}
    +
    \bm S_{af}\hat{\bm f},
    \label{eq:nonlin_bidirectional_one_pass_map}
\end{equation}
where \(\hat{\bm f}\) concatenates the loss reservoirs introduced along the
waveguide, and $\bm S_{af}$ follows the concatenation convention in
Eq.~\eqref{eq:nonlin_concatenated_reservoir_blocks}. Thus, \(\bm U\) is used before the port rearrangement and
\(\bm S\) afterward.

\paragraph{Single-bus ring closure.}

For the unidirectional ring modules, one-pass propagation remains in transfer
form. The propagation and coupler relations are
\begin{subequations}
\label{eq:nonlin_single_bus_boundary_relations}
    \begin{align}
        \hat{\bm\Psi}_{\mathrm{final}}
        &=
        \bm U_{aa}\hat{\bm\Psi}_{\mathrm{init}}
        +
        \bm U_{af}\hat{\bm f},
        \\
        \hat{\bm\Psi}_{\mathrm{init}}
        &=
        \bm R\hat{\bm\Psi}_{\mathrm{in}}
        +
        \bm T\bm E\hat{\bm\Psi}_{\mathrm{final}},
        \\
        \hat{\bm\Psi}_{\mathrm{out}}
        &=
        \bm T\hat{\bm\Psi}_{\mathrm{in}}
        +
        \bm R\bm E\hat{\bm\Psi}_{\mathrm{final}}.
    \end{align}
\end{subequations}
Here, \(\bm T\) and \(\bm R\) contain the self- and cross-coupling
amplitudes of the bus--ring coupler, while \(\bm E\) contains the
frequency-dependent round-trip phase in the selected operator basis.
Defining the feedback inverse
\begin{equation}
    \bm C_{\mathrm{ring}}
    =
    \left(
        \bm I-
        \bm U_{aa}\bm T\bm E
    \right)^{-1},
    \label{eq:nonlin_single_bus_feedback_inverse}
\end{equation}
elimination of \(\hat{\bm\Psi}_{\mathrm{init}}\) and
\(\hat{\bm\Psi}_{\mathrm{final}}\) gives the closed device map
\begin{equation}
    \hat{\bm\Psi}_{\mathrm{out}}
    =
    \bm M_{aa}\hat{\bm\Psi}_{\mathrm{in}}
    +
    \bm M_{af}\hat{\bm f},
    \label{eq:nonlin_single_bus_external_map}
\end{equation}
with
\begin{subequations}
\label{eq:nonlin_single_bus_closed_maps}
    \begin{align}
        \bm M_{aa}
        &=
        \bm T
        +
        \bm R\bm E\bm C_{\mathrm{ring}}
        \bm U_{aa}\bm R,
        \\
        \bm M_{af}
        &=
        \bm R\bm E\bm C_{\mathrm{ring}}\bm U_{af}.
    \end{align}
\end{subequations}
The inverse \(\bm C_{\mathrm{ring}}\) sums repeated round trips~\cite{Kim2025SimulatingQuantumLight}. Thus, the
single-bus boundary converts the one-pass transfer blocks
\((\bm U_{aa},\bm U_{af})\) into the external blocks
\((\bm M_{aa},\bm M_{af})\).

\paragraph{Fabry--P\'erot closure.}

For the Fabry--P\'erot modules, the internal propagation is already in the
scattering form of Eq.~\eqref{eq:nonlin_bidirectional_one_pass_map}. The two
lossless mirrors impose
\begin{subequations}
\label{eq:nonlin_fp_boundary_relations}
    \begin{align}
        \hat{\bm\Psi}_{\mathrm{init}}
        &=
        \bm T_{\mathrm{in}}\hat{\bm\Psi}_{\mathrm{in}}
        +
        \bm R_{\mathrm{FP}}\hat{\bm\Psi}_{\mathrm{final}},
        \\
        \hat{\bm\Psi}_{\mathrm{out}}
        &=
        -\bm R_{\mathrm{FP}}^{\dagger}
        \hat{\bm\Psi}_{\mathrm{in}}
        +
        \bm T_{\mathrm{out}}\hat{\bm\Psi}_{\mathrm{final}}.
    \end{align}
\end{subequations}
The matrices \(\bm T_{\mathrm{in}}\) and \(\bm T_{\mathrm{out}}\)
contain the mode-dependent mirror transmissions.
\(\bm R_{\mathrm{FP}}\) exchanges the propagation directions and contains
the signal- and idler-resonance phases used by the corresponding module.
Defining
\begin{equation}
    \bm C_{\mathrm{FP}}
    =
    \left(
        \bm I-
        \bm R_{\mathrm{FP}}\bm S_{aa}
    \right)^{-1},
    \label{eq:nonlin_fp_feedback_inverse}
\end{equation}
the boundary closure gives
\begin{equation}
    \hat{\bm\Psi}_{\mathrm{out}}
    =
    \bm M_{aa}\hat{\bm\Psi}_{\mathrm{in}}
    +
    \bm M_{af}\hat{\bm f},
    \label{eq:nonlin_fp_external_map}
\end{equation}
where~\cite{Sorensen2025ResonantSPDC}
\begin{subequations}
\label{eq:nonlin_fp_closed_maps}
    \begin{align}
        \bm M_{aa}
        &=
        \bm T_{\mathrm{out}}
        \bm S_{aa}
        \bm C_{\mathrm{FP}}
        \bm T_{\mathrm{in}}
        -
        \bm R_{\mathrm{FP}}^{\dagger},
        \\
        \bm M_{af}
        &=
        \bm T_{\mathrm{out}}
        \left(
            \bm I
            +
            \bm S_{aa}
            \bm C_{\mathrm{FP}}
            \bm R_{\mathrm{FP}}
        \right)
        \bm S_{af}.
    \end{align}
\end{subequations}
Setting all mirror reflectivities to zero gives
\(\bm R_{\mathrm{FP}}=\bm 0\) and recovers the open bidirectional-waveguide
map. This closure therefore follows the hierarchy
\(\bm U_k\rightarrow\bm S_{aa}\rightarrow\bm M_{aa}\).

\paragraph{Four-port ring closure with backscattering.}

The backscattering module constructs separate forward and backward one-pass
maps and combines them into the direction-block-diagonal scattering matrices
\(\bm S_{aa}\) and \(\bm S_{af}\). Its internal propagation relation is
\begin{equation}
    \hat{\bm\Psi}_{\mathrm{final}}
    =
    \bm S_{aa}\hat{\bm\Psi}_{\mathrm{init}}
    +
    \bm S_{af}\hat{\bm f}_{\mathrm{pr}},
    \label{eq:nonlin_backscattering_one_pass_map}
\end{equation}
where \(\hat{\bm f}_{\mathrm{pr}}\) contains the direction-resolved
propagation-loss inputs. Linear forward--backward mixing is introduced only
through the effective four-port boundary. Starting from the port order
\((\mathrm{ring}+ ,\mathrm{bus}+ ,\mathrm{ring}- ,\mathrm{bus}-)\), the
boundary is regrouped into internal and external sectors as
\begin{equation}
    \begin{pmatrix}
        \hat{\bm\Psi}_{\mathrm{init}}\\
        \hat{\bm\Psi}_{\mathrm{out}}
    \end{pmatrix}
    =
    \begin{pmatrix}
        \bm C_{11} & \bm C_{12}\\
        \bm C_{21} & \bm C_{22}
    \end{pmatrix}
    \begin{pmatrix}
        \hat{\bm\Psi}_{\mathrm{final}}\\
        \hat{\bm\Psi}_{\mathrm{in}}
    \end{pmatrix}
    +
    \begin{pmatrix}
        \bm N_{11} & \bm N_{12}\\
        \bm N_{21} & \bm N_{22}
    \end{pmatrix}
    \begin{pmatrix}
        \hat{\bm f}_{\mathrm{cav}}\\
        \hat{\bm f}_{\mathrm{ch}}
    \end{pmatrix}.
    \label{eq:nonlin_four_port_boundary}
\end{equation}
Here, \(\bm C\) is the effective coupler matrix, including the
direction-dependent cavity phases, and \(\bm N\) contains its loss-coupling
blocks. Define
\begin{equation}
    \bm F_{\mathrm{BS}}
    =
    \bm S_{aa}
    \left(
        \bm I-
        \bm C_{11}\bm S_{aa}
    \right)^{-1}.
    \label{eq:nonlin_backscattering_feedback_matrix}
\end{equation}
Eliminating the internal fields gives
\begin{equation}
    \hat{\bm\Psi}_{\mathrm{out}}
    =
    \bm M_{aa}\hat{\bm\Psi}_{\mathrm{in}}
    +
    \bm M_{af}^{\mathrm{pr}}\hat{\bm f}_{\mathrm{pr}}
    +
    \bm M_{af}^{\mathrm{cav}}\hat{\bm f}_{\mathrm{cav}}
    +
    \bm M_{af}^{\mathrm{ch}}\hat{\bm f}_{\mathrm{ch}},
    \label{eq:nonlin_backscattering_external_map}
\end{equation}
where
\begin{subequations}
\label{eq:nonlin_backscattering_closed_maps}
    \begin{align}
        \bm M_{aa}
        &=
        \bm C_{22}
        +
        \bm C_{21}\bm F_{\mathrm{BS}}\bm C_{12},
        \\
        \bm M_{af}^{\mathrm{pr}}
        &=
        \bm C_{21}
        \left(
            \bm I+
            \bm F_{\mathrm{BS}}\bm C_{11}
        \right)
        \bm S_{af},
        \\
        \bm M_{af}^{\mathrm{cav}}
        &=
        \bm N_{21}
        +
        \bm C_{21}\bm F_{\mathrm{BS}}\bm N_{11},
        \\
        \bm M_{af}^{\mathrm{ch}}
        &=
        \bm N_{22}
        +
        \bm C_{21}\bm F_{\mathrm{BS}}\bm N_{12}.
    \end{align}
\end{subequations}
The superscripts \(\mathrm{pr}\), \(\mathrm{cav}\), and \(\mathrm{ch}\)
label propagation loss and the cavity-side and channel-side coupler losses,
respectively. If the four-port coupler is lossless, \(\bm N=\bm 0\) and the
latter two reservoir maps vanish.

For every closed device, the reservoir inputs may be collected into the
common external form
\begin{equation}
    \hat{\bm\Psi}_{\mathrm{out}}
    =
    \bm M_{aa}\hat{\bm\Psi}_{\mathrm{in}}
    +
    \sum_r
    \bm M_{af}^{(r)}\hat{\bm f}_r.
    \label{eq:nonlin_complete_device_map}
\end{equation}
The generic reservoir label $r$ includes the propagation-loss reservoirs
indexed by $k$ and, where applicable, the cavity-side and channel-side
boundary-loss reservoirs. The unsubscripted block $\bm M_{af}$ and vector
$\hat{\bm f}$ denote the corresponding horizontal and vertical
concatenations, respectively.
Together with all retained loss blocks, the closed map preserves the
commutator signature in Eq.~\eqref{eq:nonlin_optical_commutator_signature},
\begin{equation}
    \bm M_{aa}\bm J\bm M_{aa}^{\dagger}
    +
    \sum_r
    \bm M_{af}^{(r)}
    \bm J_{f,r}
    \left(\bm M_{af}^{(r)}\right)^{\dagger}
    =
    \bm J.
    \label{eq:nonlin_complete_device_canonicality}
\end{equation}
All resonator closures above require the corresponding feedback inverse to be nonsingular.

\clearpage
\refstepcounter{section}
\section*{Supplementary Note \thesection.\\Solid-state single-photon-emitter model}
\addcontentsline{toc}{section}{\thesection \quad Solid-state single-photon-emitter model}
\label{sec:sse}

Here we define the reduced source model used to describe the conditional state of a single photon emitted from a solid-state emitter. Within the single-emission approximation, the photon is completely characterized by its first-order coherence, equivalently its single-photon density matrix, which resolves its temporal and spectral degrees of freedom. The model incorporates experimentally accessible parameters describing radiative decay, pure dephasing, spectral diffusion and emission-time jitter as distinct physical processes. The formulation is independent of a particular material platform or emitter type, and the resulting single-photon state can be supplied directly to the photonic input--output models implemented in PhoQuS.


\subsection{Reduced source model and single-emission wavepacket}

We describe the photon delivered to a selected optical mode rather than the microscopic dynamics of the emitter and its environment. The emitter is modeled as a localized two-level system~\cite{Kiraz2004QuantumDot} that emits a photon through spontaneous radiative decay. The model considers one selected optical transition and a fixed collected spatial and polarization mode, while the emission probability and vacuum and multiphoton contributions are treated separately.

Following standard descriptions of photons emitted by such systems, the single-shot state (indexed by $i$) and its normalized temporal wavepacket are expressed as
\begin{subequations}
    \begin{align}
        \ket{1}_i &= \int dt\, \psi_i(t) \hat{a}^\dagger(t) \ket{0}, \\
        \psi_i(t) &= \cfrac{1}{\sqrt{\tau_r}}\Theta(t-t_i)
        \exp\left[-\cfrac{t-t_i}{2\tau_r}
        -\mathrm{i}\omega_i(t-t_i)
        -\mathrm{i}\phi_i(t-t_i)\right].
    \end{align}
    \label{eq:sse_wavepacket}
\end{subequations}
The wavepacket is normalized as $\int dt\,|\psi_i(t)|^2=1$. Here, $\Theta(x)$ is the Heaviside function enforcing causality, and $\tau_r$ is the radiative population lifetime. The stochastic phase trajectory $\phi_i(t)$, shot-dependent emission frequency $\omega_i$, and emission-start time $t_i$ describe pure dephasing, spectral diffusion, and time jitter, respectively. These processes act on different timescales and affect distinct aspects of the emitted photon's spectro-temporal coherence, as described below~\cite{Kambs2018Indistinguishability}.

\textbf{Pure dephasing} describes rapid fluctuations of the optical phase $\phi_i(t)$ within an individual emission event, arising from interactions between the emitter and its local environment, such as coupling to lattice phonons. We assume Markovian phase fluctuations characterized by~\cite{IlesSmith2017Phonons}
\begin{equation}
    \label{eq:sse_markovian_pd}
    \left\langle
    e^{\mathrm{i}\left[\phi_i(t+\tau)-\phi_i(t)\right]}
    \right\rangle_{\mathrm{PD}}
    =
    e^{-\Gamma^\ast|\tau|},
\end{equation}
where $\Gamma^\ast$ is the pure-dephasing rate and $\langle\cdots\rangle_{\mathrm{PD}}$ denotes averaging over stochastic phase realizations. Pure dephasing leaves the radiative population decay unchanged but suppresses coherence between different temporal components of the emitted wavepacket. Together with radiative decay, it gives the homogeneous optical-coherence decay rate $\Gamma_h=1/(2\tau_r)+\Gamma^\ast$. Consequently, averaging over the stochastic phase fluctuation converts the realization-resolved pure wavepacket into a mixed single-photon state and produces homogeneous broadening of the emission spectrum.

\textbf{Spectral diffusion} describes slow fluctuations of the optical transition frequency between repeated excitation events, arising from slowly varying changes in the local environment of the emitter, such as charge or electric-field fluctuations. Within each emission event, $\omega_i$ is assumed to remain approximately constant, while its value varies from shot to shot according to a normalized probability distribution $p_\omega(\omega_i)$. Ensemble averaging therefore mixes photons with different carrier frequencies, resulting in inhomogeneous spectral broadening and suppression of the ensemble-averaged temporal coherence.

\textbf{Time jitter} describes shot-to-shot fluctuations in the onset of photon emission. The emission-start time $t_i$ is sampled from a normalized probability distribution $p_t(t_i)$, such that each individual photon retains the same causal radiative envelope but is displaced in time by a different amount. Such timing uncertainty may arise, for example, from the excitation process or from carrier capture and relaxation preceding radiative emission. Ensemble averaging over $t_i$ therefore broadens the photon arrival-time distribution and introduces temporal-mode mixing in the ensemble-averaged single-photon state.

\subsection{Single-photon density kernel}
\label{subsec:sse_density_kernel}

Within the single-photon sector, the spectro-temporal state of the emitted photon can be completely described by its density kernel in a continuous temporal or spectral basis. We adopt the correlation ordering and define density kernel with single-photon density operator as
\begin{subequations}
    \begin{align}
    \rho(x, x') &= \bra{x'}\hat\rho\ket{x}, \\
    \hat\rho &= \iint dx dx' \rho(x,x')\hat a^\dagger(x')\ket{0}\bra{0}\hat a(x),
    \end{align}
    \label{eq:sse_density_kernel_def}
\end{subequations}
where $\ket{x} = \hat a^\dagger(x)\ket{0} \ \left( x\in\{\omega, t\}\right)$. With this ordering, the density kernel $\rho(x,x')$ is identical to the first-order coherence function $G^{(1)}(x,x')$, defined as~\cite{Gao2019OptimalFiltering}
\begin{equation}
    \rho(x,x') = G^{(1)}(x,x') = \left\langle \hat a^\dagger(x)\hat a(x') \right\rangle.
    \label{eq:sse_density_g1_equivalence}
\end{equation}
The equivalence in Eq.~\eqref{eq:sse_density_g1_equivalence} provides a direct physical interpretation of the density kernel. In the time domain, the diagonal $\rho(t,t)$ gives the temporal probability density of photon, while the off-diagonal elements quantify the coherence between different temporal components of the emitted photon. Likewise, in the frequency domain, $\rho(\omega,\omega)$ gives the normalized power spectral density, whereas the off-diagonal terms describe coherence between different two colors. The full density kernel therefore contains both the intensity distribution and the spectro-temporal coherence of the single photon. 

For a fixed set of stochastic quantities $\left(\phi_i, \omega_i, t_i\right)$ associated with the $i$th emission event, the density kernel corresponds to a pure single-photon state. From Eqs.~\eqref{eq:sse_density_g1_equivalence} and \eqref{eq:sse_wavepacket}, its time-domain representation is
\begin{equation}
    \rho(t, t'|\phi_i, \omega_i, t_i) = \psi_i^*(t)\psi_i(t'). 
    \label{eq:sse_density_kernel_fully_resolved}
\end{equation}

However, averaging over the unresolved stochastic quantities converts the density kernel into a mixed state. The phase fluctuations responsible for pure dephasing are first averaged while the shot-dependent emission frequency and start time are held fixed. Spectral diffusion and time jitter are subsequently incorporated by averaging over the shot-to-shot variations of $\omega_i$ and $t_i$. We therefore introduce two density kernels,
\begin{subequations}
    \begin{align}
        \rho_\text{int}(t,t'|\omega_i,t_i) &= \left\langle \rho(t,t'|\phi_i,\omega_i,t_i)\right\rangle_\mathrm{PD}, \\
        \rho(t,t') &= \left\langle \left\langle \rho_\text{int}(t,t'|\omega_i, t_i)\right\rangle\right\rangle_\mathrm{SD,TJ}. 
    \end{align}
    \label{eq:sse_density_kernel_avg}
\end{subequations}
Here $\rho_\text{int}$ denotes the intrinsic density kernel of single-shot emission conditioned on fixed emission frequency and start time, after averaging over the fast pure-dephasing dynamics. The ensemble density kernel $\rho$ additionally incorporates the shot-to-shot fluctuations associated with spectral diffusion and time jitter. Except in the absence of the corresponding fluctuations, these stochastic averages generally convert the initially pure density kernel into a mixed single-photon state.

\textbf{Intrinsic density kernel} can be expressed analytically using the Markovian phase statistics in Eq.~\eqref{eq:sse_markovian_pd}:
\begin{equation}
    \rho_\mathrm{int}(t,t'|\omega_i,t_i) =
    \frac{1}{\tau_r}
    \Theta(t-t_i)\Theta(t'-t_i)
    \exp\left[-\frac{t+t'-2t_i}{2\tau_r}-\Gamma^\ast|t-t'|+\mathrm{i}\omega_i(t-t')\right].
    \label{eq:sse_density_kernel_int}
\end{equation}
The diagonal of Eq.~\eqref{eq:sse_density_kernel_int} retains the exponential radiative population decay, whereas pure dephasing suppresses the off-diagonal coherence as the temporal separation $|t-t'|$ increases. The frequency-domain expression is just Fourier transform of Eq.~\eqref{eq:sse_density_kernel_int},

\begin{equation}
\begin{aligned}
    \rho_\mathrm{int}(\omega,\omega'|\omega_i,t_i)
    &=
    \frac{1}{2\pi}\iint dt\,dt'\,
    \rho_\mathrm{int}(t,t'|\omega_i,t_i)
    e^{-\mathrm{i}(\omega t-\omega't')}
    \\
    &=
    \frac{e^{-\mathrm{i}(\omega-\omega')t_i}}
    {2\pi\tau_r[1/\tau_r+\mathrm{i}(\omega-\omega')]}
    \frac{2\Gamma_h+\mathrm{i}(\omega-\omega')}
    {\Gamma_h^2+\mathrm{i}(\omega-\omega')\Gamma_h
    +(\omega-\omega_i)(\omega'-\omega_i)}.
\end{aligned}
\label{eq:sse_density_kernel_int_freq}
\end{equation}

where $\Gamma_h = 1/(2\tau_r) + \Gamma^\ast$. Its diagonal components $\rho_\text{int}(\omega,\omega|\omega_i,t_i)$ describe the intrinsic power spectral density,
\begin{equation}
    \mathcal S_\text{int}(\omega|\omega_i,t_i) = \cfrac{\Gamma_h/\pi}{\Gamma_h^2 +(\omega-\omega_i)^2},
    \label{eq:sse_spectrum_int}
\end{equation}
which is a Lorentzian centered at $\omega_i$ with a FWHM of $2\Gamma_h$ in angular frequency. Hence $\Gamma_h$ is the homogeneous optical-coherence decay rate determined by radiative decay and pure dephasing. 

\textbf{Ensemble density kernel} is obtained by averaging intrinsic density kernel over the distributions of the shot-dependent emission frequency and start time,
\begin{equation}
    \rho(t, t') = \iint d\omega_i dt_i p(\omega_i, t_i)\rho_\mathrm{int}(t, t'|\omega_i, t_i), 
    \label{eq:sse_density_kernel_ens}
\end{equation}
where $p(\omega_i, t_i)$ is their joint probability density. In the following, we assume that spectral diffusion and time jitter are statistically independent: $p(\omega_i, t_i) = p_\omega(\omega_i)p_t(t_i)$. The ensemble density kernel in the time domain can be explicitly written as
\begin{subequations}
    \begin{align}
        \rho(t,t') &= \frac{1}{\tau_r} \exp\left[-\frac{t+t'}{2\tau_r}-\Gamma^\ast|t-t'|\right]
        \mathcal J_t\!\left(\min(t,t')\right) \chi_\omega(t-t'),
        \\
        \mathcal J_t(m)&=\int_{-\infty}^{m}dt_i\,p_t(t_i)e^{t_i/\tau_r} \ \left(m=\min(t,t')\right),
        \\
        \chi_\omega(\Delta t)&=\int d\omega_i\,p_\omega(\omega_i)e^{\mathrm{i}\omega_i\Delta t} \ \left(\Delta t=t-t'\right).
    \end{align}
    \label{eq:sse_density_kernel_ensemble_integral}
\end{subequations}
Here, $\mathcal J_t$ is the exponentially weighted cumulative factor associated with time jitter, and $\chi_\omega$ is the characteristic function of the shot-to-shot emission frequency distribution. Equivalently, the ensemble density kernel in the frequency domain is
\begin{subequations}
    \begin{align}
        \rho(\omega,\omega')&=\frac{2\Gamma_h+\mathrm{i}(\omega-\omega')}{2\pi\tau_r[1/\tau_r+\mathrm{i}(\omega-\omega')]}\chi_t(\omega-\omega')\mathcal K_\omega(\omega,\omega'),
        \\
        \chi_t(\Delta\omega)&=\int dt_i\,p_t(t_i)e^{-\mathrm{i}\Delta\omega t_i}
        \ \left(\Delta\omega=\omega-\omega'\right),
        \\
        \mathcal K_\omega(\omega,\omega')&=\int d\omega_i\,p_\omega(\omega_i)
        \frac{1}{\Gamma_h^2+\mathrm{i}(\omega-\omega')\Gamma_h+(\omega-\omega_i)(\omega'-\omega_i)}.
    \end{align}
    \label{eq:sse_density_kernel_ensemble_integral_freq}
\end{subequations}
In this representation, $\chi_t$ is the characteristic function of the emission-start-time distribution, while $\mathcal K_\omega$ describes the spectral response averaged over spectral diffusion.

Commonly used distributions for $p_x(x_i) \ \left(x\in\{\omega, t\}\right)$ is deterministic limit, $p_x(x_i) = \delta(x_i-\bar x)$, or Gaussian jitter, $p_x(x_i) = (\sqrt{2\pi}\sigma_x)^{-1}\exp\left[-(x_i-\bar x)^2/(2\sigma_x^2)\right]$. Under those conditions, the averaging factors $\mathcal J_t, \chi_t, \chi_\omega$, and $\mathcal K_\omega$ can be expressed analytically, and summarized in Table.~\ref{tab:sse_integrals}.

\begin{table}[!htbp]
\caption{Analytic averaging factors for deterministic and Gaussian temporal and spectral fluctuations.}
\label{tab:sse_integrals}
\centering
\small
\renewcommand{\arraystretch}{1.8}
\vspace{0.6em}
\begin{tabularx}{\textwidth}{
    @{}
    >{\centering\arraybackslash}p{0.20\textwidth}
    >{\centering\arraybackslash}X
    >{\centering\arraybackslash}X
    @{}
}
\toprule
\textbf{Temporal jitter} & \textbf{Deterministic} & \textbf{Gaussian}
\\
\midrule\midrule

$\mathcal J_t(m)$ & $\displaystyle e^{\bar t/\tau_r}\Theta(m-\bar t)$ &
$\displaystyle e^{\bar t/\tau_r+\sigma_t^2/(2\tau_r^2)} \Phi\left(\frac{m-\bar t-\sigma_t^2/\tau_r}{\sigma_t}\right)$
\\

$\chi_t(\Delta\omega)$&$\displaystyle e^{-\mathrm{i}\bar t\Delta\omega} $ &
$\displaystyle e^{-\mathrm{i}\bar t\Delta\omega -\sigma_t^2\Delta\omega^2/2}$
\\

\midrule
\textbf{Spectral diffusion}&\textbf{Deterministic}&\textbf{Gaussian}
\\
\midrule\midrule

$\chi_\omega(\Delta t)$ & $\displaystyle e^{\mathrm{i}\bar\omega\Delta t}$ &
$\displaystyle e^{\mathrm{i}\bar\omega\Delta t -\sigma_\omega^2\Delta t^2/2}$
\\

$\mathcal K_\omega(\omega,\omega')$ & $\displaystyle \frac{1}{\Gamma_h^2
+\mathrm{i}\Delta\omega\Gamma_h +(\omega-\bar\omega)(\omega'-\bar\omega)}$
& $\displaystyle \sqrt{\frac{\pi}{2}} \frac{w(z_1)+w(z_2)} {\sigma_\omega\left(2\Gamma_h+\mathrm{i}\Delta\omega\right)}$
\\

\bottomrule
\end{tabularx}

\vspace{0.6em}
\footnotesize
$m=\min(t,t')$, $\Delta t=t-t'$, and $\Delta\omega=\omega-\omega'$.
For Gaussian spectral diffusion,
$z_1=(\omega'-\bar\omega+\mathrm{i}\Gamma_h)/(\sqrt{2}\sigma_\omega)$
and
$z_2=(\bar\omega-\omega+\mathrm{i}\Gamma_h)/(\sqrt{2}\sigma_\omega)$,
with
$w(z)=e^{-z^2}\operatorname{erfc}(-\mathrm{i}z)$.
\end{table}
For instance of Gaussian spectral diffusion without temporal jitter, the ensemble power spectral density becomes a Voigt profile, corresponding to the convolution of homogeneous Lorentzian broadening with Gaussian inhomogeneous broadening,
\begin{equation}
    \mathcal S(\omega) = \int d\omega_i \cfrac{\Gamma_h/\pi}{\Gamma_h^2+(\omega-\omega_i)^2}\times \cfrac{1}{\sqrt{2\pi\sigma_\omega^2}}\exp\left[-\cfrac{(\omega_i-\bar\omega)^2}{2\sigma_\omega^2}\right]. 
\end{equation}

\subsection{Spectro-temporal modes and Indistinguishability}

The ensemble density kernel generally describes a mixed single-photon state and can be decomposed into a set of orthonormal spectro-temporal modes. As a single-photon density operator, $\hat\rho$ is Hermitian $\left(\hat\rho^\dagger=\hat\rho\right)$, positive semidefinite $(\hat\rho\succeq0)$, and normalized to unit trace $(\mathrm{Tr}[\hat\rho]=1)$. Its spectral decomposition therefore gives the density kernel as
\begin{equation}
    \rho(x, x') = \sum_k \lambda_k \varphi_k^*(x)\varphi_k(x'),
    \label{eq:sse_density_eigendecomposition}
\end{equation}
with nonnegative eigenvalues, $\lambda_k\ge0,\ \sum_k\lambda_k=1$, and orthonormal spectro-temporal modes, $\int dx \varphi_k^*(x)\varphi_l(x) = \delta_{kl}$. The eigenvalues $\{\lambda_k\}$ quantify the statistical occupation of the corresponding spectro-temporal modes. While a pure single photon occupies a single mode with $\lambda_0=1$, stochastic averaging of pure dephasing, spectral diffusion, or time jitter generally distributes the state over multiple orthogonal modes. 

For two photons independently emitted from the same stationary ensemble, the single-photon indistinguishability is given by the purity of the ensemble density operator~\cite{Gao2019OptimalFiltering},
\begin{equation}
    \mathcal I=\mathrm{Tr}\!\left[\hat\rho^2\right]=\sum_k \lambda_k^2 =\iint dx\,dx'\,\left|\rho(x,x')\right|^2.
    \label{eq:sse_indistinguishability}
\end{equation}
Here, $0<\mathcal I\leq1$, with $\mathcal I=1$ corresponding to a pure single-mode photon. For independent two single-photon emission events from the same source, $\mathcal I$ corresponds to the ideal Hong--Ou--Mandel indistinguishability~\cite{Branczyk2017HOM} assuming perfect matching of all other degrees of freedom~\cite{Santori2002,Thoma2016Dephasing,Pollet2026NearIdentical}.

\subsection{Implementation and interface example}

The solid-state-emitter module in PhoQuS is initialized by the radiative lifetime
$\tau_r$, pure-dephasing rate $\Gamma^\ast$, and models for temporal and spectral
jitter. The latter are represented by \texttt{TemporalJitter} and
\texttt{FrequencyJitter} objects, respectively, each of which can describe a
deterministic, Gaussian, or user-defined probability distribution. The
homogeneous coherence-decay rate is then internally evaluated as
$\Gamma_h=\Gamma^\ast+1/(2\tau_r)$.

Given a uniform angular-frequency grid $\{\omega_n\}$ or temporal grid $\{t_n\}$, \texttt{get\_rho\_time} and \texttt{get\_rho\_freq} evaluate the continuous density kernel derived in Sec.~\ref{subsec:sse_density_kernel} and construct its quadrature-weighted discrete representation,
\begin{subequations}
    \begin{align}
        \left[\mathbf R_\omega\right]_{nm} &= \rho(\omega_n,\omega_m)\Delta\omega,\\
        \left[\mathbf R_t\right]_{nm} &= \rho(t_n,t_m)\Delta t.
    \end{align}
    \label{eq:sse_phoqus_density_kernel}
\end{subequations}
For sufficiently resolved and wide grid, these matrices satisfy $\text{Tr}(\mathbf R_x)\simeq 1$. For deterministic and Gaussian jitter models, the averaging factors entering the density kernel are evaluated using the analytic expressions summarized in Table~\ref{tab:sse_integrals}. For a user-defined jitter distribution, the probability density is supplied on a numerical grid, either directly as an array or through a callable function, and corresponding averaging integrals are evaluated numerically following Eqs.~\eqref{eq:sse_density_kernel_ensemble_integral} and ~\eqref{eq:sse_density_kernel_ensemble_integral_freq}. 

\begin{table}[h]
\caption{Summary and implementation of reduced solid-state emitter model}
\label{tab:sse_interface}
\centering
\small
\begin{tabularx}{\textwidth}{
    @{}
    >{\centering\arraybackslash}p{0.20\textwidth}
    >{\centering\arraybackslash}p{0.20\textwidth}
    >{\raggedright\arraybackslash}X
    @{}
}
\toprule
\textbf{Model parameters} & \textbf{Object} & \textbf{Definition} \\
\midrule\midrule

$\tau_r$ & \texttt{SSE} & Radiative population lifetime of the selected transition. \\

$\Gamma^\ast$ & \texttt{SSE} & Markovian pure-dephasing rate of the optical coherence. \\

$\Gamma_h=(2\tau_r)^{-1}+\Gamma^\ast$ & \texttt{SSE} & Homogeneous optical-coherence decay rate. \\

$p_\omega(\omega_i)$ & \texttt{FrequencyJitter} & Normalized carrier frequency probability distribution by spectral diffusion. \\

$p_t(t_i)$ & \texttt{TemporalJitter} & Normalized emission-start time probability distribution by temporal jitter.\\

\midrule
\textbf{Output} & \textbf{Routine} & \textbf{Definition} \\
\midrule\midrule

$\mathbf{R}_\omega$ & \texttt{get\_rho\_freq} & Quadrature-weighted frequency-domain density matrix, $[\mathbf R_\omega]_{nm}=\rho(\omega_n,\omega_m)\Delta\omega$. \\

$\mathbf{R}_t$ & \texttt{get\_rho\_time} & Quadrature-weighted time-domain density matrix, $[\mathbf R_t]_{nm}=\rho(t_n,t_m)\Delta t$. \\

$\{\lambda_k,\mathbf v_k\},\,\mathcal I$ & \texttt{decomp\_rho\_freq} or \texttt{decomp\_rho\_time}
& Spectro-temporal eigenvalues and eigenmodes obtained from $\mathbf R_\omega$ or $\mathbf R_t$,
together with the indistinguishability
$\mathcal I=\sum_k\lambda_k^2$. \\

\bottomrule
\end{tabularx}
\end{table}

The quadrature-weighted density matrix can subsequently be diagonalized using \texttt{decomp\_rho\_time} or \texttt{decomp\_rho\_freq}. These routines return the eigenvalues $\{\lambda_k\}$ and corresponding discretized spectro-temporal eigenmodes $\{\mathbf v_k\}$, and evaluate the single-photon indistinguishability as $\mathcal I = \sum_k \lambda_k^2$, consistent with Eq.~\eqref{eq:sse_indistinguishability}. The temporal and frequency-domain routines provide equivalent representations of the same single-photon state and therefore converge to the same eigenvalue spectrum and indistinguishability for sufficiently resolved numerical grids.

The complete workflow of the solid-state-emitter module, from the physical source parameters and jitter models to the discretized density matrix, spectro-temporal modes, and indistinguishability, is summarized in Table~S2.

The quadrature-weighted density matrix provides a direct interface to the Bogoliubov modules implemented in PhoQuS. As an example, we consider mode-selective quantum frequency conversion, where a single photon enters the input channel $i$ and is converted to the output channel $s$. For a photon-number-preserving transformation, the frequency-resolved input--output relation for the $s$--channel can be written as~\cite{Chiriano2025Purifying}:
\begin{equation}
    \hat{\mathbf a}_s^{\mathrm{out}} 
    =
    \mathbf S_{si}\hat{\mathbf a}_i^\mathrm{in} +
    \mathbf S_{ss}\hat{\mathbf a}_s^\mathrm{in} +
    \sum_\ell \mathbf L_{s\ell}\hat{\mathbf f}_\ell^\mathrm{in},
    \label{eq:sse_qfc_io}
\end{equation}
where $\mathbf S_{si}$ is the frequency-resolved conversion matrix from channel $i$ to $s$. 

For an SSE photon injected only into channel $i$, while channel $s$ and all auxiliary loss channels are initially in vacuum,  the normally ordered correlation matrix receives a contribution only from the first term in Eq.~\eqref{eq:sse_qfc_io}. The one photon block of the reduced output state in channel $s$ is therefore represented by the unnormalized density matrix
\begin{equation}
    \mathbf R_s^{\mathrm{out}}
    =
    \mathbf S_{si}^{\ast}
    \mathbf R_i^{\mathrm{in}}
    \mathbf S_{si}^{\mathsf T}.
\end{equation}
Accordingly, the precomputed emitter state $\mathbf R_i^\mathrm{in}$ and the Bogoliubov transfer matrix $\mathbf S_{si}$ can be directly combined without resolving the microscopic emitter dynamics within the subsequent photonic simulation. The resulting matrix corresponds to the output correlation kernel $\left[\mathbf R_s^\mathrm{out}\right]_{nm} = \Delta\omega\braket{\hat a_s^{\mathrm{out},\dagger}(\omega_n)\hat a_s^\mathrm{out}(\omega_m)}$. 

The output correlation kernel is un-normalized, and contains information of conversion probability and resulting quantum state purity. The trace of $\mathbf R_s^{\mathrm{out}}$ gives the probability that the input single photon is found in the converted output channel,
\begin{equation}
    \eta_{i\rightarrow s}
    =
    \mathrm{Tr}\!\left[\mathbf R_s^{\mathrm{out}}\right].
    \label{eq:sse_qfc_conversion_probability}
\end{equation}
Conditioned on successful conversion, the normalized output state is
\begin{equation}
    \widetilde{\mathbf R}_s^{\mathrm{out}}
    =
    \frac{\mathbf R_s^{\mathrm{out}}}{\eta_{i\rightarrow s}},
    \label{eq:sse_qfc_conditional_density}
\end{equation}
and the conditional output quantum state purity is
\begin{equation}
    \mathcal P_{\mathrm{out}}
    =
    \mathrm{Tr}\!\left[
        \left(\widetilde{\mathbf R}_s^{\mathrm{out}}\right)^2
    \right].
    \label{eq:sse_qfc_output_purity}
\end{equation}

\subsection{Scope and limitations}

The reduced SSE model is intended to describe the conditional spectro-temporal
state of a single photon using experimentally accessible source parameters,
rather than the microscopic emitter dynamics. The principal assumptions and
limitations of the present formulation are summarized in
Table~\ref{tab:sse_scope}. Extensions beyond these assumptions can be
incorporated by replacing the corresponding source kernel or probability model
before interfacing the resulting state with the PhoQuS photonic modules.

\begin{table}[!htbp]
\caption{Scope and limitations of the reduced solid-state-emitter model.}
\label{tab:sse_scope}
\centering
\small
\renewcommand{\arraystretch}{1.35}

\begin{tabularx}{\textwidth}{
    @{}
    >{\raggedright\arraybackslash}p{0.24\textwidth}
    >{\raggedright\arraybackslash}p{0.34\textwidth}
    >{\raggedright\arraybackslash}X
    @{}
}
\toprule
\textbf{Aspect} & \textbf{Present treatment} & \textbf{Limitation} \\
\midrule

Photon-number sector
&
Conditional single-photon state with $\mathrm{Tr}(\hat\rho)=1$.
&
Vacuum probability, source brightness, multiphoton emission, and re-excitation
are not included in the density kernel.
\\

Emitter transition
&
One selected optical transition coupled to a fixed spatial and polarization mode.
&
Multiple transitions, polarization mixing, and which-path degrees of freedom
are not explicitly modeled.
\\

Radiative decay
&
Single exponential population decay characterized by $\tau_r$.
&
Multiexponential relaxation or explicit microscopic emitter--cavity dynamics
are not described.
\\

Pure dephasing
&
Markovian phase fluctuations characterized by a constant rate $\Gamma^\ast$.
&
Non-Markovian bath dynamics, phonon sidebands, and frequency-dependent
dephasing are not explicitly included.
\\

Spectral diffusion
&
Emission frequency $\omega_i$ is constant within each emission event and
varies between events according to $p_\omega(\omega_i)$.
&
Frequency fluctuations occurring on timescales comparable to or shorter than
a single emission event are outside the quasi-static approximation.
\\

Temporal jitter
&
The emission-start time $t_i$ varies between events according to $p_t(t_i)$.
&
Microscopic excitation, carrier-capture, and relaxation dynamics generating
the jitter are not explicitly resolved.
\\

Jitter correlations
&
Temporal jitter and spectral diffusion are treated as statistically independent,
$p(\omega_i,t_i)=p_\omega(\omega_i)p_t(t_i)$.
&
Correlated frequency--time fluctuations are not included in the current
implementation.
\\

\bottomrule
\end{tabularx}
\end{table}

\clearpage
\refstepcounter{section}
\section*{Supplementary Note \thesection.\\Gaussian computational framework}
\addcontentsline{toc}{section}{\thesection \quad Gaussian computational framework}
\label{sec:gaussian-framework}

This section describes the common Gaussian framework through which the source
and device models in Supplementary Notes~1 and~2 are assembled into photonic
circuits and connected to measurement. Under the undepleted-pump approximation,
the nonlinear modules of Supplementary Note~1 produce Gaussian states or
Gaussian channels. These outputs, together with passive component models, are
composed at the level of first and second field moments. Photon counting is
applied only after the frequency-resolved circuit propagation and provides the
non-Gaussian conditioning required for heralded-state calculations. The mixed
single-photon state of Supplementary Note~2 is incorporated into the same
workflow through an auxiliary Gaussian embedding.

\subsection{Complex covariance representation and Gaussian channels}
\label{subsec:gaussian-representation}

The sampled operators in Eq.~\eqref{eq:nonlin_discrete_psi_operators} obey a
grid-weighted commutation relation. For each optical band, direction, or
transverse mode, we therefore define canonical frequency-bin operators by
\begin{equation}
    \hat a_{j,n}
    =
    \sqrt{\Delta\omega}\,
    \hat\Psi_{j,n},
    \qquad
    \left[
        \hat a_{j,n},
        \hat a_{l,m}^{\dagger}
    \right]
    =
    \delta_{jl}\delta_{nm}.
    \label{eq:gaussian_canonical_bins}
\end{equation}
All canonical modes are collected into
$\hat{\bm a}=(\hat a_1,\ldots,\hat a_K)^{\mathsf T}$ and represented in the
doubled complex basis
\begin{equation}
    \hat{\bm\xi}
    =
    \begin{pmatrix}
        \hat{\bm a}\\
        \hat{\bm a}^{\dagger}
    \end{pmatrix},
    \qquad
    \left[
        \hat{\bm\xi},
        \hat{\bm\xi}^{\dagger}
    \right]
    =
    \bm K,
    \qquad
    \bm K
    =
    \begin{pmatrix}
        \bm I & 0\\
        0 & -\bm I
    \end{pmatrix}.
    \label{eq:gaussian_doubled_basis}
\end{equation}
The calculations considered here have zero displacement. Their Gaussian state
is therefore specified completely by the complex covariance matrix
\begin{equation}
    V_{uv}
    =
    \frac{1}{2}
    \left\langle
        \hat\xi_u\hat\xi_v^{\dagger}
        +
        \hat\xi_v^{\dagger}\hat\xi_u
    \right\rangle,
    \qquad
    \bm V
    =
    \begin{pmatrix}
        \bm N+\bm I/2 & \bm M\\
        \bm M^{*} & \bm N^{*}+\bm I/2
    \end{pmatrix},
    \label{eq:gaussian_complex_covariance}
\end{equation}
where
$N_{uv}=\langle\hat a_v^{\dagger}\hat a_u\rangle$ and
$M_{uv}=\langle\hat a_u\hat a_v\rangle$. Thus,
$\bm V_{\mathrm{vac}}=\bm I/2$. With
\begin{equation}
    \bm T
    =
    \frac{1}{\sqrt{2}}
    \begin{pmatrix}
        \bm I & \bm I\\
        -\mathrm{i}\bm I & \mathrm{i}\bm I
    \end{pmatrix},
    \qquad
    \hat{\bm r}
    =
    \bm T\hat{\bm\xi},
    \label{eq:gaussian_complex_quadrature_transform}
\end{equation}
the usual quadrature covariance is
$\bm V_{xp}=\bm T\bm V\bm T^{\dagger}$. The physicality condition can be
written directly in the complex basis as
\begin{equation}
    \bm V+\frac{1}{2}\bm K
    \succeq
    0.
    \label{eq:gaussian_covariance_physicality}
\end{equation}

A zero-displacement Gaussian channel is represented by a pair of matrices
$(\bm X,\bm Y)$ acting as
\begin{equation}
    \bm V_{\mathrm{out}}
    =
    \bm X
    \bm V_{\mathrm{in}}
    \bm X^{\dagger}
    +
    \bm Y.
    \label{eq:gaussian_channel_map}
\end{equation}
Here, $\bm X$ contains the coherent optical transformation and $\bm Y$ is the
noise covariance contributed by inaccessible input sectors. The channel is
physical when
\begin{equation}
    \bm Y
    +
    \frac{1}{2}
    \left(
        \bm K_{\mathrm{out}}
        -
        \bm X\bm K_{\mathrm{in}}\bm X^{\dagger}
    \right)
    \succeq
    0.
    \label{eq:gaussian_channel_physicality}
\end{equation}
This form provides the common computational interface used for both nonlinear
devices and passive lossy components. If channel 1 is followed by channel 2,
their exact composition is
\begin{subequations}
\label{eq:gaussian_channel_composition}
    \begin{align}
        \bm X_{2\circ1}
        &=
        \bm X_2\bm X_1,
        \\
        \bm Y_{2\circ1}
        &=
        \bm X_2\bm Y_1\bm X_2^{\dagger}
        +
        \bm Y_2.
    \end{align}
\end{subequations}
Independent source states are assembled by direct sum, followed by a
permutation into the global rail ordering.

\subsection{Source and nonlinear-device interfaces}
\label{subsec:gaussian-source-interface}

The input--output maps in Supplementary Note~1 are expressed in
module-dependent operator bases that may contain both annihilation and creation
operators. Appending the Hermitian-conjugate relation and permuting the result
into the ordering of Eq.~\eqref{eq:gaussian_doubled_basis} converts the complete
device map in Eq.~\eqref{eq:nonlin_complete_device_map} to
\begin{equation}
    \hat{\bm\xi}_{\mathrm{out}}
    =
    \bm X_{\mathrm{NL}}
    \hat{\bm\xi}_{\mathrm{in}}
    +
    \sum_r
    \bm L_r
    \hat{\bm\xi}_{f,r},
    \label{eq:gaussian_nonlinear_lifted_map}
\end{equation}
where
\begin{subequations}
\label{eq:gaussian_nonlinear_lift}
    \begin{align}
        \bm X_{\mathrm{NL}}
        &=
        \bm P_{\mathrm{out}}
        \left(
            \bm M_{aa}
            \oplus
            \bm M_{aa}^{*}
        \right)
        \bm P_{\mathrm{in}}^{\mathsf T},
        \\
        \bm L_r
        &=
        \bm P_{\mathrm{out}}
        \left(
            \bm M_{af}^{(r)}
            \oplus
            \bm M_{af}^{(r)*}
        \right)
        \bm P_{f,r}^{\mathsf T}.
    \end{align}
\end{subequations}
The matrices $\bm P$ only reorder operator sectors. For open waveguides and
open bidirectional devices, the corresponding completed one-pass maps
$(\bm U_{aa},\bm U_{af})$ or $(\bm S_{aa},\bm S_{af})$ replace the closed-device
blocks in Eq.~\eqref{eq:gaussian_nonlinear_lift}.

If the optical inputs and the loss reservoirs are uncorrelated, Eq.~\eqref{eq:gaussian_nonlinear_lifted_map}
defines the Gaussian channel
\begin{equation}
    \bm V_{\mathrm{out}}
    =
    \bm X_{\mathrm{NL}}
    \bm V_{\mathrm{in}}
    \bm X_{\mathrm{NL}}^{\dagger}
    +
    \bm Y_{\mathrm{NL}},
    \qquad
    \bm Y_{\mathrm{NL}}
    =
    \sum_r
    \bm L_r
    \bm V_{f,r}
    \bm L_r^{\dagger}.
    \label{eq:gaussian_nonlinear_channel}
\end{equation}
For vacuum reservoirs, $\bm V_{f,r}=\bm I/2$. In particular, a spontaneous
nonlinear source driven from optical vacuum has
\begin{equation}
    \bm V_{\mathrm{src}}
    =
    \frac{1}{2}
    \bm X_{\mathrm{NL}}
    \bm X_{\mathrm{NL}}^{\dagger}
    +
    \bm Y_{\mathrm{NL}}.
    \label{eq:gaussian_nonlinear_vacuum_output}
\end{equation}
This conversion retains the full finite-gain Bogoliubov transformation and the
vacuum noise introduced by propagation and boundary loss. Number-preserving
QFC modules are represented by a vanishing anomalous block in
$\bm X_{\mathrm{NL}}$, whereas SPDC and SFWM modules generally produce nonzero
normal and anomalous correlations in Eq.~\eqref{eq:gaussian_complex_covariance}.

The conditional single-photon state in Supplementary Note~2 is non-Gaussian
and cannot itself be represented by a covariance matrix. To combine it with
the same circuit and photon-counting backend, it is embedded exactly as the
one-photon-heralded sector of a larger Gaussian state. Let the discretized
emitter density kernel in Eq.~\eqref{eq:sse_phoqus_density_kernel} have the
eigendecomposition
\begin{equation}
    \mathbf R_{\omega}
    =
    \mathbf W\mathbf\Lambda\mathbf W^{\dagger},
    \qquad
    \mathbf\Lambda
    =
    \operatorname{diag}
    (\lambda_1,\lambda_2,\ldots),
    \qquad
    \sum_\nu\lambda_\nu=1.
    \label{eq:gaussian_sse_eigendecomposition}
\end{equation}
Because Supplementary Note~2 adopts the correlation ordering
$R_{nm}=\langle\hat a_n^{\dagger}\hat a_m\rangle$, the associated physical
creation supermodes are
\begin{equation}
    \hat b_\nu^{\dagger}
    =
    \sum_n
    W_{n\nu}^{*}
    \hat a_n^{\dagger},
    \qquad
    \hat\rho_{\mathrm{SSE}}
    =
    \sum_\nu
    \lambda_\nu
    \hat b_\nu^{\dagger}
    \lvert0\rangle\langle0\rvert
    \hat b_\nu.
    \label{eq:gaussian_sse_supermodes}
\end{equation}
One auxiliary herald mode $\hat h_\nu$ is introduced for each occupied
supermode, and a product of two-mode squeezed vacua is prepared with
\begin{equation}
    \tanh^2 r_\nu
    =
    q\lambda_\nu,
    \qquad
    0<q<\lambda_{\max}^{-1}.
    \label{eq:gaussian_sse_embedding_gain}
\end{equation}
Explicitly, the Gaussian parent state is
\begin{equation}
    \lvert G_{\mathrm{SSE}}\rangle
    =
    \bigotimes_\nu
    \left[
        \sqrt{1-q\lambda_\nu}
        \sum_{p=0}^{\infty}
        \left(q\lambda_\nu\right)^{p/2}
        \lvert p\rangle_{b_\nu}
        \lvert p\rangle_{h_\nu}
    \right],
    \qquad
    \hat\rho_G^{\mathrm{SSE}}
    =
    \lvert G_{\mathrm{SSE}}\rangle
    \langle G_{\mathrm{SSE}}\rvert.
    \label{eq:gaussian_sse_parent_state}
\end{equation}
If $\Pi_h^{(1)}$ projects the auxiliary system onto exactly one photon in total,
without resolving which $h_\nu$ contains it, the Gaussian parent state
$\hat\rho_G^{\mathrm{SSE}}$ satisfies
\begin{equation}
    \operatorname{Tr}_h
    \left[
        \Pi_h^{(1)}
        \hat\rho_G^{\mathrm{SSE}}
    \right]
    =
    q
    \prod_\nu
    \left(
        1-q\lambda_\nu
    \right)
    \hat\rho_{\mathrm{SSE}}.
    \label{eq:gaussian_sse_exact_recovery}
\end{equation}
Thus, $q$ controls only the unconditioned Gaussian occupation and heralding
probability; after normalization, the recovered single-photon density matrix
is independent of $q$. Loss and mode mixing in either the emitter or herald
path can be inserted as Gaussian channels before the final projection.

\subsection{Frequency-resolved circuit propagation}
\label{subsec:gaussian-circuit-propagation}

After source construction, each optical mode is associated with a spatial
rail, a transverse-mode label, and an internal frequency grid. The global
annihilation vector is ordered first by these optical modes and then by their
internal spectral modes. Passive components act as
\begin{equation}
    \hat{\bm a}_{\mathrm{out}}
    =
    \bm R
    \hat{\bm a}_{\mathrm{in}}
    +
    \bm R_v
    \hat{\bm v},
    \qquad
    \bm X_R
    =
    \bm R\oplus\bm R^{*},
    \label{eq:gaussian_passive_operator_map}
\end{equation}
where $\hat{\bm v}$ denotes inaccessible vacuum inputs. When only the optical
ports are retained, a passive transformation satisfying
$\bm R\bm R^{\dagger}\preceq\bm I$ becomes
\begin{equation}
    \bm V_{\mathrm{out}}
    =
    \bm X_R
    \bm V_{\mathrm{in}}
    \bm X_R^{\dagger}
    +
    \frac{1}{2}
    \left(
        \bm I
        -
        \bm X_R\bm X_R^{\dagger}
    \right).
    \label{eq:gaussian_passive_channel}
\end{equation}
The additive term vanishes for a unitary optical transformation and restores
the vacuum covariance for a lossy channel.

Frequency-preserving circuit elements are applied independently at each sampled
frequency. For a component connecting a set of compatible rail and transverse
modes,
\begin{equation}
    \hat{\bm a}_{\mathrm{out}}(\omega_n)
    =
    \bm R(\omega_n)
    \hat{\bm a}_{\mathrm{in}}(\omega_n).
    \label{eq:gaussian_frequency_resolved_component}
\end{equation}
This form includes wavelength-dependent directional couplers and compact-model
scattering matrices. Propagation and programmed phase shifts correspond to
diagonal elements
$R_{jj}(\omega_n)=\exp[\mathrm{i}\phi_j(\omega_n)]$, while attenuation is
included through a contractive $\bm R(\omega_n)$. Layout-dependent propagation
phases and component responses are supplied by the electromagnetic and
fabrication models described in the subsequent sections.

Only matching transverse modes are mixed by a component. If a participating
rail does not contain a matching mode, the missing input is appended in vacuum
before the transformation. When two participating modes use different but
overlapping frequency grids, their spectral bases are interpolated onto a
common grid, orthonormalized, and completed to full unitary bases. The state is
expanded by the complementary vacuum modes before Eq.~\eqref{eq:gaussian_frequency_resolved_component}
is applied. This preserves the canonical commutation relations after the grid
synchronization rather than treating the interpolated basis as a nonunitary
mode map.

Sources and ancillary vacua are combined by direct sum, and rail permutations
place their covariance blocks in the circuit ordering. The transformations of
Eqs.~\eqref{eq:gaussian_channel_composition} and
\eqref{eq:gaussian_passive_channel} are then applied sequentially through the
network. Analyzer rotations used to define a measurement basis are treated as
ordinary passive circuit components. All propagation is performed in the full
frequency-resolved representation. The rail-local spectral reduction described
in Section~\ref{sec:spectral-reduction} is applied only to the final covariance
matrix before photon counting.

\subsection{Frequency-unresolved photon counting and conditional states}
\label{subsec:gaussian-photon-counting}

The detectors distinguish spatial rails but do not identify which internal spectral or transverse mode contains a detected photon. A detector on rail $r$ therefore measures the total photon number
\begin{equation}
\hat n_r=\sum_{\mu=1}^{K_r}\hat a_{r\mu}^{\dagger}\hat a_{r\mu},\qquad K=\sum_{r=1}^{R}K_r,
\label{eq:gaussian_rail_number}
\end{equation}
where $K_r$ is the number of internal modes on that rail. A threshold detector distinguishes $n_r=0$ from $n_r\geq1$, whereas a photon-number-resolving (PNR) detector distinguishes the individual values of $n_r$. Both measurements sum over the unresolved internal modes.

The input to the detector calculation is the final zero-displacement complex covariance $\bm V$, ordered as in Eq.~\eqref{eq:gaussian_doubled_basis}. We use
\begin{equation}
\bm Q=\bm V+\frac{1}{2}\bm I_{2K},\qquad\bm\Pi_K=
\begin{pmatrix}
0&\bm I_K\\ 
\bm I_K&0
\end{pmatrix},
\qquad\bm A=\bm\Pi_K(\bm I_{2K}-\bm Q^{-1}).
\label{eq:gaussian_detection_matrices}
\end{equation}
The matrix $\bm\Pi_K$ exchanges the annihilation and creation sectors, and $\bm A=\bm A^{\mathsf T}$ is the complex symmetric adjacency matrix entering the hafnian expressions below. The probability that all $K$ modes are in vacuum is
\begin{equation}
p_0=(\det\bm Q)^{-1/2}.
\label{eq:gaussian_vacuum_probability}
\end{equation}

\subsubsection{Threshold detection}

A rail produces no click only if all of its internal modes are in vacuum. For a set of rails $\mathcal S$, let $P_0(\mathcal S)$ be the probability that every rail in this set produces no click, regardless of the occupations elsewhere. This probability is calculated from the reduced covariance of those modes.

Let $\mathcal C$ and $\mathcal Z$ be the clicked and unclicked rail sets. The desired event requires vacuum on $\mathcal Z$ and excludes vacuum on each rail in $\mathcal C$. Inclusion--exclusion gives
\begin{equation}
P_{\mathrm{th}}(\mathcal C,\mathcal Z)=\sum_{\mathcal T\subseteq\mathcal C}(-1)^{|\mathcal T|}P_0(\mathcal Z\cup\mathcal T),\qquad P_0(\mathcal S)=(\det\bm Q_{\mathcal S})^{-1/2}.
\label{eq:gaussian_threshold_probability}
\end{equation}
Here $\bm Q_{\mathcal S}$ is the principal submatrix containing both annihilation and creation indices of every internal mode on $\mathcal S$, and $P_0(\varnothing)=1$.

For example, with two clicked rails, one starts with $P_0(\mathcal Z)$, subtracts the vacuum event for each clicked rail, and adds back the event in which both are in vacuum because it was subtracted twice. The general sum applies the same rule to every subset of $\mathcal C$ and includes all positive photon numbers on the clicked rails.

\subsubsection{Photon-number-resolving detection}

For a PNR outcome $\bm n=(n_1,\ldots,n_R)$, many internal-mode occupations can produce the same measured rail counts. For example, one photon on a rail with two internal modes can arise from occupations $(1,0)$ or $(0,1)$. Their probabilities must be added. More generally,
\begin{equation}
P_{\mathrm{PNR}}(\bm n)=\sum_{\substack{\bm s\\ \sum_{\mu=1}^{K_r}s_{r\mu}=n_r\;\text{for all }r}}\langle\bm s|\hat\rho_G|\bm s\rangle,
\label{eq:gaussian_pnr_internal_sum}
\end{equation}
where $\bm s$ specifies the occupation of every internal mode.

PhoQuS evaluates this sum using the $\Lambda$-blocked loop-hafnian framework
of Bulmer \textit{et al.}~\cite{Bulmer2026}. It assigns an auxiliary variable
$z_r$ to each rail, with each photon on that rail contributing one power of
$z_r$. All internal modes on the same rail receive the same variable:
\begin{equation}
\bm\Lambda(\bm z)=\bigoplus_{r=1}^{R}z_r\bm I_{K_r},\qquad\bm D(\bm z)=\bm\Lambda(\bm z)\oplus\bm\Lambda(\bm z).
\label{eq:gaussian_grouping_matrix}
\end{equation}
For zero displacement, the resulting generating function is
\begin{equation}
G(\bm z)=\frac{1}{\sqrt{\det[\bm I_{2K}-\bm D(\bm z)\bm\Pi_K\bm A]}},\qquad P_{\mathrm{PNR}}(\bm n)=p_0[\bm z^{\bm n}]G(\bm z).
\label{eq:gaussian_grouped_pnr}
\end{equation}
The notation $[\bm z^{\bm n}]$ selects the coefficient of $z_1^{n_1}\cdots z_R^{n_R}$ in the expansion about $\bm z=0$. Because the variables are shared within each rail, this coefficient already contains the sum over all compatible internal-mode occupations.

The numerical calculation proceeds in two steps. First, for the requested total photon number $d=\sum_r n_r$, a second auxiliary variable $t$ is used
to select terms containing exactly $d$ photons:
\begin{equation}
G_d(\bm z)=[t^d]G(t\bm z)=[t^d]\exp\!\left[\sum_{k=1}^{d}\frac{t^k}{2k}\operatorname{Tr}\left\{[\bm D(\bm z)\bm\Pi_K\bm A]^k\right\}\right].
\label{eq:gaussian_pnr_total_degree}
\end{equation}
Thus, $G_d$ is a polynomial whose terms all have total degree $d$. The implementation forms its coefficients from the matrix traces and the finite-order expansion of the exponential. Terms above order $d$ cannot contribute to $[t^d]$ and are not needed.

Second, the required distribution of these $d$ photons among the rails is selected by evaluating $G_d$ at roots of unity. Writing $z_{r,j}=\exp[2\pi ij/(n_r+1)]$, the probability is
\begin{equation}
P_{\mathrm{PNR}}(\bm n)=\frac{p_0}{\prod_r(n_r+1)}\sum_{j_1=0}^{n_1}\cdots\sum_{j_R=0}^{n_R}\left(\prod_{r=1}^{R}z_{r,j_r}^{-n_r}\right)G_d(z_{1,j_1},\ldots,z_{R,j_R}).
\label{eq:gaussian_pnr_root_sum}
\end{equation}
The phase factors cancel the unwanted powers of each variable. Together with the fixed total degree $d$, they leave only the coefficient associated with $\bm n$. This finite sum evaluates Eq.~\eqref{eq:gaussian_pnr_internal_sum} without explicitly listing its internal-mode photon allocations.

\subsubsection{Off-diagonal density-matrix elements and conditional states}

Photon-number probabilities are diagonal elements of a density matrix. Reconstructing a conditional state also requires its off-diagonal elements,
which describe coherences between different Fock states. For resolved occupation vectors $\bm p$ and $\bm q$, the Gaussian-state matrix element is
\begin{equation}
\langle\bm p|\hat\rho_G|\bm q\rangle=\frac{p_0\,\operatorname{haf}(\bm A_{\bm q\oplus\bm p})}{\sqrt{\prod_{\mu=1}^{K}p_\mu!q_\mu!}}.
\label{eq:gaussian_resolved_fock_element}
\end{equation}
In this convention, row and column $\mu$ of $\bm A$ are repeated $q_\mu$ times, while row and column $K+\mu$ are repeated $p_\mu$ times. Zero occupations omit the corresponding indices. The first half therefore corresponds to the ket and the second to the bra, as in Ref.~\cite{Bulmer2026}. The hafnian sums products of pair weights from the resulting repeated matrix.

For heralding, divide the modes into the measured rails $\mathcal H$ and the unmeasured modes whose state is to be reconstructed. Any other modes can first be traced out by selecting the relevant covariance submatrix. Let $\bm h$ be the measured rail counts and let $\bm p,\bm q$ now specify Fock occupations only of the retained, unmeasured modes. Then
\begin{equation}
\widetilde\rho_{\bm p,\bm q}^{(\bm h)}=\sum_{\substack{\{s_{r\mu}\}\\
\sum_{\mu=1}^{K_r}s_{r\mu}=h_r\;(r\in\mathcal H)}}\langle\{s_{r\mu}\},\bm p|
\hat\rho_G|\{s_{r\mu}\},\bm q\rangle.
\label{eq:gaussian_conditional_density_elements}
\end{equation}
For each possible internal occupation of the heralding rails, the same occupation appears in the bra and ket. Those contributions are summed because the detectors do not distinguish them. The retained occupations $\bm p$ and $\bm q$ remain independent, so their coherences are preserved. Unresolved heralding therefore performs an incoherent sum over the compatible detected occupations.

The grouped off-diagonal construction of Ref.~\cite{Bulmer2026} first collects the measured-mode indices into blocks corresponding to the heralding rails. The indices associated with the retained occupations are then repeated and permuted so that matrix elements with unequal bra and ket occupations can be evaluated using the $\Lambda$-blocked hafnian formulation. This calculation sums over the unresolved herald occupations, with the vacuum prefactor and Fock-state normalization taken from the original Gaussian state and the specified occupations. It produces each unnormalized conditional matrix element directly. Summing its diagonal gives the heralding probability, and dividing by that probability gives the conditional state:
\begin{equation}
p_{\bm h}=\operatorname{Tr}\widetilde{\bm\rho}^{(\bm h)},\qquad\bm\rho^{(\bm h)}=\widetilde{\bm\rho}^{(\bm h)}/p_{\bm h}.
\label{eq:gaussian_conditional_normalization}
\end{equation}
The trace equals the full heralding probability when the complete retained-mode Fock basis is included. With a finite Fock cutoff, it contains only the
probability mass represented within that cutoff, and normalizing the finite
matrix gives the conditional state restricted to that subspace. If several herald patterns are accepted, their unnormalized matrices are summed before normalization. Gaussian source and circuit propagation is therefore followed by a final photon-counting condition that can produce a non-Gaussian state.

\clearpage
\refstepcounter{section}
\section*{Supplementary Note \thesection.\\Electromagnetic simulations and photonic compact models}
\addcontentsline{toc}{section}{\thesection \quad Electromagnetic simulations and photonic compact models}
\label{sec:compact-models}

This section describes how electromagnetic responses are converted into reusable complex transformations evaluated during circuit simulation.

\subsection{Single-waveguide propagation model}
\label{sec:single-waveguide-model}

Propagation along a waveguide is evaluated by integrating the local propagation constant along its physical path. The effective index $n_{\mathrm{eff},s}$ and group index $n_{\mathrm{g},s}$ of segment $s$ are obtained from precomputed eigenmode data at the reference angular frequency $\omega_0$. The propagation constant is expanded to first order as
\begin{equation}
\beta_s(\omega)\simeq\frac{\omega_0}{c}n_{\mathrm{eff},s}+\frac{\omega-\omega_0}{c}n_{\mathrm{g},s}+\frac{\omega}{c}\Delta n_s,
\label{eq:path_local_beta}
\end{equation}
where $c$ is the vacuum speed of light and $\Delta n_s$ is an optional programmed index shift. For segment lengths $\Delta L_s$, the accumulated phase is
\begin{equation}
\phi_{\mathrm{path}}(\omega)=\sum_s\beta_s(\omega)\Delta L_s,
\label{eq:path_accumulated_phase}
\end{equation}
and the waveguide transformation is
\begin{equation}
\hat a(\omega)\longmapsto e^{i\phi_{\mathrm{path}}(\omega)}\hat a(\omega).
\label{eq:path_phase_transformation}
\end{equation}
Thus, $n_{\mathrm{eff}}$ determines the carrier phase and $n_{\mathrm{g}}$ determines the group delay. The programmed index shift is taken to be frequency independent over the modeled band. This model follows a prescribed local guided mode, neglecting coupling between transverse modes. Propagation loss can be handled separately by applying lumped loss after propagation, but didn't consider in this paper.

\subsubsection{Eigenmode database}
\label{sec:eigenmode_database}

The database stores sampled parameter vectors together with the effective and group indices of each calculated mode. For the waveguide model used here, each vector has seven entries,
\begin{equation}
\mathbf p=(\lambda,w,\theta_{\mathrm{sw}},h_{\mathrm{etch}},h_{\mathrm{film}},|\kappa|,\vartheta),
\label{eq:waveguide-database-parameters}
\end{equation}
representing wavelength, top width, sidewall angle, etch depth, film thickness, curvature magnitude and in-plane propagation orientation. Samples are assembled from parameter grids and evaluated using the finite-difference eigenmode solver in Ansys Lumerical MODE. Curvature is represented through the bent-waveguide solver, with straight propagation treated as the large-radius limit. The solver outputs $n_{\mathrm{eff}}$, $n_{\mathrm{g}}$ and the TE polarization fraction. Modes are assigned to TE or TM families using a TE polarization fraction of at least $0.5$ or below $0.5$, respectively, and stored separately by mode index.

\subsubsection{Interpolation of modal indices}

The propagation calculations use a separate local fit for each stored
wavelength slice. A query is assigned to a slice when its wavelength agrees with the stored value within a tolerance of $1~\mathrm{nm}$. For each modal index $f\in\{n_{\mathrm{eff}},n_{\mathrm{g}}\}$, samples with invalid or out-of-range values, defined here as values outside $[1,3.5]$, are excluded. The remaining geometric and orientation coordinates are normalized using their sampled ranges within the selected wavelength slice. Numerically constant coordinates are omitted.

For a query point $\mathbf q$ in the normalized parameter space, the $k=64$
nearest valid samples are selected according to their Euclidean distance.
Let $\mathcal N_k(\mathbf q)$ denote this neighborhood and let
$\Delta\mathbf x_j=\mathbf x_j-\mathbf q$. A query-centered quadratic
polynomial, containing the constant, linear, quadratic and cross terms, is fitted as
\begin{equation}
\mathbf c_f(\mathbf q)=\operatorname*{arg\,min}_{\mathbf c}\sum_{j\in\mathcal N_k(\mathbf q)}w_j^2\left[f_j-\mathbf v(\Delta\mathbf x_j)^{T}\mathbf c\right]^2,\qquad\widehat f(\mathbf q)=c_{f,0}(\mathbf q),
\label{eq:waveguide-index-fit}
\end{equation}
where $\mathbf v(\Delta\mathbf x)$ contains the centered polynomial basis. Because the polynomial is centered at the query, its fitted intercept $c_{f,0}$ is the predicted modal index.

The initial spatial weights are
\begin{equation}
w_j^{(0)}=\exp\left[-\left(\frac{d_j}{r_{\mathbf q}}\right)^2\right],\qquad d_j=\|\mathbf x_j-\mathbf q\|_2,\qquad r_{\mathbf q}=\max_{j\in\mathcal N_k(\mathbf q)}d_j .
\end{equation}
Four fitting passes are performed. Between passes, the weights are updated cumulatively using Tukey's bisquare function applied to median-centered residuals, with cutoff $4.685$ and a robust scale equal to $1.4826$ times the median absolute deviation. The effective- and group-index surfaces are fitted independently using the same procedure. Repeated queries are cached. The neighborhood size was fixed before the circuit simulations and verified against paired direct-FDE calculations along representative curved paths. Frequency dependence during propagation is then supplied by Eq.~\eqref{eq:path_local_beta}.

\subsubsection{Mapping a waveguide path to database parameters}

Each spatial rail carries a \texttt{gdsfactory} path and its current absolute terminal position. A new waveguide section is specified by path pieces with endpoint widths and a transverse-mode assignment. Its local vertices $\mathbf q_s$ are rotated by the terminal orientation and translated to the terminal position, giving absolute vertices $\mathbf r_s=\mathbf t+\mathbf R\mathbf q_s$. Adjacent vertices define
\begin{equation}
\Delta L_s=\|\mathbf r_{s+1}-\mathbf r_s\|_2,\qquad\mathbf r_{s+1/2}=\frac{\mathbf r_s+\mathbf r_{s+1}}{2}.
\label{eq:path_segment_geometry}
\end{equation}
Widths are assigned by linear interpolation between the endpoint values along the sampled vertices and averaged at segment midpoints. Curvature is evaluated from the path and resampled at those midpoints. The segment tangent sets the propagation orientation, including the placement rotation; the implementation uses curvature magnitude and folds orientation into $[0,90^\circ]$ according to the symmetry assumed by the index model.

The virtual-wafer fields of Section~\ref{sec:virtual-wafer} are evaluated at each absolute midpoint to form the local parameters,
\begin{equation}
p_{j,s}=p_{j,s}^{(0)}+\mu_j+\widetilde{\delta p}_j(\mathbf r_{s+1/2}).
\label{eq:path_segment_parameter}
\end{equation}
Only linewidth and film thickness receive nonzero wafer deviations in the fabrication-aware benchmark. The resulting parameter vectors select the modal indices used in Eq.~\eqref{eq:path_local_beta}. After applying the accumulated phase to the covariance matrix, the rail path and terminal position are updated for the next circuit operation.

\begin{figure}[h]
\centering
\includegraphics[width=1.0\textwidth]{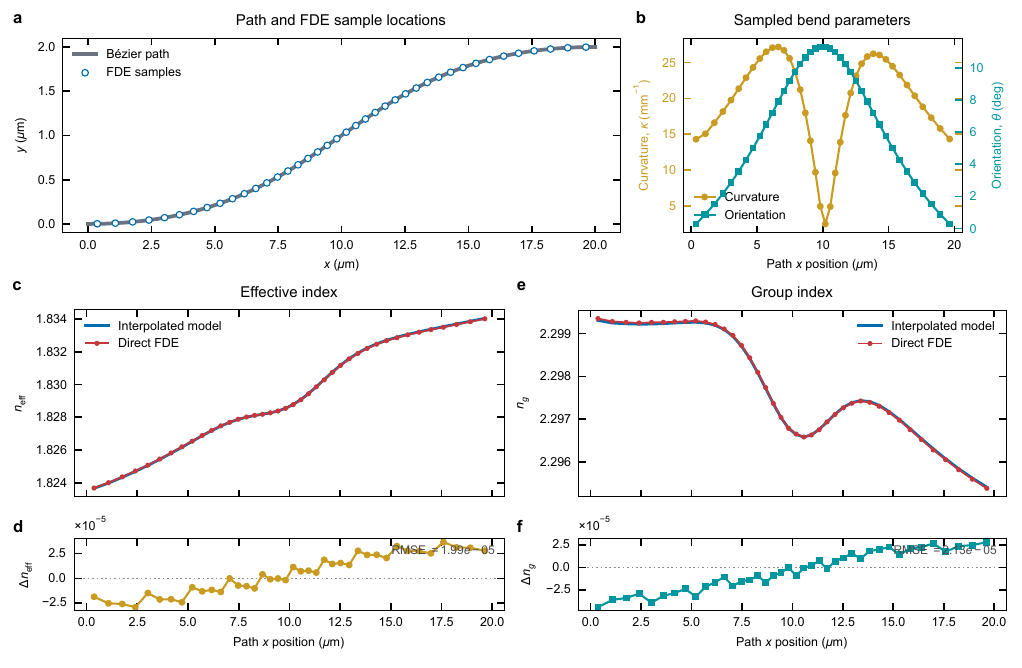}
\caption{\textbf{Index sampling example of a B\'ezier path.} 
\textbf{a}, Spatial profile of the B\'ezier path and the corresponding sampling points. 
\textbf{b}, Curvature and orientation of the B\'ezier path evaluated at each sampling point. 
\textbf{c}, Comparison of the sampled effective index between the interpolated model and values directly calculated from the FDE solver at the sampling points. 
\textbf{d}, Sampling error in the effective index, defined as $\Delta n_\mathrm{eff} = n_\mathrm{eff,model} - n_\mathrm{eff,FDE}.$ 
\textbf{e}, Comparison of the sampled effective group index between the interpolated model and values directly calculated from the FDE solver at the sampling points. 
\textbf{f}, Sampling error in the group index, defined as $\Delta n_\mathrm{g} = n_\mathrm{g,model} - n_\mathrm{g,FDE}.$}

\label{fig:index_error}
\end{figure}

For example, a cubic B\'ezier connection from $(0,0)$ to $(L_x,L_y)$ uses control points $(0,0)$, $(L_x/2,0)$, $(L_x/2,L_y)$ and $(L_x,L_y)$. Sampling this curve supplies the segment lengths, curvatures, orientations and wafer coordinates required by the same propagation model. Fig.~\ref{fig:index_error} demonstrates the sampling of a B\'ezier path example ($L_x=20\,\mu\mathrm{m}$, $L_y=2\,\mu\mathrm{m}$) with the linewidth linearly increasing from $0.95\,\mu\mathrm{m}$ to $1.05\,\mu\mathrm{m}$ along the path.

\subsection{Photonic device compact model}

Complex modal transmission coefficients are calculated using Ansys Lumerical eigenmode expansion (EME) simulations. Broadband responses are obtained using the built-in perturbative wavelength sweep, which assumes that the mode profiles do not change significantly over the sampled wavelength range. The resulting complex responses are converted into reusable compact models that return transmission amplitudes and phases at the requested wavelength and local fabrication parameters.

\subsubsection{Design and fabrication parameter sampling}

For each nominal device design, the sampled fabrication parameters are waveguide linewidth $w$ and film thickness $h$. The deviations $\Delta w\in\{-50,0,50\}~\mathrm{nm}$ and $\Delta h\in\{-25,0,25\}~\mathrm{nm}$ form a $3\times3$ parameter grid. At each grid point, the complex response is sampled at 501 spectral points spanning $1500$--$1600~\mathrm{nm}$, corresponding to a wavelength spacing of approximately $0.2~\mathrm{nm}$.

The spectral sampling must resolve the phase accumulated between adjacent points. For a approximated response with path length $L$ and group index $n_{\mathrm g}$,
\begin{equation}
|\Delta\phi|\simeq \frac{2\pi n_{\mathrm g}L}{\lambda^2}\Delta\lambda<\pi,\qquad\Delta\lambda<\frac{\lambda^2}{2n_{\mathrm g}L}.
\label{eq:compact-model-phase-sampling}
\end{equation}
For $L=1~\mathrm{mm}$, $n_{\mathrm g}=2.5$ and $\lambda=1550~\mathrm{nm}$, the upper bound is $0.4805~\mathrm{nm}$; a $0.2~\mathrm{nm}$ interval gives an estimated phase increment of $1.31~\mathrm{rad}$. This provides a sampling criterion for the propagation phase of the non-resonant devices considered here. Rapid phase changes near transmission zeros or resonant device require separate attention.

\subsubsection{Interpolation of complex responses}

Interpolation is performed separately on the amplitude and phase of each modal transmission coefficient. Let $S_{ba}(\lambda,\mathbf p)$ denote the coefficient from input mode $a$ to output mode $b$, with $\mathbf p=(h,w)$. A fixed reference wavelength $\lambda_*$ is chosen as the stored spectral sample nearest $1550~\mathrm{nm}$. At each fabrication-grid point, the response is decomposed as
\begin{equation}
\begin{split}
A_{ba}(\lambda,\mathbf p)&=|S_{ba}(\lambda,\mathbf p)|,\\
u_{ba}(\mathbf p)&=\frac{S_{ba}(\lambda_*,\mathbf p)}{|S_{ba}(\lambda_*,\mathbf p)|},\\
\delta\phi_{ba}(\lambda,\mathbf p)&=\operatorname{unwrap}_{\lambda}\arg\!\left[S_{ba}(\lambda,\mathbf p)u_{ba}^{*}(\mathbf p)\right].
\end{split}
\label{eq:compact-model-phase-decomposition}
\end{equation}
This phase representation requires nonzero transmission amplitude at the reference wavelength. Unwrapping proceeds separately towards longer and shorter wavelengths from $\lambda_*$, with $\delta\phi_{ba}(\lambda_*,\mathbf p)=0$. The reference phase is retained separately, so that this operation does not remove fabrication-dependent phase shifts at the reference wavelength.

Reference-phase branches are first selected on the original $3\times3$ thickness--linewidth grid and then calibrated using complex transmission data at four additional off-grid parameter points. At each original grid point, the initial reference phase is written as
\begin{equation}
\phi_{*,ba}^{(0)}(\mathbf p_j)=\arg u_{ba}(\mathbf p_j)+2\pi k_{ba,j},\qquad k_{ba,j}\in\mathbb Z.
\label{eq:compact-model-reference-branches}
\end{equation}
The initial integer offsets are selected jointly across the grid using a smoothness criterion. With the central offset fixed to zero and adjacent offsets constrained to differ by at most one, a two-stage mixed-integer linear optimization first minimizes changes in the phase slope along and between the two parameter directions, then minimizes the phase increments between adjacent points. Parameter intervals are normalized by their mean grid spacing.

This initial assignment is refined using the off-grid data. The calibrated reference phases are
\begin{equation}
\phi_{*,ba}(\mathbf p_j)=\phi_{*,ba}^{(0)}(\mathbf p_j)+2\pi c_{g,j},\qquad c_{g,j}\in\{-1,0,1\},
\label{eq:compact-model-calibrated-branches}
\end{equation}
where $g$ labels a group of transmission coefficients sharing the same correction. The central correction is fixed to zero.

Let $\ell_j(\mathbf p)$ denote the tensor-product quadratic Lagrange weight associated with original grid point $j$. The correction applied at an off-grid point is therefore
\begin{equation}
\Delta\phi_g(\mathbf p)=2\pi\sum_{j=1}^{9}\ell_j(\mathbf p)c_{g,j}.
\label{eq:compact-model-branch-correction}
\end{equation}
Although the corrections are integer multiples of $2\pi$ at the original grid nodes, their interpolated values need not be. They can therefore change the predicted phase between grid points without changing the stored complex transmission at those nodes.

For each group, all $3^8=6561$ correction assignments are evaluated. The selected assignment minimizes the mean squared phase difference between the corrected model and the off-grid complex transmission data, with each difference wrapped into $[-\pi,\pi)$:
\begin{equation}
J_g(\mathbf c_g) =\frac{1}{N_g}\sum_{(ba,m,\nu)\in\mathcal I_g}\left[\operatorname{wrap}_{[-\pi,\pi)}\left\{\arg\left[\widehat S_{ba}^{(0)}(\lambda_\nu,\mathbf p_m)S_{ba}^{\mathrm{cal}}(\lambda_\nu,\mathbf p_m)^*\right]
+\Delta\phi_g(\mathbf p_m)\right\}\right]^2.
\label{eq:compact-model-branch-calibration-objective}
\end{equation}
Here $\widehat S_{ba}^{(0)}$ is the model before calibration, $S_{ba}^{\mathrm{cal}}$ is the stored off-grid response, and $\mathcal I_g$ contains the valid channel, parameter-point and wavelength combinations, with $N_g=|\mathcal I_g|$. The calculation uses the native off-grid samples within the model's wavelength range, rather than values interpolated from the off-grid data. Samples are included only when both transmission amplitudes exceed $10^{-9}$.

Calibration modifies only the reference-phase branches. The amplitude interpolants and reference-relative spectral phases remain unchanged. The added phase is independent of wavelength at a fixed fabrication point, so it does not change the spectral phase slope of an individual calibrated coefficient.

The four additional parameter points constrain branch selection but do not establish a unique physical phase winding over the fabrication domain. Because they are used for calibration, comparisons at these points measure calibration agreement rather than independent validation. Additional parameter points not used in branch selection would be required for an independent assessment of the interpolated phase.

Once the phase branches have been fixed, three real-valued quantities are interpolated separately: the amplitude $A_{ba}$, the wavelength-dependent phase change $\delta\phi_{ba}$ and the reference phase $\phi_{*,ba}$. The first two depend on $(\lambda,h,w)$, whereas the reference phase depends only on $(h,w)$.

Interpolation along each coordinate uses a quadratic polynomial through three consecutive grid samples. For a query coordinate $x$ and selected sample coordinates $x_0,x_1,x_2$, the Lagrange weights are
\begin{equation}
L_j^{(x)}(x)=\prod_{\substack{m=0\\m\neq j}}^{2}\frac{x-x_m}{x_j-x_m},\qquad j=0,1,2.
\label{eq:compact-model-lagrange-weights}
\end{equation}
A sampled scalar quantity $f$ is then interpolated as $\widehat f(x)=\sum_{j=0}^{2}L_j^{(x)}(x)f(x_j)$. At a sample coordinate, its own weight is one and the other weights are zero, so the stored value is reproduced.

For the three-dimensional quantities $f\in\{A_{ba},\delta\phi_{ba}\}$, the weights are multiplied across the wavelength, thickness and linewidth coordinates:
\begin{equation}
\widehat f(\lambda,h,w)=\sum_{i,j,k=0}^{2}L_i^{(\lambda)}(\lambda)L_j^{(h)}(h)L_k^{(w)}(w)f(\lambda_i,h_j,w_k).
\label{eq:compact-model-tensor-interpolation}
\end{equation}
This is equivalent to applying one-dimensional quadratic interpolation successively along the three coordinates. Because the fabrication grid contains three samples along each parameter direction, all nine fabrication-grid points are used. Together with three consecutive wavelength samples selected for the query, each evaluation uses 27 stored values. Near the wavelength boundaries, the selected samples remain within the stored grid.

The reference phase is interpolated over the fabrication grid alone:
\begin{equation}
\widehat\phi_{*,ba}(h,w)=\sum_{j,k=0}^{2}L_j^{(h)}(h)L_k^{(w)}(w)\phi_{*,ba}(h_j,w_k).
\label{eq:compact-model-reference-interpolation}
\end{equation}
Finally, the interpolated reference phase and wavelength-dependent phase change are added and combined with the interpolated amplitude:
\begin{equation}
\widehat S_{ba}(\lambda,\mathbf p)=\widehat A_{ba}(\lambda,\mathbf p)\exp\!\left\{i\left[\widehat\phi_{*,ba}(\mathbf p)+\widehat{\delta\phi}_{ba}(\lambda,\mathbf p)
\right]\right\}.
\label{eq:compact-model-reconstruction}
\end{equation}
The phase branches are fixed during model construction and are not recalculated from the query wavelengths.


\subsection{Directional-coupler compact model}
\label{sec:dc-compact-model}

The directional coupler (DC) is divided into an input transition, a uniform coupling region and an output transition. The transitions are characterized by EME simulations, whereas propagation through the uniform region is calculated from its symmetric and antisymmetric supermodes. This decomposition allows the coupling length to be varied without repeating the EME transition simulations.

For each sampled film thickness and linewidth, EME simulations provide the complex transmission coefficients from each input waveguide mode to the two coupling-region supermodes. These coefficients form the transition matrix
\begin{equation}
T_{\mathrm{in}}=
\begin{pmatrix}
t_{1s} & t_{2s}\\
t_{1a} & t_{2a}
\end{pmatrix},
\label{eq:dc-input-transition}
\end{equation}
where $t_{j\mu}$ connects input waveguide $j$ to supermode $\mu\in\{s,a\}$. Each coefficient is interpolated using the amplitude--phase procedure described above. For the corresponding reversed output transition, reciprocity gives $T_{\mathrm{out}}=T_{\mathrm{in}}^{T}$ under the same modal normalization and consistent port phase conventions.

The effective and group indices of both supermodes are sampled on the same film-thickness and linewidth grid. Their interpolated models are constructed using the procedure in Section~\ref{sec:single-waveguide-model}. Propagation through a coupling region of length $L_{\mathrm{DC}}$ is represented by
\begin{equation}
P_{\mathrm{DC}}(\omega,\mathbf p)=
\begin{pmatrix}
e^{i\beta_s(\omega,\mathbf p)L_{\mathrm{DC}}} & 0\\
0 & e^{i\beta_a(\omega,\mathbf p)L_{\mathrm{DC}}}
\end{pmatrix},
\label{eq:dc-supermode-propagation}
\end{equation}
with
\begin{equation}
\beta_\mu(\omega,\mathbf p)=\frac{\omega_0}{c}n_{\mathrm{eff},\mu}(\omega_0,\mathbf p)+\frac{\omega-\omega_0}{c}n_{\mathrm g,\mu}(\omega_0,\mathbf p),\qquad \mu\in\{s,a\}.
\label{eq:dc-supermode-beta}
\end{equation}
Here $\mathbf p=(h,w)$ and $\omega_0$ is the expansion frequency.

The complete forward transmission matrix is
\begin{equation}
S_{\mathrm{DC}}=T_{\mathrm{out}}P_{\mathrm{DC}}T_{\mathrm{in}}.
\label{eq:dc-compact-response}
\end{equation}
The output amplitudes therefore include interference between the symmetric and antisymmetric propagation channels. The model retains these two supermodes and neglects multiple reflections between the transition regions.

For each fabrication-parameter query, the matrices are combined on the original EME wavelength grid. The amplitude and reference-relative unwrapped phase of each resulting port-to-port coefficient are then linearly interpolated to the wavelengths required for circuit propagation.

\subsection{Rapid adiabatic coupler compact model}
\label{sec:rac-compact-model}

The rapid adiabatic coupler (RAC) is simulated as a complete device using EME. At each sampled film thickness and linewidth, excitation of each input waveguide mode provides the complex transmission coefficients into the two output waveguide modes. These coefficients form a frequency-dependent $2\times2$ forward transmission matrix for each nominal RAC design.

Each coefficient is converted into a compact model using the amplitude--phase interpolation procedure described above. The reference-phase branches are selected jointly over the fabrication grid before interpolation. A circuit query therefore returns the complete RAC transmission response at the specified wavelengths, film thickness and linewidth.

Additional waveguides used for port alignment or delay matching are modeled separately using the path-propagation model of Section~\ref{sec:single-waveguide-model}. Their propagation phases are composed with the EME-derived RAC response according to the circuit layout.

\subsection{Off-grid comparison after phase-branch calibration}

We compared the compact-model responses with responses obtained from directly sampled electromagnetic data at four off-grid parameter combinations, $\Delta w\in\{-25,25\}~\mathrm{nm}$ and $\Delta h\in\{-12.5,12.5\}~\mathrm{nm}$. For the DC, the reference response is assembled from the off-grid transition coefficients and supermode indices using the same model construction described above. For the RACs, the reference consists of the directly simulated port-to-port transmission coefficients.

\begin{figure}[h]
\centering
\includegraphics[width=1.0\textwidth]{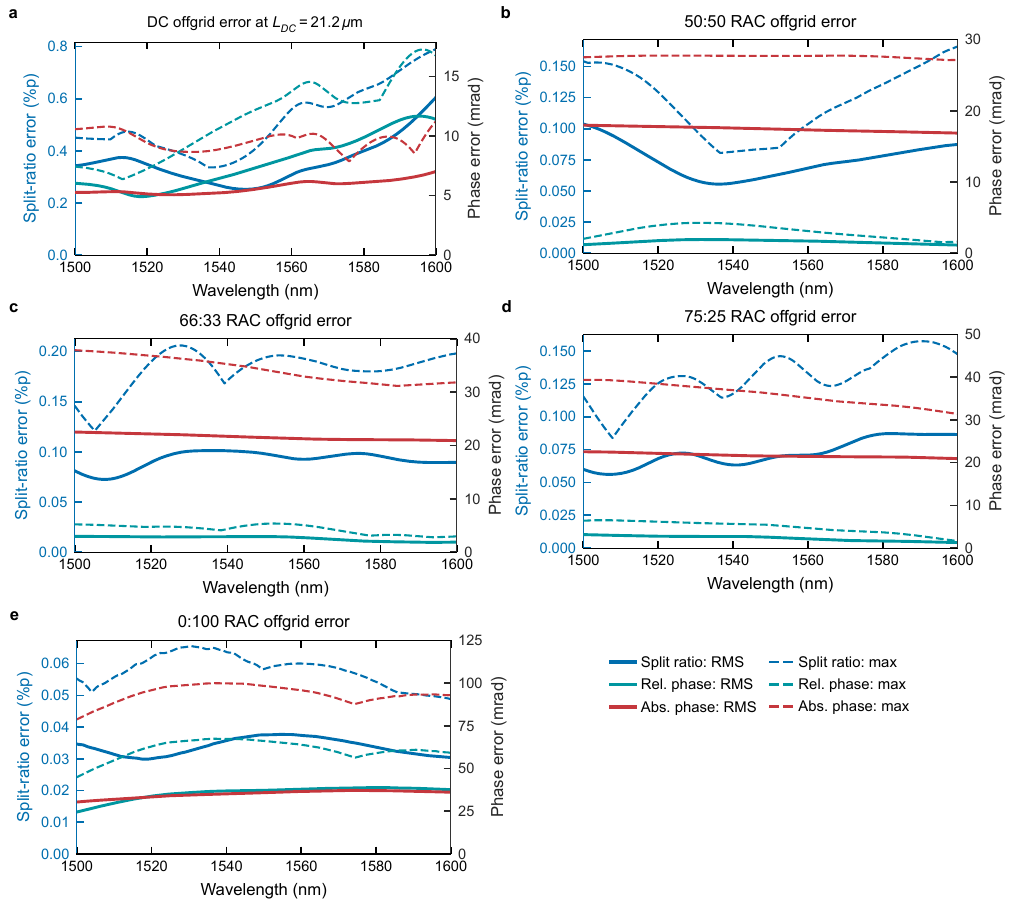}
\caption{\textbf{Off-grid residuals after phase-branch calibration.}
Differences between compact-model responses and responses obtained from directly sampled off-grid electromagnetic data for the DC and RAC models. Blue curves show splitting-ratio errors in percentage points, with the splitting ratio defined as the power in one output divided by the total power in the two outputs for a given input. Red curves show phase errors of individual complex transmission coefficients, and teal curves show errors in the relative phase between the two outputs for the same input. Phase differences are wrapped to the interval $(-\pi,\pi]$. Solid and dashed curves denote the root-mean-square and maximum absolute errors, respectively, over the four off-grid parameter combinations and the corresponding transmission channels at each wavelength. These parameter points were also used for phase-branch calibration; the plotted residuals therefore do not constitute an independent test of phase interpolation at unsampled points.}
\label{fig:offgrid}
\end{figure}

Figure~\ref{fig:offgrid} reports the residual differences in splitting ratio, individual transmission phase and relative phase between the two outputs. Because these off-grid data were also used to calibrate the reference-phase branches, the comparison characterizes the residual errors after calibration rather than providing an independent validation of the interpolated phase. It also does not measure agreement with a fabricated device.

Agreement at these parameter points does not exclude alternative phase continuations or larger errors at unsampled points between them and the original grid. Subsequent circuit calculations assume that the selected phase continuation remains applicable throughout the sampled parameter domain; uncertainty associated with alternative branch assignments is not propagated into the reported circuit results.

\subsection{Phase shifter and MZI models}
\label{sec:phase-shifter-mzi-models}

A phase shifter is modeled as a waveguide section with a prescribed effective-index change $\Delta n$, taken to be independent of frequency over the simulated bandwidth. For a uniform control section of length $L_{\mathrm h}$, the additional phase is
\begin{equation}
\Delta\phi(\omega)=\frac{\omega L_{\mathrm h}}{c}\Delta n =\frac{2\pi L_{\mathrm h}}{\lambda}\Delta n.
\label{eq:phase-shifter-index-change}
\end{equation}
This phase is added to the propagation phase calculated from the local waveguide geometry. Although $\Delta n$ is frequency independent, the induced phase is linear in frequency and therefore also changes the group delay by $L_{\mathrm h}\Delta n/c$. The control parameter specifies an optical index change; no electrical or thermal actuator model is used.

A tunable Mach--Zehnder interferometer (MZI) is assembled from two coupler models, the intervening waveguide arms and an input phase section. Its forward transmission matrix is
\begin{equation}
S_{\mathrm{MZI}}(\omega)=C_2(\omega)D_{\mathrm{arm}}(\omega)C_1(\omega)D_{\mathrm{in}}(\omega),
\label{eq:mzi-compact-response}
\end{equation}
where $C_1$ and $C_2$ are the DC or RAC transmission matrices, with port ordering chosen according to their physical orientation. The diagonal matrices $D_{\mathrm{in}}$ and $D_{\mathrm{arm}}$ contain the propagation phases of the input sections and internal arms, respectively. Independent index changes on one input rail and one internal arm provide the two control parameters.

Each coupler is evaluated using the fabrication parameters sampled at its own layout position, while waveguide phases are accumulated along their paths. Consequently, the two couplers and arms can experience different local fabrication variations. The assembled model retains the wavelength dependence of both the coupler responses and the controlled propagation phases. The control lengths and optimization protocol used in the GHZ benchmark are specified in Section~\ref{sec:ghz-benchmark}.

\clearpage
\refstepcounter{section}
\section*{Supplementary Note \thesection.\\Virtual-wafer generation and layout-aware parameter assignment}
\addcontentsline{toc}{section}{\thesection \quad Virtual-wafer generation and layout-aware parameter assignment}
\label{sec:virtual-wafer}

Fabrication variation in a photonic circuit is generally correlated over space, so components placed near one another cannot be treated as statistically independent instances. Layout-aware simulation therefore requires a spatial process model whose local values can be assigned to physical component coordinates \cite{Lu2017,Bogaerts2019,Xing2023}. For the controlled benchmark considered here, linewidth and film-thickness deviations were represented by stationary Gaussian-correlated random fields. This model captures the prescribed variation amplitude and correlation length but does not attempt to reproduce the full hierarchical decomposition of wafer-scale, die-scale, layout-dependent and residual random variation observed in foundry data \cite{Xing2023}.

\subsection{Statistical model of fabrication variation}

For process parameter $j$, the local physical value at in-plane position $\mathbf{r}=(x,y)$ was written as

\begin{equation}
p_j(\mathbf{r})=p_{j,0}+\mu_j+\delta p_j(\mathbf{r}),
\label{eq:local_process_parameter}
\end{equation}

where $p_{j,0}$ is the nominal design value, $\mu_j$ is a spatially uniform offset and $\delta p_j(\mathbf{r})$ is a zero-mean within-wafer fluctuation. The fluctuation was modeled as a stationary and isotropic Gaussian random field satisfying

\begin{equation}
\mathbb{E}\left[\delta p_j(\mathbf{r})\right]=0
\end{equation}

and

\begin{equation}
\mathbb{E}\left[\delta p_j(\mathbf{r})\delta p_j(\mathbf{r}+\mathbf{d})\right]=\sigma_j^2\exp\left[-\left(\frac{\lVert\mathbf{d}\rVert}{\ell_j}\right)^2\right],
\label{eq:process_covariance}
\end{equation}

where $\sigma_j$ is the prescribed within-wafer standard deviation and $\ell_j$ is the $1/e$ spatial correlation length.

Two process parameters were varied in the fabrication-aware benchmark. The linewidth deviation had $\sigma_{\mathrm{w}}=10~\mathrm{nm}$ and $\ell_{\mathrm{w}}=0.5~\mathrm{mm}$, whereas the film-thickness deviation had $\sigma_{\mathrm{t}}=5~\mathrm{nm}$ and $\ell_{\mathrm{t}}=1.0~\mathrm{mm}$. Separate pseudorandom realizations were generated for the two parameters, and no cross-correlation between linewidth and thickness variation was imposed.

The generator also permits a spatially uniform wafer-to-wafer offset drawn from $\mathcal{N}(0,\sigma_j'^2)$. This contribution was disabled in the present benchmark by setting $\sigma_j'=0$.

The source covariance matrices were fixed for each source choice and were not sampled from these process fields. The virtual-wafer model therefore represents passive-device linewidth and thickness variation only.

\subsection{Generation of Gaussian-correlated random fields}

The correlated fields were generated using the Fourier-filtering approach introduced for layout-aware photonic-circuit simulation in Ref.~\cite{Lu2017}. For each process parameter, an uncorrelated real-valued random array was first generated on a two-dimensional Cartesian grid,

\begin{equation}
z_j(x_m,y_n)
\sim
\mathcal{N}(0,1).
\end{equation}

The array was convolved with the isotropic Gaussian kernel

\begin{equation}
g_j(\mathbf{r})\propto\exp\left[-\frac{2\lVert\mathbf{r}\rVert^2}{\ell_j^2}\right].
\label{eq:gaussian_filter}
\end{equation}

The multiplicative normalization of $g_j$ does not affect the final field because the convolved map is subsequently standardized. The discrete convolution was evaluated using two-dimensional fast Fourier transforms,

\begin{equation}
m_j=\mathcal{F}^{-1}\left[\mathcal{F}[z_j]\cdot\mathcal{F}[g_j]\right],
\label{eq:fft_convolution}
\end{equation}

where $\mathcal{F}$ and $\mathcal{F}^{-1}$ denote the discrete Fourier transform and its inverse.

For spatially white input noise, the covariance of the filtered field is proportional to the autocorrelation of the filter kernel. The Gaussian kernel in equation~\eqref{eq:gaussian_filter} gives

\begin{equation}
\int g_j(\mathbf{u})g_j(\mathbf{u}+\mathbf{d})\mathrm{d}^2\mathbf{u}\propto\exp\left[-\left(\frac{\lVert\mathbf{d}\rVert}{\ell_j}\right)^2\right],
\end{equation}

so that $\ell_j$ corresponds to the $1/e$ correlation length of the resulting process field rather than the width of the filtering kernel itself.

The finite-grid result was standardized separately for each process parameter,

\begin{equation}
\delta p_j(x_m,y_n)
=
\sigma_j
\frac{
m_j(x_m,y_n)-\overline{m}_j
}{
s[m_j]
},
\label{eq:field_normalization}
\end{equation}

where $\overline{m}_j$ and $s[m_j]$ are the spatial sample mean and sample standard deviation of the convolved map. Equation~\eqref{eq:field_normalization} enforces the prescribed zero mean and standard deviation on the finite computational domain while retaining its spatial correlation structure. The linewidth and thickness realizations were generated using fixed pseudorandom seeds of 50,000 and 100,000, respectively, so that every circuit configuration was evaluated on the same stored process fields.

\subsection{Grid discretization and interpolation}

Both process fields were generated over the square domain

\begin{equation}
-20~\mathrm{mm}
\leq x,y \leq
20~\mathrm{mm}
\end{equation}

using $N=251$ grid points along each coordinate. The resulting grid spacing was

\begin{equation}
\Delta x=\Delta y
=
\frac{40~\mathrm{mm}}{N-1}
=
0.16~\mathrm{mm}.
\end{equation}

Because equation~\eqref{eq:fft_convolution} was evaluated without zero padding, the discrete Fourier transform implements circular convolution and therefore imposes periodic boundary conditions on the $40\times40~\mathrm{mm}^2$ computational domain. The RAC--MZI circuit used in the benchmark had a footprint of $30{,}000~\mu\mathrm{m}\times270~\mu\mathrm{m}$. Its longitudinal extent began at the left boundary of the domain and ended $10~\mathrm{mm}$ before the opposite boundary. The shortest separation introduced by periodic wrapping between longitudinally separated parts of the layout was therefore $10~\mathrm{mm}$, ten times the largest correlation length used, $\ell_{\max}=1.0~\mathrm{mm}$. For the prescribed correlation function $C(r)=\exp[-(r/\ell)^2]$, the corresponding artificial correlation is bounded by $\exp(-100)$, and is therefore negligible. The periodic boundary condition consequently does not introduce appreciable correlations between physically separated parts of the circuit layout.

After field generation, each discrete process field was converted into a continuous coordinate-dependent function using bicubic tensor-product spline interpolation. The resulting interpolants, $\widetilde{\delta p}_{\mathrm{w}}(x,y)$ and $\widetilde{\delta p}_{\mathrm{t}}(x,y)$, supplied the linewidth and film thickness deviations, respectively, at arbitrary coordinates along each placed circuit layout.

\begin{figure}[h]
\centering
\includegraphics[width=1.0\textwidth]{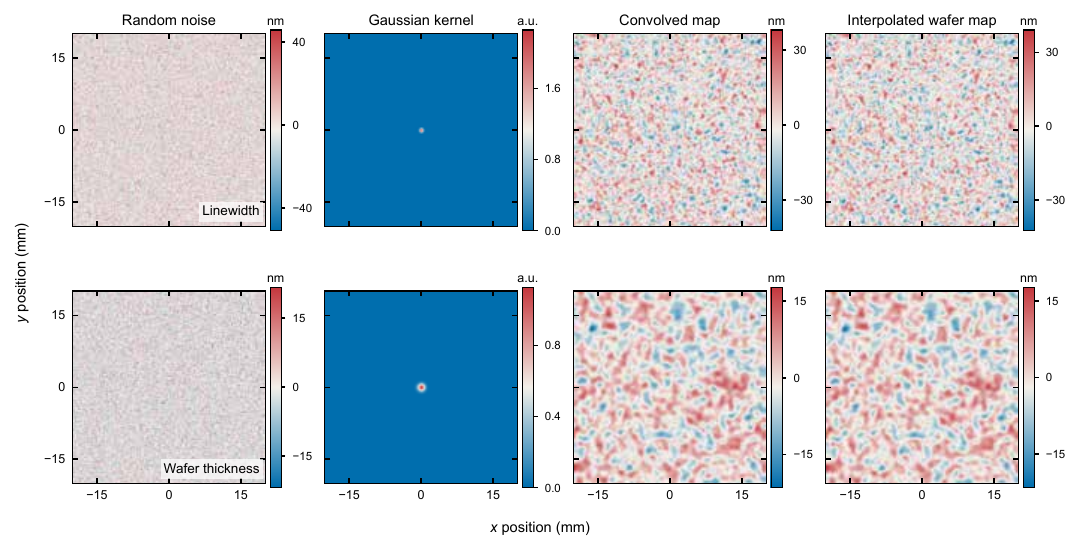}
\caption{
\textbf{Generation of spatially correlated virtual-wafer process fields.}
From left to right, the columns show the uncorrelated Gaussian input, Gaussian filtering kernel, standardized FFT-convolved field and bicubic-spline-interpolated field. The upper row shows linewidth variation with a standard deviation of 10~nm and a correlation length of 0.5~mm. The lower row shows film-thickness variation with a standard deviation of 5~nm and a correlation length of 1.0~mm. Both fields were generated on a 251$\times$251 grid covering a 40~mm$\times$40~mm domain. The larger spatial features in the film-thickness map result from its longer correlation length.
}

\label{fig:wafer_map_visualization}
\end{figure}

\subsection{Mapping layout components to wafer fields}
\label{sec:layout-wafer-mapping}

Each component model described in Section~\ref{sec:compact-models} was parameterized by the local linewidth and film thickness. To assign these parameters to a placed circuit, the logical netlist was first converted into a physical coordinate representation. The simulation state stored the absolute terminal position of each optical rail, while each compact component stored the relative coordinates of its input and output ports. For port $q$ of component $i$, the absolute coordinate was obtained as

\begin{equation}
\mathbf{r}_{i,q}=\mathbf{t}_i+\mathbf{R}_i\mathbf{r}_{i,q}^{(\mathrm{rel})},
\label{eq:absolute_port_coordinate}
\end{equation}

where $\mathbf{r}_{i,q}^{(\mathrm{rel})}$ is the port coordinate in the component-local reference frame, $\mathbf{R}_i$ represents the orientation or reflection applied during placement and $\mathbf{t}_i$ is the translation required to align the component input ports with the current rail terminals. Port-coordinate consistency was checked when connecting a multiport component to ensure that all of its input ports coincided with their corresponding rail terminals.

\subsubsection{Representative-point sampling of compact components}

The compact directional-coupler and RAC models were generated under the assumption that the geometric parameters are uniform within each simulated device. Their layout-dependent responses were therefore evaluated using one representative process coordinate per placed component. This coordinate was taken to be the centroid of the component ports,

\begin{equation}
\overline{\mathbf{r}}_i=
\frac{1}{N_{p,i}}
\sum_{q=1}^{N_{p,i}}
\mathbf{r}_{i,q},
\label{eq:component_sampling_coordinate}
\end{equation}

where $N_{p,i}$ is the number of ports of component $i$. For the delay-matched RAC implementation, the equivalent geometric centre was determined from the total longitudinal extent of the component and the midpoint between its two input rails.

The process parameters assigned to component $i$ were then

\begin{equation}
p_{i,j}=p_{i,j}^{(0)}+\mu_j+\widetilde{\delta p}_{j}\left(\overline{\mathbf{r}}_i\right),
\label{eq:component_local_parameter}
\end{equation}

where $p_{i,j}^{(0)}$ is the nominal parameter used to construct the compact model. In the present benchmark, $j\in\{\mathrm{w},\mathrm{t}\}$ denotes linewidth and film thickness, respectively, and $\mu_j=0$. The resulting local parameter vector was supplied to the corresponding interpolated compact model to obtain its frequency-dependent scattering matrix,

\begin{equation}
\mathbf{S}_i(\omega)=\mathbf{S}_i\left(\omega;p_{i,\mathrm{w}},p_{i,\mathrm{t}},\mathbf{u}_i\right),
\label{eq:layout_dependent_compact_response}
\end{equation}

where $\mathbf{u}_i$ denotes any additional design or control parameters of the component. The evaluated complex scattering matrix, including its wavelength-dependent amplitude and phase, was then assigned to that component instance in the circuit.

This single-point assignment is expected to be accurate when the component footprint is small compared with the shortest process correlation length. It does not resolve gradients occurring within an extended compact-device region. This limitation is most relevant to structures such as RACs whose longitudinal extent may no longer be negligible compared with $\ell_{\mathrm{w}}$. The fabrication-aware results should therefore be interpreted under a locally uniform compact-component approximation.

A distributed treatment could instead divide an extended device into longitudinal sections and evaluate the local modes and inter-section overlap matrices using fabrication-parameter-dependent datasets. Dataset-based eigenmode-expansion methods provide an efficient route to such a model by reusing precomputed modal fields and overlap integrals \cite{Song2025}. Incorporating this distributed within-device variation would, however, require a different compact-model construction and was outside the scope of the present benchmark.

\subsubsection{Path-resolved sampling of connecting waveguides}

Routing waveguides and interferometer arms represented explicitly as paths were treated with finer spatial resolution. Each path was stored as an ordered sequence of vertices and transformed from its local coordinate system to the absolute layout coordinate system using the terminal position and orientation of the corresponding rail. Consecutive vertices $\mathbf{r}_s$ and $\mathbf{r}_{s+1}$ defined segment $s$, with length and midpoint

\begin{equation}
\Delta L_s=\left\lVert\mathbf{r}_{s+1}-\mathbf{r}_{s}
\right\rVert,\qquad
\mathbf{r}_{s+1/2}=\frac{\mathbf{r}_{s+1}+\mathbf{r}_{s}}{2}.
\label{eq:path_segment_geometry}
\end{equation}

The virtual-wafer fields were evaluated at every segment midpoint,

\begin{equation}
p_{j,s}=p_{j,s}^{(0)}+\mu_j+\widetilde{\delta p}_{j}
\left(\mathbf{r}_{s+1/2}\right).
\label{eq:path_segment_parameter}
\end{equation}

The local parameter vector included the linewidth, film thickness, etch depth, sidewall angle, curvature and propagation orientation, although only width and thickness received nonzero virtual-wafer deviations in the present study. The effective and group indices of each guided mode were subsequently evaluated from the interpolated waveguide model,

\begin{equation}
n_{\mathrm{eff},s}=n_{\mathrm{eff}}\left(\mathbf{p}_s\right),\qquad
n_{\mathrm{g},s}=n_{\mathrm{g}}\left(\mathbf{p}_s\right).
\label{eq:path_local_indices}
\end{equation}

Around the reference angular frequency $\omega_0$, the segment propagation constant was approximated to first order as

\begin{equation}
\beta_s(\omega)\simeq\frac{\omega_0}{c}n_{\mathrm{eff},s}+\frac{\omega-\omega_0}{c}n_{\mathrm{g},s}+\frac{\omega}{c}\Delta n_s,
\label{eq:path_local_beta}
\end{equation}

where $\Delta n_s$ denotes an additional programmed refractive-index shift when a thermo-optic phase control was present. The total frequency-dependent phase accumulated along the path was

\begin{equation}
\phi_{\mathrm{path}}(\omega)=
\sum_s\beta_s(\omega)\Delta L_s.
\label{eq:path_accumulated_phase}
\end{equation}

The corresponding waveguide transformation was applied as

\begin{equation}
\hat{a}(\omega)\longmapsto\exp\left[i\phi_{\mathrm{path}}(\omega)\right]\hat{a}(\omega).
\label{eq:path_phase_transformation}
\end{equation}

Thus, connecting waveguides retained the accumulated effects of spatially varying effective index, group index and programmed phase along their physical trajectories, whereas precomputed compact devices used a single locally sampled parameter vector. After each propagation or compact-component operation, the terminal coordinates and path records of the affected rails were updated before placement of the next netlist element.

\subsection{Layout-dependent process assignment in a representative RAC--MZI}
\label{sec:layout-mapping}

To illustrate how the virtual-wafer fields were assigned to a physically placed circuit, one universal RAC--MZI cell from the programmable layout was evaluated at a representative position on the fixed virtual-wafer realization. The cell comprised an input heater, two RACs and an internal heater. This example is intended to visualize the coordinate-to-parameter assignment described above rather than to constitute an additional quantum-circuit benchmark. In particular, it shows how the same spatial process fields were converted into path-resolved waveguide phases and locally evaluated compact-device responses before quantum-state propagation or control optimization.

For each translated placement, the virtual-wafer fields were evaluated directly at the representative coordinates of the compact components and at the midpoints of the waveguide-path segments, as described in Section~\ref{sec:layout-wafer-mapping}. These local values were supplied without prior spatial averaging to the corresponding compact-device and waveguide models. Compact components therefore received one local linewidth and film thickness pair, whereas explicit waveguide paths retained the sequence of process deviations encountered along their physical trajectories.

\begin{figure}[h]
\centering
\includegraphics[width=1.0\textwidth]{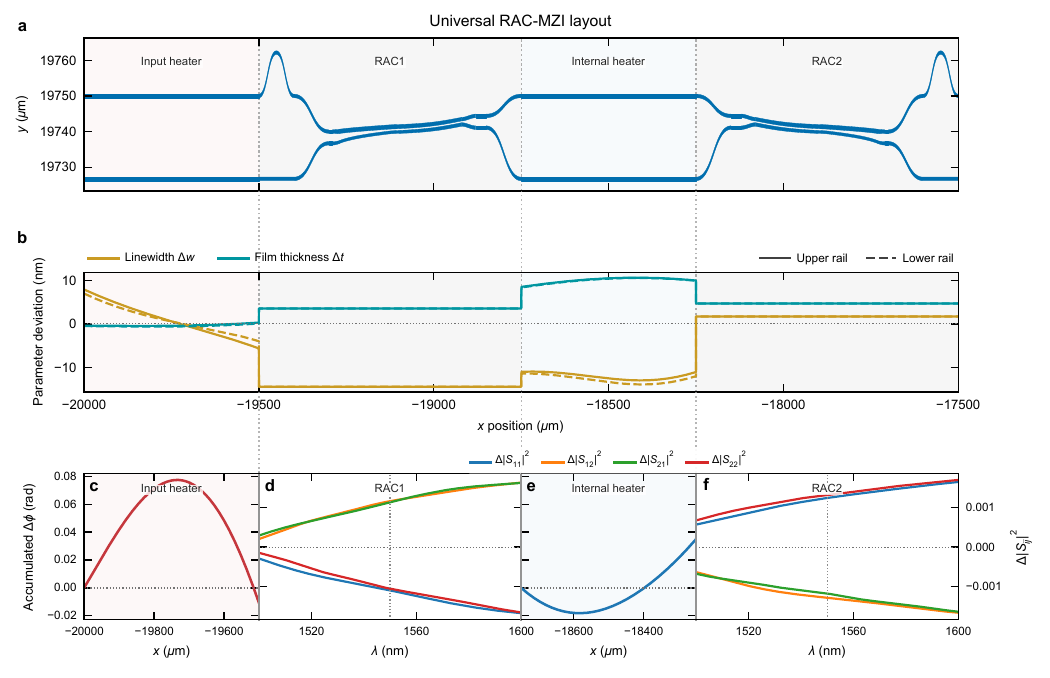}
\caption{
\textbf{Layout-dependent process assignment in a representative universal RAC--MZI.}
\textbf{a}, Physical layout of one universal RAC--MZI cell comprising an input heater, two RACs and an internal heater at the selected virtual-wafer position.
\textbf{b}, Linewidth and film thickness deviations assigned along the longitudinal coordinate of the cell. Solid and dashed curves denote the upper and lower rails, respectively. The heater paths were sampled at successive segment midpoints, whereas each compact RAC was assigned a single pair of linewidth and film thickness values evaluated at its representative component coordinate. The plateaus and step changes at the component boundaries therefore reflect the transition between path-resolved and representative-point sampling rather than discontinuities in the underlying process fields. Vertical grey guides and background shading associate the component regions across the three rows.
\textbf{c}, Accumulated differential phase between the upper and lower rails through the input heater section.
\textbf{d}, Wavelength-dependent power-response deviations of RAC1 relative to its nominal response.
\textbf{e}, Accumulated differential phase between the upper and lower rails through the internal heater section.
\textbf{f}, Wavelength-dependent power-response deviations of RAC2 relative to its nominal response.
Panels \textbf{c} and \textbf{e} use the left y-axis, whereas panels \textbf{d} and \textbf{f} use the right y-axis. The phase differences in \textbf{c} and \textbf{e} were evaluated at $1550~\mathrm{nm}$ and initialized to zero at the beginning of each heater section. The shared legend above the third row applies only to \textbf{d} and \textbf{f}; the blue, orange, green and red curves denote $\Delta|S_{11}|^2$, $\Delta|S_{12}|^2$, $\Delta|S_{21}|^2$ and $\Delta|S_{22}|^2$, respectively, where $\Delta|S_{ij}|^2=|S_{ij}^{\mathrm{mapped}}|^2-|S_{ij}^{\mathrm{nom}}|^2$. Vertical dotted lines in \textbf{d} and \textbf{f} indicate $1550~\mathrm{nm}$, and horizontal dotted lines indicate zero differential phase or zero power-response deviation.
}

\label{fig:layout-mapping}
\end{figure}

Figure~\ref{fig:layout-mapping} visualizes the resulting assignment. The physical geometry of the RAC--MZI cell is shown together with the linewidth and film thickness deviations assigned along its longitudinal coordinate. Within the heater sections, the virtual-wafer fields were evaluated at successive waveguide-segment midpoints, so that the sampled parameters vary continuously along each rail. Each RAC was instead evaluated using a single linewidth--thickness pair sampled at its representative component coordinate. The plateaus and step changes at the heater--RAC boundaries therefore arise from the change between path-resolved and representative-point sampling and do not indicate discontinuities in the underlying virtual-wafer fields. The lower panels show how the assigned process parameters enter the corresponding component transformations. For the input and internal heater sections, the local propagation constants were integrated along the two rails according to equation~\eqref{eq:path_accumulated_phase}. At the reference wavelength, these propagation constants are determined by the local effective indices.

\begin{equation}
\Delta\phi_h(x;\omega_{\mathrm{t}})=\sum_{s\leq x}\left[\beta_{\mathrm{u},s}(\omega_{\mathrm{t}})\Delta L_{\mathrm{u},s}-\beta_{\mathrm{l},s}(\omega_{\mathrm{t}})\Delta L_{\mathrm{l},s}\right],
\label{eq:accumulated_differential_phase}
\end{equation}

where $h$ denotes the input or internal heater section and each curve is initialized to zero at the beginning of the corresponding section. The differential phase was evaluated at the reference wavelength of $1550~\mathrm{nm}$. For this representative placement, the accumulated differential phase at the ends of the input and internal heater sections was approximately $-0.0107~\mathrm{rad}$ and $0.0349~\mathrm{rad}$, respectively. The programmed phase modulator induced index shifts were set to zero in this illustration, so these differential phases arise from the mapped fabrication variations and the physical waveguide paths rather than from active tuning.

For the compact RACs, the locally sampled linewidth and film thickness values were supplied to the corresponding frequency-dependent compact models. The power-response deviation plotted for RAC $i$ was defined as

\begin{equation}
\Delta\left|S_{mn}^{(i)}(\lambda)\right|^2=\left|S_{mn}^{(i)}\left(\lambda;p_{i,\mathrm{w}},p_{i,\mathrm{t}}\right)\right|^2-\left|S_{mn,\mathrm{nom}}^{(i)}(\lambda)\right|^2,
\label{eq:layout_mapped_sparameter_change}
\end{equation}

where $S_{mn,\mathrm{nom}}^{(i)}$ denotes the response of the same nominal RAC model without virtual-wafer deviations. Although RAC1 and RAC2 belong to the same nominal component family, their different physical coordinates produce different locally sampled parameter pairs and hence different spectral responses. This example therefore directly connects a continuous virtual-wafer realization to the distributed phase evolution of explicit waveguides and the spectral response of compact multiport devices.

\clearpage
\refstepcounter{section}
\section*{Supplementary Note \thesection.\\Output-stage spectral supermode reduction}
\addcontentsline{toc}{section}{\thesection \quad Output-stage spectral supermode reduction}
\label{sec:spectral-reduction}

This section describes the rail-local spectral reduction used before
frequency-unresolved photon counting.  The reduction is performed only after
the source state has undergone all circuit and measurement-basis
transformations in the full frequency-resolved representation.  It therefore
acts on the final Gaussian state rather than on the source model or on an
intermediate circuit state.

Let the output contain $R$ spatial rails and let rail $r$ contain
$K_r^{\mathrm{full}}$ orthonormal internal spectral modes.  A detector that
resolves the spatial rail but not these internal modes measures the total rail
occupation
\begin{equation}
\hat n_r=\sum_{\mu=1}^{K_r^{\mathrm{full}}}\hat a_{r\mu}^{\dagger}\hat a_{r\mu}.
\label{eq:reduction_rail_number}
\end{equation}
Equation~\eqref{eq:reduction_rail_number} is invariant under any passive unitary change of spectral basis within the same rail.  This freedom is used to concentrate the output correlations into a small number of rail-local spectral supermodes before the photon-counting probabilities are evaluated.

\subsection{Rail-local spectral basis from the covariance matrix}

The present implementation is formulated for zero-displacement Gaussian states. With $\hat{\boldsymbol a}=(\hat{\boldsymbol a}_1^{T},\ldots, \hat{\boldsymbol a}_R^{T})^{T}$ and $K_{\mathrm{tot}}=\sum_r K_r^{\mathrm{full}}$, the complex covariance matrix is written as
\begin{equation}
V=\begin{pmatrix}
N+\frac{1}{2}I & M \\
M^{*} & N^{*}+\frac{1}{2}I
\end{pmatrix},
\label{eq:reduction_complex_covariance}
\end{equation}
where
\begin{equation}
N_{ij}=\langle \hat a_j^{\dagger}\hat a_i\rangle,
\qquad
M_{ij}=\langle \hat a_i\hat a_j\rangle.
\label{eq:reduction_normal_anomalous}
\end{equation}
Here, $N$ is the vacuum-subtracted normal correlation matrix and $M$ is the anomalous correlation matrix. Partitioning the internal modes by rail defines blocks $N_{rs}$ and $M_{rs}$ of size $K_r^{\mathrm{full}}\times K_s^{\mathrm{full}}$.

For each rail, PhoQuS constructs the covariance-feature matrix
\begin{equation}
C_r=
\begin{bmatrix}
N_{r1} & \cdots & N_{rR} &
M_{r1} & \cdots & M_{rR}
\end{bmatrix}
\label{eq:reduction_feature_matrix}
\end{equation}
and its Hermitian positive-semidefinite Gram matrix
\begin{equation}
G_r=C_r C_r^{\dagger}=\sum_{s=1}^{R}
\left(
N_{rs}N_{rs}^{\dagger}+M_{rs}M_{rs}^{\dagger}
\right).
\label{eq:reduction_gram_matrix}
\end{equation}
The use of both terms in Eq.~\eqref{eq:reduction_gram_matrix} is important: $N$ describes phase-insensitive occupation and coherence, whereas $M$ describes phase-sensitive pair correlations.  Correlations with every output rail, including the same rail, contribute to the basis of rail $r$.

The eigendecomposition
\begin{equation}
G_r=U_r\Lambda_r U_r^{\dagger},
\qquad
\lambda_{r1}\geq\lambda_{r2}\geq\cdots\geq0,
\label{eq:reduction_gram_eigendecomposition}
\end{equation}
orders the rail-local spectral supermodes by decreasing covariance-feature weight. If $U_{r,K}$ contains the first $K$ columns of $U_r$, then
\begin{equation}
U_{r,K}=\underset{X^{\dagger}X=I_K}{\operatorname{arg\,max}}
\left\|X^{\dagger}C_r\right\|_{\mathrm F}^{2}=
\underset{X^{\dagger}X=I_K}{\operatorname{arg\,min}}
\left\|\left(I-XX^{\dagger}\right)C_r\right\|_{\mathrm F}^{2}.
\label{eq:reduction_ky_fan_objective}
\end{equation}
Thus, for a specified local rank $K$, the retained basis minimizes the total squared magnitude of the normal and anomalous covariance features lying outside the selected rail-local subspace. The objective does not include the vacuum identity in Eq.~\eqref{eq:reduction_complex_covariance}, which contains no information for ranking spectral modes.

Combining the rail-local eigenvector matrices gives
\begin{equation}
\mathcal U=\bigoplus_{r=1}^{R}U_r,\qquad\hat{\boldsymbol b}=\mathcal U^{\dagger}\hat{\boldsymbol a}.
\label{eq:reduction_global_basis}
\end{equation}
The associated transformation in the complex basis and the rotated covariance matrix are
\begin{equation}
\mathcal S=\mathcal U^{\dagger}\oplus\mathcal U^{T},\qquad
V'=\mathcal S V\mathcal S^{\dagger}.
\label{eq:reduction_covariance_rotation}
\end{equation}
Because $\mathcal U$ is block diagonal in the spatial-rail partition, this transformation preserves every total rail-number operator in Eq.~\eqref{eq:reduction_rail_number}.

For retained dimensions $\boldsymbol K=(K_1,\ldots,K_R)$, the first $K_r$ transformed modes of each rail are retained.  The remaining modes are traced out. Gaussian partial trace is implemented by selecting from $V'$ the rows and columns associated with the retained annihilation and creation operators. The resulting principal covariance submatrix, denoted $V_{\mathrm{red}}$, is therefore the exact Gaussian marginal of the retained supermodes; the approximation enters only when photons occupying the discarded supermodes are omitted from the subsequent frequency-unresolved count.

\subsection{Selection of the retained dimension}

After ordering the spectral supermodes using the covariance-Gram construction, PhoQuS determines how many modes should be retained. Let
\begin{equation}
\boldsymbol K=(K_1,\ldots,K_R)
\end{equation}
denote a candidate dimension vector, where $K_r$ is the number of leading supermodes retained in rail $r$. The discarded subsystem $\mathcal D(\boldsymbol K)$ contains all remaining modes from all rails. The truncation tolerance is imposed on this complete discarded subsystem, rather than independently on each rail.

Two quantities are evaluated directly from the rotated covariance matrix $V'$. Let $V_{\mathcal D}$ denote the covariance matrix of the discarded subsystem. For a zero-displacement Gaussian state, the probability that all discarded modes are in vacuum is
\begin{equation}
p_{0,\mathcal D}(\boldsymbol K)=\frac{1}{\sqrt{\det\left(V_{\mathcal D}+\frac{1}{2}I_{2|\mathcal D|}\right)}}.
\label{eq:reduction_tail_vacuum_probability}
\end{equation}

The corresponding tail-nonvacuum probability is
\begin{equation}
\epsilon_{\mathrm{PNR}}(\boldsymbol K)=1-p_{0,\mathcal D}(\boldsymbol K).
\label{eq:reduction_tail_nonvacuum}
\end{equation}
This quantity is the probability that at least one discarded mode is non-vacuum. The full and reduced frequency-unresolved photon-counting outcomes are identical whenever the discarded subsystem is in vacuum.

The second quantity measures the fraction of the mean photon population carried by the discarded modes:
\begin{equation}
\epsilon_{\bar n}(\boldsymbol K)=\frac{\bar n_{\mathcal D}}{\bar n_{\mathrm{tot}}},\qquad\bar n_{\mathcal D}=\sum_{i\in\mathcal D}\left(V'_{ii}-\frac{1}{2}\right),\qquad\bar n_{\mathrm{tot}}=\sum_{i=1}^{K_{\mathrm{tot}}}\left(V'_{ii}-\frac{1}{2}\right),
\label{eq:reduction_tail_mean}
\end{equation}
where the diagonal elements are taken from the annihilation block of $V'$. If the total mean photon number is numerically zero, $\epsilon_{\bar n}$ is defined to be zero. Unlike $\epsilon_{\mathrm{PNR}}$, the quantity $\epsilon_{\bar n}$ is not a bound on the full photon-count distribution; it instead prevents a truncation from discarding a substantial fraction of the mean optical population.

The combined detector-oriented selection metric is
\begin{equation}
\epsilon_{\mathrm{red}}(\boldsymbol K)=\max\{{\epsilon_{\mathrm{PNR}}(\boldsymbol K),\epsilon_{\bar n}(\boldsymbol K)\}}.
\label{eq:reduction_combined_metric}
\end{equation}

For a prescribed tolerance $\tau$, a candidate dimension vector is accepted only when

\begin{equation}
\epsilon_{\mathrm{red}}(\boldsymbol K)\leq\tau.
\end{equation}

Thus, both the tail-nonvacuum probability and the discarded mean-photon fraction must be below the specified tolerance.

The candidate dimension vectors are examined along a nested sequence. Starting from the minimum permitted number of modes in each rail, the algorithm adds one supermode at a time. If the current dimension vector is $\boldsymbol K^{(\ell)}$, the rail receiving the next mode is
\begin{equation}
r_{\ell}^{*}=\underset{r:K_r^{(\ell)}<K_r^{\mathrm{full}}}{\operatorname{arg\,max}}\lambda_{r,K_r^{(\ell)}+1},
\label{eq:reduction_next_rail}
\end{equation}
where $\lambda_{r,K_r^{(\ell)}+1}$ is the largest Gram eigenvalue not yet retained in rail $r$. The dimension vector is then updated as
\begin{equation}
\boldsymbol K^{(\ell+1)}=\boldsymbol K^{(\ell)}+\boldsymbol e_{r_{\ell}^{*}},
\label{eq:reduction_dimension_update}
\end{equation}
where $\boldsymbol e_r$ increases only the retained dimension of rail $r$.

Comparing Gram eigenvalues across rails therefore determines how the available spectral dimensions are allocated among the rails. A rail containing substantial multimode covariance weight may receive several additional modes before a nearly single-mode rail receives another.

For every candidate $\boldsymbol K^{(\ell)}$, the quantities $\epsilon_{\mathrm{PNR}}$ and $\epsilon_{\bar n}$ are recomputed from the corresponding discarded marginal of $V'$. The first candidate satisfying Eq.~\eqref{eq:reduction_combined_metric} is selected. For a physical covariance matrix, both metrics are non-increasing along this sequence because each step removes one mode from the discarded subsystem. If no modes are discarded, both metrics are defined to be zero and the unreduced state is recovered.

\begin{figure*}[t]
\centering
\includegraphics[width=\textwidth]{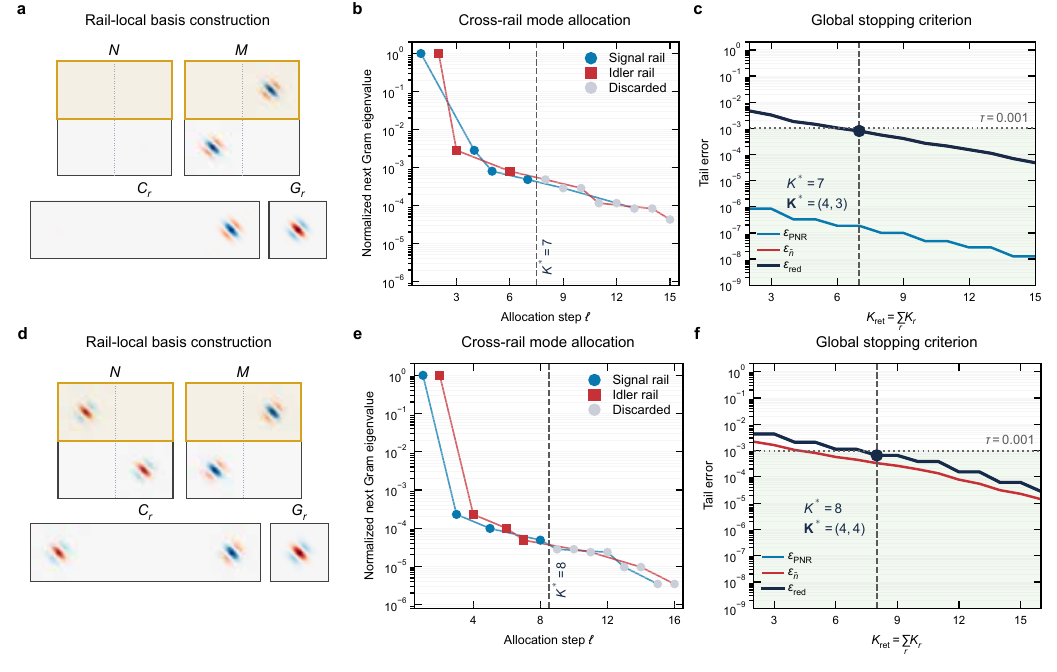}
\caption{\textbf{Rail-local covariance-Gram basis construction and automatic selection of the retained spectral dimension.}
The upper row corresponds to a lower-occupation SPDC output generated with mean signal-rail photon number $\bar n_s=1.82\times10^{-4}$. The lower row shows a higher-occupation output obtained with $\bar n_s=1.97$. Each signal and idler rail initially contains 51 spectral modes.
(\textbf{a},\textbf{d}) Construction of the covariance-feature matrix $C_r=[N_{r1}\ \cdots\ N_{rR}\ M_{r1}\ \cdots\ M_{rR}]$ and the corresponding Gram matrix $G_r=C_rC_r^\dagger$ for the signal rail. The highlighted block rows of $N$ and $M$ are concatenated to form $C_r$. The panels display the real parts of the matrices; $N$, $M$, and $C_r$ share a common normalization within each row, whereas $G_r$ is normalized separately. The color scales are therefore not intended for absolute comparison between the two pump conditions.
(\textbf{b},\textbf{e}) Nested cross-rail allocation of spectral supermodes. After retaining the prescribed minimum of one mode per rail, each subsequent mode is assigned to the rail having the largest available next Gram eigenvalue. Colored markers denote modes retained at the selected cutoff, whereas gray markers show subsequent modes that are discarded. The eigenvalues are normalized by the largest allocation score in each case.
(\textbf{c},\textbf{f}) Tail-nonvacuum probability $\epsilon_{\mathrm{PNR}}$, discarded mean-photon fraction $\epsilon_{\bar n}$, and their maximum $\epsilon_{\mathrm{red}}$, plotted against the total retained dimension $K_{\mathrm{ret}}=\sum_rK_r$. The dotted horizontal line denotes the tolerance $\tau=10^{-3}$, and the shaded region satisfies the combined stopping criterion. The first accepted candidate is $K_{\mathrm{ret}}^\ast=7$, with $\boldsymbol K^\ast=(4,3)$ in low-gain example, whereas $K_{\mathrm{ret}}^\ast=8$, with $\boldsymbol K^\ast=(4,4)$ in high-gain example. The stopping point is determined by $\epsilon_{\bar n}$ in the lower-occupation case and by $\epsilon_{\mathrm{PNR}}$ in the higher-occupation case.}
\label{fig:spectral_reduction_construction}
\end{figure*}

Figure~\ref{fig:spectral_reduction_construction} illustrates the complete basis-construction and dimension-selection procedure for a two-mode SPDC output represented by 51 spectral modes per rail. The upper row corresponds to the mean signal-rail photon number is $\bar n_s=1.82\times10^{-4}$. The lower row corresponds to $\bar n_s=1.97$. In the low-gain case, the anomalous block $M$ dominates the covariance features, whereas the normal block $N$ becomes appreciable at the higher gain, illustrating why both contributions are included in $G_r$.

The two tail quantities also play complementary roles. In the low-gain case, $\epsilon_{\bar n}$ determines the stopping point because $\epsilon_{\mathrm{PNR}}$ is suppressed by the small absolute photon population. At the higher occupation, $\epsilon_{\mathrm{PNR}}$ becomes the limiting quantity. For the tolerance $\tau=10^{-3}$, example happen to yield

$$
K_{\mathrm{ret}}^\ast=\sum_r K_r^\ast=7,\qquad\boldsymbol K^\ast=(4,3).
$$
for low-gain case and
$$
K_{\mathrm{ret}}^\ast=\sum_r K_r^\ast=8,\qquad\boldsymbol K^\ast=(4,4).
$$
for high-gain case.

This common retained dimension is specific to the examples shown; in general, both the total dimension and its allocation among rails are determined from the final output covariance matrix.

\subsection{Scope and limitations}

The covariance-Gram basis has a precise but restricted optimality property. Equation~\eqref{eq:reduction_ky_fan_objective} minimizes the one-sided Frobenius residual of the vacuum-subtracted normal and anomalous covariance features within the class of passive, rail-local rank-$K$ subspaces. It does not directly minimize the error of a selected photon-counting probability, post-selected state or protocol-level observable. Nor does it generally minimize the Frobenius distance between the full covariance matrix and a rank-reduced reconstruction.

The criterion determines where truncation should stop by testing the exact Gaussian marginal of the discarded modes. Moreover, the Gram-eigenvalue allocation in Eq.~\eqref{eq:reduction_next_rail} is greedy with respect to retained covariance-feature weight; it is not guaranteed to produce the globally smallest total dimension that satisfies the two tail conditions.

The present construction assumes zero displacement. A displaced Gaussian state can carry appreciable photon-counting weight in a spectral direction that is not represented by the covariance-Gram basis. Extending the method to such states requires the first moments to be included in both the basis-ranking objective and the discarded-subsystem vacuum probability.

Finally, PhoQuS performs the reduction only after the final spectral mixing and measurement-basis transformation. The method reduces the internal spectral dimension supplied to the photon-counting backend, but it does not reduce the cost of the preceding full-resolution propagation or change the asymptotic complexity of exact photon counting.

\clearpage
\refstepcounter{section}
\section*{Supplementary Note \thesection.\\Frequency-resolved Gottesman-Kitaev-Preskill (GKP) state generation}
\addcontentsline{toc}{section}{\thesection \quad Frequency-resolved GKP-state generation example}
\label{sec:gkp-example}

This example extends the discrete-variable benchmarks to the preparation of an approximate Gottesman--Kitaev--Preskill (GKP) state. We replace the single-mode squeezing operations of a fixed reference circuit with frequency-resolved SPDC transformations and compare direct pumping with cascaded squeezing. The comparison tests whether matching the squeezing strength alone is sufficient to reproduce the heralded state when the spectro-temporal modes also depend on gain.

\subsection{Reference circuit and source model}

The reference is the three-mode GKP preparation circuit in The Walrus gallery~\cite{WalrusGKP,Tzitrin2020GKP}. Three input single-mode squeezed vacua pass through three beam splitters, followed by a squeezing operation on the third output. The signed squeezing parameters are
\begin{equation}
(r_1,r_2,r_3,r_4)=(-1.38155,-1.21670,0.77988,0.19580),
\label{eq:gkp-reference-squeezing}
\end{equation}
with the convention $\hat S(r)=\exp[r(\hat a^2-\hat a^{\dagger 2})/2]$. In chronological order, the beam splitters act on logical rail pairs $(0,1)$, $(1,2)$ and $(0,1)$, with
\begin{equation}
\begin{split}
\boldsymbol\theta&=(1.04182,0.87702,0.90244),\\
\boldsymbol\phi&=(1.48354,1.69629,-0.24252).
\end{split}
\label{eq:gkp-interferometer-parameters}
\end{equation}
Here the annihilation-operator transformation is
\begin{equation}
B(\theta,\phi)=
\begin{pmatrix}
\cos\theta&-e^{-i\phi}\sin\theta\\
e^{i\phi}\sin\theta&\cos\theta
\end{pmatrix}.
\label{eq:gkp-beamsplitter}
\end{equation}
Detection of $(5,7)$ photons in the first two outputs heralds the approximate GKP state in the third. The abstract reference assigns one bosonic mode to each rail. Its heralded state, rather than an ideal infinite-energy GKP state, defines the comparison reference.

The frequency-resolved source uses the lithium-tantalate SPDC framework of Section~\ref{sec:nonlinear-propagation}, with a $5~\mathrm{mm}$, $2.8~\mu\mathrm{m}$-wide waveguide and an optimized poling profile targeting a Gaussian longitudinal interaction envelope. An unchirped Gaussian pump is centered at $775~\mathrm{nm}$, with envelope $\widetilde\beta_p(t)\propto\exp[-\sigma_\omega^2t^2/2]$ and $\sigma_\omega=4.7\times10^{12}~\mathrm{rad\,s^{-1}}$. Signal and idler are centered at $1550~\mathrm{nm}$ and each span a nominal $40~\mathrm{nm}$ interval at $100~\mathrm{GHz}$ spacing, giving 51 frequency-bin modes per field. The pump has 501 temporal samples, and each nonlinear waveguide uses 500 propagation cells. The
source-kernel coupling parameter is set to 28. Fabrication variation, propagation loss, pump self-phase modulation and cross-phase modulation are disabled.

Signal and idler are generated in the $\mathrm{TE}_{00}$ and $\mathrm{TE}_{20}$ transverse modes, respectively. Before their interference, these modes are assumed to be ideally separated and mapped onto a common transverse mode in the two beam-splitter input rails. So loss and spectral distortion associated with this mode splitting and conversion are not included.

For direct pumping, the energy of each source is calibrated so that its dominant squeezing parameter matches $|r_j|$ in Eq.~\eqref{eq:gkp-reference-squeezing}. Specifically, $r_{\max}=\operatorname{arsinh}(s_{\max})$, where $s_{\max}$ is the largest singular value of the pair-generating Bogoliubov block. The four pump energies are $830.21442$, $671.29767$, $305.60659$ and $21.01973~\mathrm{pJ}$, respectively.

\subsection{Mode matching and gain-dependent delay}

For a matched pair of signal and idler spectro-temporal modes, a balanced beam splitter converts two-mode squeezed vacuum into two single-mode squeezed vacuum with opposite squeezing axes. Constant phase shifts align these axes with the signed parameters in Eq.~\eqref{eq:gkp-reference-squeezing}. Residual spectral or temporal mismatch hinders this factorization and can leave a selected output mode mixed.

A central source of temporal mismatch is group-velocity walk-off. A spectral phase linear in frequency translates a temporal envelope without changing its shape; different phase slopes for signal and idler therefore produce a relative arrival-time shift. Such shifts leave the spectral intensities and source Schmidt coefficients unchanged, but can reduce the mutual mode overlap at the beam splitter. Spectral purity and signal--idler mode matching are thus distinct requirements. Occupation of additional spectral modes also affects frequency-unresolved heralding~\cite{Bulmer2026}.

The initial compensation is defined in the pump frame. Writing $\Delta\beta_{1,\mu}=v_{g,\mu}^{-1}-v_{g,p}^{-1}$ for $\mu\in\{s,i\}$, the compensating transformation is
\begin{equation}
P_\mu(a_\mu L):\quad\hat a_\mu(\omega)\longmapsto
e^{-i(\omega-\bar\omega_\mu)\tau_\mu}\hat a_\mu(\omega), \qquad \tau_\mu=a_\mu L\Delta\beta_{1,\mu}.
\label{eq:gkp-walkoff-compensation}
\end{equation}
This transformation shifts the temporal envelope by $-\tau_\mu$. In the low-gain limit, a uniform effective nonlinear coupling over an interaction length $L$ gives an effective generation position at the waveguide midpoint. The temporal centers of the signal and idler output wavepackets are therefore shifted relative to the pump by the walk-off accumulated over a distance $L/2$. The direct-pump baseline therefore uses $a_s=a_i=1/2$ before the TMSV-to-SMSV beam splitter. The corresponding pump-frame temporal walk-off delays, $\tau_\mu^{(0)}=L\Delta\beta_{1,\mu}/2$, are $-0.475246~\mathrm{ps}$ and $0.486519~\mathrm{ps}$ for signal and idler, respectively; $P_\mu(L/2)$ applies the opposite spectral phase. This compensation corrects temporal shifts without altering the spectral amplitudes. It does not, however, guarantee mode matching when nonlinear propagation changes the mode shapes or their phase slopes at higher gain. All beam splitters and compensators are modeled as ideal, lossless operations.

We first choose fixed spectral modes against which the squeezing phases will be defined. To obtain these modes, we apply $P(L/2)$ and a balanced beam splitter to the output of the low-gain SPDC stage used in the TWOC cascade. This stage has dominant squeezing parameter  $r_{\mathrm{ref}}=0.05$ and pump energy $1.37924~\mathrm{pJ}$.

For each of the two beam-splitter outputs, we construct the Gram matrix $G_r$ from its normal and anomalous correlations, as defined in Eq.~\eqref{eq:reduction_gram_matrix}. The normalized eigenvector with the largest eigenvalue is the leading spectral supermode on that rail. We denote the reference mode for the output under consideration by $u^{\mathrm{ref}}$.

An eigenvector has an arbitrary overall complex phase. We fix this phase by making its value at the center frequency real and positive. These reference modes are then kept fixed. They are used when choosing the phase shifts that align the input squeezing axes with the signed parameters in Eq.~\eqref{eq:gkp-reference-squeezing}. They also fix the phase convention of the spectral modes used to represent the final conditional states. Choosing this convention does not change a mode's temporal intensity or apply a physical delay.

For each of the three input sources, we next determine the signal and idler delays before their interference at the balanced beam splitter. The relevant modes at this stage are the dominant output Schmidt modes of the SPDC source, before compensation. For vacuum input, they are obtained from the anomalous correlation matrix
\begin{equation}
(M_{si})_{pq}=\left\langle\hat a_{s,p}^{\mathrm{out}}\hat a_{i,q}^{\mathrm{out}}
\right\rangle,
\end{equation}
where $p$ and $q$ label frequency bins. Writing $M_{si}=U_M\operatorname{diag}(m_k)V_M^\dagger$ with the singular values ordered from largest to smallest,
the dominant modes are $u_s=(U_M)_{:,0}$ and $u_i=(V_M^*)_{:,0}$. Both vectors have unit discrete norm.

The delays are chosen to make these two modes overlap as closely as possible and to align the resulting squeezed pulses with the pump at $t=0$. Under Eq.~\eqref{eq:gkp-walkoff-compensation}, the signal and idler mode amplitudes acquire the factors $e^{-i(\omega-\bar\omega)\tau_s}$ and $e^{-i(\omega-\bar\omega)\tau_i}$, respectively. Their squared overlap therefore depends only on $\delta\tau=\tau_s-\tau_i$. We determine this relative compensation by minimizing
\begin{equation}
J_{\mathrm{source}}(\delta\tau)=1-\left|\sum_p u_{s,p}^*u_{i,p}e^{i(\omega_p-\bar\omega)\delta\tau}\right|^2,\qquad\delta\tau=\tau_s-\tau_i,
\label{eq:gkp-source-delay-objective}
\end{equation}
where $\bar\omega$ is the common signal and idler center frequency.

Maximizing the signal--idler overlap within each source fixes their relative delay, but not their timing relative to the pump. The squeezed pulses prepared by different sources can therefore arrive at different times even when the signal and idler modes within each source are well matched. Since all sources share the same pump time reference, we align the prepared pulses to the pump center. We determine their timing from the temporal center of each mode, defined as the intensity-weighted mean time,
\begin{equation}
T[u]=\frac{\int t|\widetilde u(t)|^2\,dt}{\int |\widetilde u(t)|^2dt},\qquad
\widetilde u(t)=\sum_p u_p e^{-i(\omega_p-\bar\omega)t}.
\label{eq:gkp-mode-time-centroid}
\end{equation}
Let $t_p=0$ be the center of the pump envelope. Since a compensation delay $\tau_\mu$ shifts the corresponding temporal profile by $-\tau_\mu$, we initially set $\tau_c=\frac{T[u_s]+T[u_i]}{2}-t_p,\qquad \tau_s=\tau_c+\frac{\delta\tau}{2},\qquad\tau_i=\tau_c-\frac{\delta\tau}{2}.$
This choice places the average of the signal and idler temporal centers at the pump center while preserving the optimized overlap.

We then check the timing of the squeezed pulses produced by the balanced beam splitter. After the beam splitter and the phase shifts that align the squeezing axes, we calculate the leading spectral supermode on each output rail from its Gram matrix. If both temporal centers have a remaining common shift from $t_p$,
we add that shift to both $\tau_s$ and $\tau_i$. This adjustment leaves $\delta\tau$ and the signal--idler overlap unchanged. We verify that both output supermodes are centered at $t_p$ within the numerical tolerance. The final values of $\tau_s$ and $\tau_i$ include this adjustment; no additional delay operation is applied after the beam splitter.

The inline squeezing operation requires a different calculation because its input is the state leaving the interferometer, rather than vacuum. This operation acts on the third logical rail together with an additional rail initially in vacuum. The latter is the auxiliary rail used to implement single-mode squeezing with an SPDC stage and balanced beam splitters.

For this operation, we independently vary the signal and idler compensation delays before and after the SPDC stage, giving four delay parameters. In chronological order, the operation consists of an input phase rotation, $\mathrm{BS}^\dagger$, the compensation before the SPDC stage, the SPDC stage, the compensation after it, $\mathrm{BS}$, and an output phase rotation. The input and output phase rotations are constant across frequency.

To choose these delays, we first identify the spectral mode that actually enters the inline operation. We construct $G_r$ for the third logical rail from the full covariance matrix immediately after the interferometer. The normalized eigenvector with the largest eigenvalue of $G_r$ defines the incident spectral supermode $v$.

The shape of $v$ is therefore determined by the propagated state, not replaced by the low-gain reference mode. Its overall phase is set relative to the same reference mode $u^{\mathrm{ref}}$ used for the input squeezers, by requiring
$(u^{\mathrm{ref}})^\dagger v$ to be real and positive. This fixes the phase convention without changing the spectral shape or timing of the incident mode.

Within the inline operation, let $j=0$ denote the third logical rail and $j=1$ the auxiliary vacuum rail. Let $\hat{\boldsymbol a}_j$ collect the annihilation operators of the frequency bins on rail $j$. For comparison with the ideal squeezing operation, we define a mode with spectral vector $v$ on each rail,
$\hat b_j=v^\dagger\hat{\boldsymbol a}_j$. The desired transformation is
\begin{equation}
\hat b_{j,\mathrm{out}}=\cosh r_4\,\hat b_{j,\mathrm{in}}-\sinh r_4\,\hat b_{j,\mathrm{in}}^\dagger.
\label{eq:gkp-inline-target}
\end{equation}
Thus, the operation should reproduce the squeezing parameter $r_4$ in the incident spectral mode without changing that mode or coupling it to the auxiliary rail.

We evaluate this requirement using the Bogoliubov transformation of the complete inline operation, including its compensation delays and phase rotations. Write this transformation as
$\hat{\boldsymbol a}_{j,\mathrm{out}}=\sum_{k=0}^{1}\left(A_{jk}\hat{\boldsymbol a}_{k,\mathrm{in}}+B_{jk}\hat{\boldsymbol a}_{k,\mathrm{in}}^\dagger\right).$
Here $A_{jk}$ and $B_{jk}$ map input frequency bins on rail $k$ to output frequency bins on rail $j$. Their projections onto the selected spectral mode are
\begin{equation}
\alpha_j=v^\dagger A_{jj}v, \qquad \beta_j=v^\dagger B_{jj}v^*.
\label{eq:gkp-inline-projected-coefficients}
\end{equation}
After adjusting the constant input and output phases, these coefficients should approach $\alpha_j=\cosh r_4$ and $\beta_j=-\sinh r_4$.

The delay optimization also accounts for changes in spectral shape and coupling between the two rails. We use
\begin{equation}
J_{\mathrm{inline}}=E_{\mathrm{coeff}}+E_{\mathrm{mode}}+0.25\,E_{\mathrm{rail}}.
\label{eq:gkp-inline-delay-objective}
\end{equation}
The first term, $E_{\mathrm{coeff}}$, measures the relative errors of $\alpha_j$ and $\beta_j$ from their target values. The second term, $E_{\mathrm{mode}}$, measures the parts of $A_{jj}v$ and $B_{jj}v^*$ outside the selected spectral mode. These parts are $A_{jj}v-\alpha_jv$ and $B_{jj}v^*-\beta_jv$; their norms are divided by $\|A_{jj}v\|$ and $\|B_{jj}v^*\|$, respectively. The third term, $E_{\mathrm{rail}}$, measures coupling between the third logical rail and the auxiliary rail through $\|A_{jk}\|_{\mathrm F}/\|A_{jj}\|_{\mathrm F}$ and $\|B_{jk}\|_{\mathrm F}/\|B_{jj}\|_{\mathrm F}$ for $k\neq j$. Here $\|\cdot\|_{\mathrm F}$ denotes the Frobenius norm. Each term is the root mean square of four contributions: one for each Bogoliubov block on each of the two rails.

The same four inline compensation delays are used for both parallel circuits and are chosen to minimize their average value of $J_{\mathrm{inline}}$. For every candidate set of delays, the constant input and output phases are recalculated using the incident spectral mode of each inline operation. The mode projections are used only to choose the delays and phases. Circuit propagation retains the full frequency basis; spectral reduction for photon counting is performed only at the final output, as described in Section~\ref{sec:spectral-reduction}.

All compensation fractions are restricted to $-0.5\leq a_\mu\leq1.5$. For each input source, we evaluate Eq.~\eqref{eq:gkp-source-delay-objective} on a grid of relative delays and refine each detected local minimum.

For the inline operation, we use the modified Powell method, starting with all four compensation fractions equal to $0.5$. The search allows at most 18 iterations or 180 objective evaluations. The initial set of delays is also evaluated separately, and the set with the lowest evaluated objective is retained.

For the three input sources, the optimized compensation fractions are
\begin{equation}
\begin{aligned}
(a_{s,1},a_{s,2},a_{s,3})&=(0.40504,0.42224,0.46412),\\
(a_{i,1},a_{i,2},a_{i,3})&=(0.40576,0.42281,0.46436).
\end{aligned}
\end{equation}
For the inline operation, the optimized fractions are
\begin{equation}
(a_{s,\mathrm{in}},a_{i,\mathrm{in}},  a_{s,\mathrm{out}},a_{i,\mathrm{out}})=(0.49938,0.50003,0.50000,0.50000).
\end{equation}
After fixing these delays and phases, we evaluate the fidelity of the resulting heralded GKP state.

\subsection{Cascaded squeezing with temporal walk-off compensation}

Temporal walk-off compensation (TWOC) aligns the interacting fields between successive nonlinear stages, allowing the contributions generated in different stages to overlap in the same temporal modes and interfere coherently~\cite{Kim2024TWOC}. Here identical SPDC stages are separated by $P(L)$, which compensates the signal and idler walk-off relative to the pump over one complete stage.

The final $P(L/2)$ compensation serves a different purpose: it corrects the nominal signal--idler delay at the cascade output before their interference at the TMSV-to-SMSV beam splitter. Thus, the interstage compensation supports coherent accumulation across stages, whereas the output compensation provides temporal mode matching for the subsequent signal--idler mixing. An inline squeezing gate also receives $P(L/2)$ at its input to align the incoming modes with the nonlinear interaction.

All stages within a cascade share the same pump energy and phase reference. A frequency-independent phase $\varphi/2$ is applied to each rail immediately before the final SPDC stage to control its phase relative to the preceding stages, without changing the compensation delays.

The squeezing grows approximately linearly with stage count, while weaker individual stages reduce the accumulated mode distortion. For the GKP circuit, every TWOC stage uses one common source with pump energy $1.37924072~\mathrm{pJ}$ and dominant squeezing $r_0=0.05000007$. Its spectral purity, defined as in Eq.~\eqref{eq:ghz-source-purity}, is $0.99078743$. Stage counts and final interstage pair phases are
\begin{equation}
\begin{aligned}
(N_1,N_2,N_3,N_4)&=(28,25,16,4),\\
(\varphi_1,\varphi_2,\varphi_3,\varphi_4)&=(-0.92501,1.27041,0.97383,0.47729)~\mathrm{rad}.
\end{aligned}
\label{eq:gkp-twoc-settings}
\end{equation}
The phases are calibrated against $r_M=\tfrac12\operatorname{arsinh}(2|\langle\hat b\hat b\rangle|)$, averaged over the dominant covariance-Gram modes of the two SMSV outputs. It equals the squeezing parameter for a pure single-mode squeezed vacuum, but does not imply that the retained state is pure. The input resources are phase aligned to shared low-gain reference modes. For the inline gate, input and output constant phases are obtained from the projections of its complete Bogoliubov map including SMSV-TMSV-SMSV conversion onto those same reference modes.

\clearpage
\refstepcounter{section}
\section*{Supplementary Note \thesection.\\Fabrication-aware GHZ-state benchmark}
\addcontentsline{toc}{section}{\thesection \quad Fabrication-aware GHZ-state benchmark}
\label{sec:ghz-benchmark}

This benchmark tests how source spectrum, circuit architecture, passive-device implementation, spatial fabrication variation and programmable control combine at the level of a three-photon GHZ observable. Two fixed source covariance matrices, two circuit architectures and two coupler implementations define eight configurations. Each configuration is evaluated at the same 80 anchor positions on one virtual-wafer realization. The source covariance is fixed within each source class; only the passive circuit models vary with position.

\subsection{Source states}

The non-resonant source is the lossless lithium-tantalate SPDC model described in Section~\ref{sec:nonlinear-propagation}. A $775~\mathrm{nm}$ Gaussian pump with pulse energy $0.1~\mathrm{pJ}$ and spectral width $\sigma_\omega=4.7\times10^{12}~\mathrm{rad\,s^{-1}}$ drives a $5~\mathrm{mm}$ poled waveguide. Signal and idler are centered at $1550~\mathrm{nm}$ and each span $40~\mathrm{nm}$ at $100~\mathrm{GHz}$ spacing, giving 51 frequency-bin modes per field.

The resonant source is a lossless single-ring spontaneous four-wave-mixing (SFWM) model. The ring circumference is $1.42~\mathrm{mm}$ and its free spectral range is $100~\mathrm{GHz}$. The pump, signal and idler resonances are $1552.41$, $1550.00$ and $1554.83~\mathrm{nm}$, respectively, and their bus-coupler power reflectivities are $0.10$, $0.02$ and $0.02$. A $10~\mathrm{pJ}$, $0.3~\mathrm{ns}$ square pump pulse is represented by 501 temporal samples. Signal and idler each span $0.03~\mathrm{nm}$ at $0.06~\mathrm{GHz}$ spacing, giving 63 frequency-bin modes per field.

The spectral purity is calculated from the singular values (Schmidt coefficients) $s_k$ of the source transfer matrix,
\begin{equation}
P_{\mathrm{spec}}=\frac{\sum_k s_k^4}{\left(\sum_k s_k^2\right)^2}.
\label{eq:ghz-source-purity}
\end{equation}
The nominal spectral purities are $0.990739$ for SPDC and $0.993378$ for SFWM. The two source models are reused unchanged at every wafer position.

\subsection{Circuit architectures and physical couplers}

The task-specific circuit is the ten-rail signal network shown in Fig.~6b. It contains eight 50:50 couplers, two $33{:}66$ couplers, two $75{:}25$ couplers, 12 crossings and two fixed $\pi$ phase shifts, together with the routing sections needed to preserve the rail order. Ten independent copies of the selected source covariance are connected to the ten signal rails and their ten partner rails. No internal circuit parameter is retuned across the wafer sweep. The only optimized quantity in the reported task-specific fidelity is the final output-local phase of the target GHZ state, as defined below.

The programmable architecture implements the same nominal signal transformation in a 12-rail rectangular Clements mesh. Two vacuum rails are inserted at input rails 4 and 7, and the unused output rails are 3 and 9. The $12\times12$ target matrix is projected to its nearest unitary before decomposition. The resulting mesh contains 66 adjacent-rail transformations arranged in 12 layers, followed by a fixed output phase layer. For rails $m$ and $n$, the convention is
\begin{equation}
T_{mn}(\theta,\phi)=
\begin{pmatrix}
    e^{i\phi}\cos\theta & -\sin\theta\\
    e^{i\phi}\sin\theta & \cos\theta
\end{pmatrix},
\label{eq:ghz-clements-transform}
\end{equation}
corresponding to a phase shift on rail $m$ followed by a real beam splitter. Each physical universal MZI comprises a $500~\mu\mathrm{m}$ input phase section, a nominal 50:50 coupler, a $500~\mu\mathrm{m}$ internal phase section and a second nominal 50:50 coupler. The second RAC is used in the reverse orientation. Inactive rails traverse physical bypass paths designed to match the nominal delay of the active MZI cell.

Both architectures use the complex, frequency- and fabrication-dependent models of Section~\ref{sec:compact-models}. In the task-specific DC circuit, the nominal coupling lengths are $21.235$, $27.785$, $11.150$ and $51.643~\mu\mathrm{m}$ for the 50:50, $33{:}66$, $75{:}25$ and crossing transformations, respectively. The programmable DC--MZI uses a $21.2~\mu\mathrm{m}$ nominal 50:50 coupling section. The RAC implementation uses the corresponding ratio-specific models in the task-specific circuit and the nominal 50:50 model in the programmable mesh. The complex responses are not projected to unitary matrices: loss, spectral phase and splitting-ratio dispersion are retained. In the circuit simulations, linewidth and film thickness are sampled from the virtual wafer as described in Section~\ref{sec:layout-wafer-mapping}.

\begin{figure}[t]
\centering
\includegraphics[width=\textwidth]{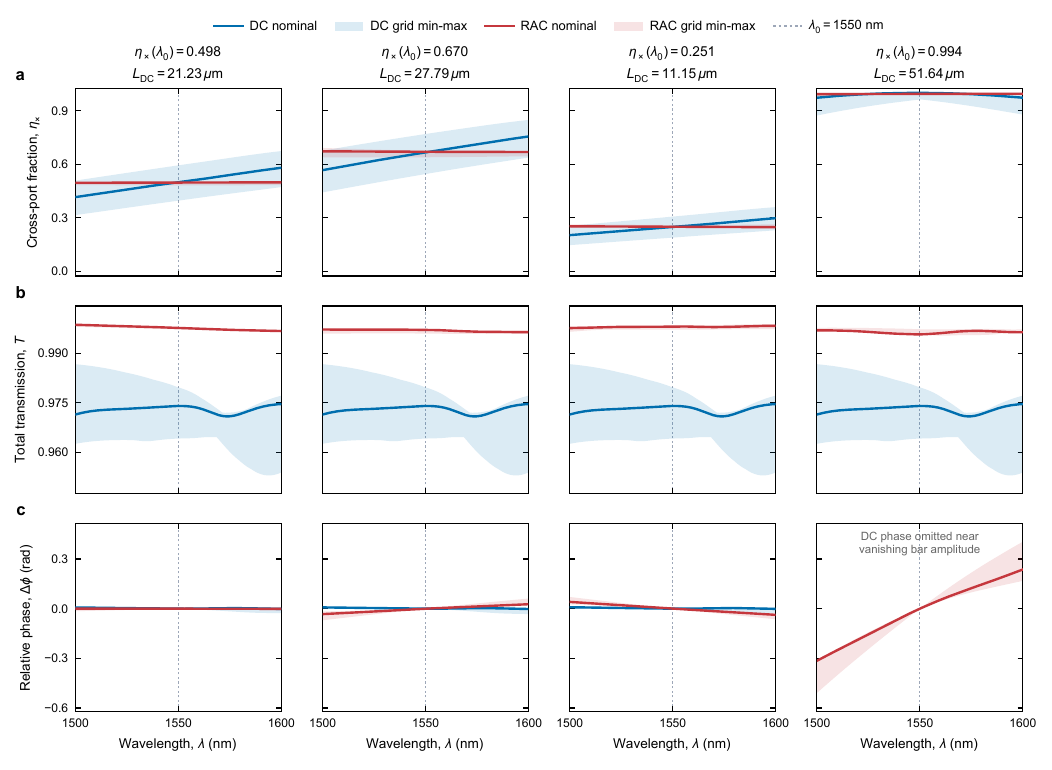}
\caption{
\textbf{Spectral responses of the task-specific DC and RAC transformations.}
Columns correspond to the nominal 50:50, 33:66, 75:25 and crossing transformations, with splitting ratios expressed in bar:cross order. The annotations give the nominal RAC cross-port fraction at $\lambda_0=1550~\mathrm{nm}$ and the DC coupling length.
\textbf{a}, Cross-port power fraction $\eta_{\times}$, defined as the sum of the cross-port powers for the two inputs divided by their summed transmitted powers.
\textbf{b}, Total transmitted power $T$, averaged over the two inputs.
\textbf{c}, Relative cross-to-bar phase deviation $\Delta\phi$ for excitation at upper input. For each parameter combination, the relative phase is unwrapped along wavelength and its own value at $\lambda_0$ is subtracted. The relative phase is omitted for the DC crossing transformation, as it can become highly sensitive to small perturbations when the bar-port amplitude is close to zero. Solid curves show nominal responses. Shaded regions span the pointwise minimum and maximum over the 81 parameter combinations on the $9\times9$ linewidth--thickness grid, excluding samples with undefined phase where applicable. They do not represent confidence intervals or the distribution over physical wafer positions. The vertical dashed line marks $\lambda_0$.
}
\label{fig:ghz-coupler-responses}
\end{figure}

Supplementary Fig.~\ref{fig:ghz-coupler-responses} compares the nominal responses of the four logical transformations with their variation over the linewidth and film thickness ranges of the virtual wafer. For each parameter, the minimum and maximum deviations are obtained by evaluating the wafer field on the wafer grid. Nine uniformly spaced values are then selected along each parameter axis, giving 81 combinations over the resulting rectangular range. These combinations are not samples of 81 physical wafer positions and do not preserve the spatial correlations between the two process fields. The DC coupling lengths are held fixed throughout this comparison. For the $33{:}66$ response shown here, the output ports of the $66{:}33$ RAC model are interchanged. The RAC models exhibit flatter cross-port fractions, whereas the total transmission and relative phase remain implementation-dependent. The circuit comparison therefore retains the full complex response rather than assigning robustness from the splitting ratio alone.

\subsection{Matched wafer sweep and programmable control}

The process fields and layout-mapping procedure are given in Section~\ref{sec:virtual-wafer}. The layout anchor is fixed at $x=-20~\mathrm{mm}$ and translated along $y$ from $19.75$ to
$-19.75~\mathrm{mm}$ in $0.5~\mathrm{mm}$ steps, giving 80 positions. All eight configurations use the same stored linewidth and film thickness fields and the same anchor coordinates. Because the DC and RAC layouts have different footprints, a matched position denotes a common wafer field and circuit anchor, not identical local samples for every component.

At each programmable-mesh position, the two effective-index shifts of every MZI are optimized at $\lambda_0=1550~\mathrm{nm}$. A component-level stage first maximizes
\begin{equation}
\mathcal{F}_2(T,S)=\frac{|\operatorname{Tr}(T^\dagger S)|^2}{2\operatorname{Tr}(S^\dagger S)}
\label{eq:ghz-local-control-objective}
\end{equation}
for each two-mode target $T$, using differential evolution followed, when needed, by Nelder--Mead refinement. A global L-BFGS-B stage then jointly optimizes the 132 phase settings against
\begin{equation}
\mathcal{F}_{U}=\frac{|\operatorname{Tr}(U_{\mathrm{tar}}^\dagger S_{\mathrm{mesh}}(\lambda_0))|^2}{12\operatorname{Tr}[S_{\mathrm{mesh}}^\dagger(\lambda_0)S_{\mathrm{mesh}}(\lambda_0)]}.
\label{eq:ghz-global-control-objective}
\end{equation}
Both effective-index shifts are bounded by $-5\times10^{-3}\leq\Delta n\leq5\times10^{-3}$. The global stage uses three starts, the first from the component-level solution and the remainder from Gaussian perturbations of that solution, with at most 1,000 L-BFGS-B iterations and a stopping target of $\mathcal{F}_{U}=0.9999$. If the target is not reached, the best encountered solution is retained. The final heater settings and optimizer status for every position are provided as Source Data.

For phase-shifter section $h\in\{\mathrm{input},\mathrm{internal}\}$ of MZI $m$, the phase-equivalent wrapped setting at $\lambda_0$ is
\begin{equation}
\widetilde{\Delta\phi}_{h,m}=\left(\frac{2\pi L_h}{\lambda_0}|\Delta n_{h,m}|\right)\bmod 2\pi,
\qquad L_h=500~\mu\mathrm{m}.
\label{eq:ghz-wrapped-phase}
\end{equation}
The aggregate wrapped programmed phase excursion is
\begin{equation}
C_\phi=\sum_m\left(\widetilde{\Delta\phi}_{\mathrm{input},m}+\widetilde{\Delta\phi}_{\mathrm{internal},m}\right).
\label{eq:ghz-phase-excursion}
\end{equation}

Here, $C_\phi$ is an aggregate control-range metric; no actuator-specific conversion from phase excursion to electrical power is applied.

\subsection{Frequency-unresolved reconstruction and GHZ fidelity}

Each tomographic basis is applied to the three dual-rail output qubits before frequency-unresolved PNR probabilities are evaluated. A logical 0 corresponds to the rail count $(1,0)$ and a logical 1 to $(0,1)$. The fixed single-photon input pattern is $[1,0,1,1,0,1,0,1,1,0]$, and the four-rail heralding pattern is $[1,1,1,0]$. In the programmable layout these patterns are embedded into the 12-rail spaces by inserting vacuum rails 4 and 7 at the input and rails 3
and 9 at the output; the logical output pairs are $(0,1)$, $(2,4)$ and $(5,6)$. For every basis, the eight accepted logical outcomes are normalized to unit total probability.

The task-specific reconstructions use the 13 settings
\begin{equation}
\begin{split}
\mathcal{B}_{\mathrm{task}}=\{&ZZZ,XXX,XYY,YXY,YYX,XYX,XYZ,\\
&YZY,ZXX,YXX,YZZ,YYZ,ZZX\},
\end{split}
\label{eq:ghz-task-tomography-settings}
\end{equation}
whereas the programmable-mesh and source-swap calculations use the eight unique settings
\begin{equation}
\mathcal{B}_{\mathrm{mesh}}=\{ZZZ,XXX,XYY,YXY,YYX,XXZ,ZZY,ZXX\}.
\label{eq:ghz-mesh-tomography-settings}
\end{equation}
Both reconstructions use informationally incomplete Pauli measurements and positivity-constrained least squares. The five settings $ZZZ$, $XXX$, $XYY$, $YXY$ and $YYX$ are common to both calculations and constrain the computational-basis populations and principal GHZ coherences. The remaining settings provide additional constraints on the reconstructed density matrix. The larger task-specific set was retained because these states showed broader departures from the target GHZ state, whereas the programmable-mesh states remained closer to the target. No full Pauli tomography was performed. All Pauli observables compatible with each setting are evaluated once. The density matrix is reconstructed by
\begin{equation}
\widehat\rho=\underset{\rho\succeq0,\,\operatorname{Tr}\rho=1}{\operatorname{arg\,min}}\sum_j\left[\operatorname{Tr}(A_j\rho)-\widehat a_j\right]^2,
\label{eq:ghz-density-reconstruction}
\end{equation}
using the CLARABEL solver. Before photon counting, each tomographic covariance matrix is compressed using the rail-local Gram construction in Section~\ref{sec:spectral-reduction} with a 0.1\% tolerance.

For the task-specific circuit, the reported fidelity is maximized only over the relative GHZ phase,
\begin{equation}
\begin{split}
|\mathrm{GHZ}_\phi\rangle&=\frac{|000\rangle+e^{i\phi}|111\rangle}{\sqrt{2}},\\
F_{\mathrm{task}}&=\max_\phi\langle\mathrm{GHZ}_\phi|\widehat\rho|\mathrm{GHZ}_\phi\rangle\\
&=\frac{\widehat\rho_{000,000}+\widehat\rho_{111,111}}{2}+|\widehat\rho_{000,111}|.
\end{split}
\label{eq:ghz-phase-optimized-fidelity}
\end{equation}
This removes one output-local phase but does not correct internal phase error or spectral distinguishability. For the programmable mesh and source-swap replays, the target is fixed to
$|\mathrm{GHZ}_{+}\rangle=(|000\rangle+|111\rangle)/\sqrt{2}$ and $F_{+}=\langle\mathrm{GHZ}_{+}|\widehat\rho|\mathrm{GHZ}_{+}\rangle$. State purity is $P=\operatorname{Tr}(\widehat\rho^2)$.

\subsection{Position-resolved fidelity and reconstructed states}

All 80 positions are retained for the task-specific circuits. For the programmable meshes, only solutions with reference-wavelength mesh fidelity $\mathcal{F}_{U}\geq0.99$ [Eq.~(7)] are included, leaving 71 DC and 61 RAC configurations. The same configurations are used for both sources within each coupler implementation. Table~\ref{tab:ghz-full-results} reports the distribution summaries used in Fig.~6e.

\begin{table}[t]
\centering
\caption{\textbf{Position-resolved GHZ fidelity distributions.}
Task-specific distributions include all 80 virtual-wafer positions. Programmable distributions include the 71 DC and 61 RAC configurations satisfying $\mathcal{F}_{U}\geq0.99$. Task-specific values are phase optimized according to Eq.~\eqref{eq:ghz-phase-optimized-fidelity}; programmable values use the fixed $|\mathrm{GHZ}_{+}\rangle$ target.}
\label{tab:ghz-full-results}
\small
\begin{tabular}{lllccc}
    \toprule
    Source & Architecture & Coupler & Median & [25th, 75th] & Range \\
    \midrule
    SPDC & Task-specific & DC  & 0.9078 & [0.8555, 0.9543] & [0.2496, 0.9849] \\
    SPDC & Task-specific & RAC & 0.6173 & [0.3845, 0.9110] & [0.1918, 0.9743] \\
    SFWM & Task-specific & DC  & 0.9155 & [0.8619, 0.9618] & [0.2525, 0.9932] \\
    SFWM & Task-specific & RAC & 0.6217 & [0.3898, 0.9174] & [0.1956, 0.9817] \\
    SPDC & Programmable & DC  & 0.9826 & [0.9807, 0.9846] & [0.9697, 0.9873] \\
    SPDC & Programmable & RAC & 0.9870 & [0.9854, 0.9877] & [0.9541, 0.9887] \\
    SFWM & Programmable & DC  & 0.9916 & [0.9899, 0.9930] & [0.9864, 0.9951] \\
    SFWM & Programmable & RAC & 0.9946 & [0.9943, 0.9956] & [0.9929, 0.9968] \\
    \bottomrule
\end{tabular}
\end{table}

At the 80 matched task-specific positions, the median paired differences $F_{\mathrm{RAC}}-F_{\mathrm{DC}}$ are
$-0.184$ for SPDC and $-0.187$ for SFWM. The DC implementation outperforms RAC at 59 of 80 positions for both sources.

For the programmable meshes, paired DC--RAC comparisons use the 54 positions at which both implementations satisfy $\mathcal{F}_{U}\geq0.99$. The median paired differences $F_{\mathrm{RAC}}-F_{\mathrm{DC}}$ are 0.00379 for SPDC and 0.00361 for SFWM.

Source comparisons use all retained configurations within
each implementation: 71 for DC and 61 for RAC. The median paired advantages $F_{\mathrm{SFWM}}-F_{\mathrm{SPDC}}$ are 0.00847 for DC and 0.00801 for RAC. Replacing SPDC with SFWM reduces the interquartile range by 21.9\% and 43.4\%, respectively.

Supplementary Fig.~\ref{fig:ghz-delay-analysis}a, b shows representative SPDC density matrices. For each coupler implementation, the displayed state is taken from the simulated position whose fidelity is closest to the corresponding SPDC median. Despite retaining high state purity, the task-specific RAC example contains substantial populations and coherences outside the target GHZ subspace. This indicates a systematic departure of the effective circuit transformation from its target, rather than fidelity loss arising primarily from mode mismatch. Together with the relative phase responses in Supplementary Fig.~\ref{fig:ghz-coupler-responses} and possible phase errors from routing waveguides as in Supplementary Fig.~\ref{fig:layout-mapping}, this systematic departure is consistent with cumulative phase and delay mismatch associated with the ratio-specific RAC sections and their larger routing footprint. The reconstruction does not, however, isolate phase from propagation loss or other circuit-level effects. The programmable examples, by contrast, remain concentrated in the $|000\rangle$ and $|111\rangle$ subspace after retuning.

\subsection{Output delay spread and source swaps}

\begin{figure}[t]
\centering
\includegraphics[width=\textwidth]{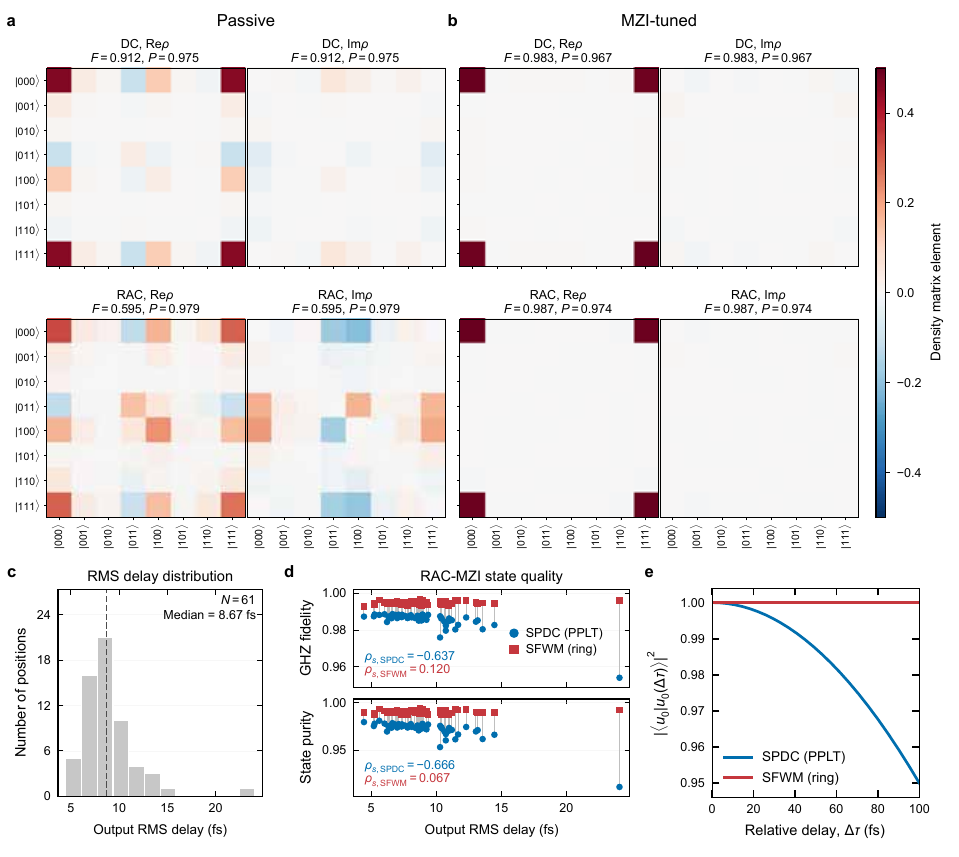}
\caption{\textbf{Reconstructed GHZ states and source-dependent sensitivity to circuit delay.}
\textbf{a,b}, Real and imaginary parts of representative reconstructed density matrices for SPDC inputs in the task-specific circuits (\textbf{a}) and programmable MZI meshes (\textbf{b}). The upper and lower rows show the DC and RAC implementations, respectively. Each state is selected from the position whose fidelity is closest to
the median of the corresponding distribution. Task-specific distributions include all 80 positions, whereas the programmable distributions include 71 DC and 61 RAC configurations satisfying $\mathcal{F}_{U}\geq0.99$. Task-specific states are displayed after the local phase adjustment. Annotations give the fidelity $F$ and purity $P=\operatorname{Tr}(\hat{\rho}^{\,2})$ of each displayed state.
\textbf{c}, Distribution of within-output RMS delays $\tau_{\mathrm{out,RMS}}$ for the 61 retained RAC meshes. The dashed line marks the median, $8.67~\mathrm{fs}$.
\textbf{d}, Reconstructed GHZ fidelity (upper) and state purity (lower) for SPDC (PPLT, blue circles) and SFWM (ring, red squares) inputs in these same meshes. Gray lines connect source pairs evaluated with identical wafer
positions, layouts and phase settings. The displayed $\rho_s$ values are source-specific Spearman rank coefficients calculated over the 61 configurations.
\textbf{e}, Squared overlap between the dominant signal spectral mode of each source and a copy shifted by a relative delay $\Delta\tau$. The modes are obtained from the individual source covariance matrices using the Gram construction of Section~\ref{sec:spectral-reduction}. This panel characterizes the sensitivity of each source mode to a relative time shift, independently of circuit propagation.}
\label{fig:ghz-delay-analysis}
\end{figure}

Supplementary Fig.~\ref{fig:ghz-delay-analysis}c summarizes the output delay spread across the 61 retained RAC meshes. The median $\tau_{\mathrm{out,RMS}}$ is $8.67~\mathrm{fs}$, with an interquartile range from $7.12$ to $10.39~\mathrm{fs}$. The observed values span $4.45$--$24.08~\mathrm{fs}$, with only one configuration exceeding $15~\mathrm{fs}$. 
It provides a circuit-level summary of how spatial fabrication variation and the resulting uncertainty on control settings affect timing, even after the central-wavelength transformation has been calibrated. These samples describe positions within one correlated wafer realization; repeating the analysis over independently generated wafer fields would extend it to a Monte Carlo assessment of variation between wafers.

The increasing SFWM fidelity advantage at larger delay spreads mainly reflects degradation of the SPDC result, rather than improvement of the SFWM result (Supplementary Fig.~\ref{fig:ghz-delay-analysis}d). For SPDC, the Spearman coefficients between delay spread and reconstructed fidelity and purity are $-0.637$ and $-0.666$, respectively. The corresponding SFWM coefficients are $0.120$ and $0.067$, indicating little monotonic association over the sampled configurations. The accompanying decrease in SPDC purity is consistent with spectral and temporal mismatch contributing to mixing of the reconstructed logical state. These results support a contribution from mode mismatch, although they do not separately determine its magnitude relative to other circuit effects.

Supplementary Fig.~\ref{fig:ghz-delay-analysis}e examines the source dependence more directly through the overlap between each source's dominant signal mode and a delayed copy. The mode $u_0$ is extracted from the individual source covariance using the Gram construction of Section~\ref{sec:spectral-reduction}, before circuit propagation. The squared overlap $\mathcal{O}(\Delta\tau)
=|\langle u_0|u_0(\Delta\tau)\rangle|^2$ decreases more rapidly for the broader SPDC mode. At $\Delta\tau=100~\mathrm{fs}$, its value is approximately $0.950$, whereas the SFWM overlap remains essentially unity. This comparison illustrates the greater sensitivity of the selected SPDC source to relative temporal shifts.

This direct overlap analysis does not reproduce the circuit fidelity and purity decrease on the same delay scale. It describes two copies of one source mode separated by a single delay, whereas $\tau_{\mathrm{out,RMS}}$ summarizes delay differences at the circuit outputs. This scalar metric does not resolve the relative arrival times of the interfering fields at individual beam splitters inside the circuit, although their effects are retained in the full frequency-resolved simulation. Consequently, $\tau_{\mathrm{out,RMS}}$ should be interpreted as a supplementary diagnostic of temporal mismatch, not as a direct predictor of fidelity or purity. Together, the source-mode overlap and the source-dependent correlations support the interpretation that residual temporal mismatch contributes to the stronger degradation observed for SPDC inputs.

\section{Gaussian boson sampling (GBS) photon counting benchmark}
\label{sec:gbs-benchmark}

\begin{figure}[t]
\centering
\includegraphics[width=\textwidth]{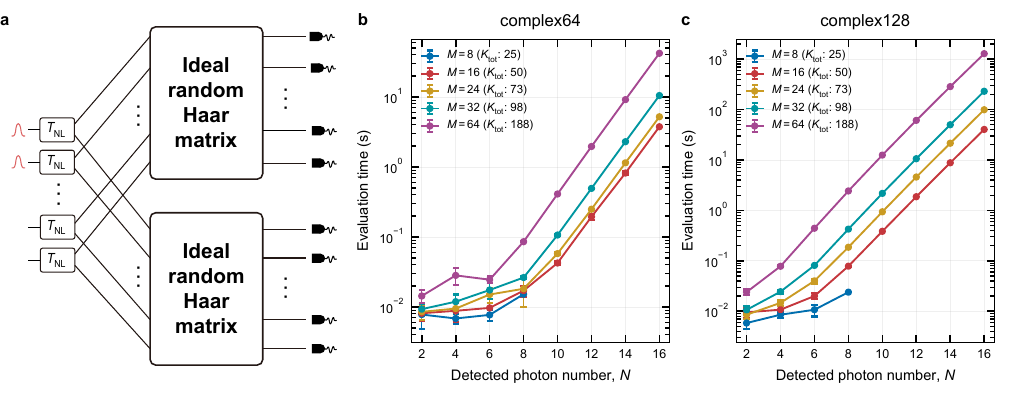}
\caption{\textbf{Gaussian boson sampling benchmark.}
\textbf{a}, Two-mode squeezer Gaussian boson sampling circuit schematic. The number of pumped sources is set to $\min(M/2,N_{\max}/2)$, with $N_{\max}=16$. Thus, for each fixed $M$, the same output state is used to evaluate detection probabilities for all photon numbers $N$ considered. $K_\mathrm{tot}$ is number of supermodes after reduction process in Sec.~\ref{sec:spectral-reduction} with tolerance $\tau=10^{-3}$.
\textbf{b}, Benchmark result on \texttt{complex64} arithmetic.
\textbf{c}, Benchmark result on \texttt{complex128} arithmetic.
}
\label{fig:gbs}
\end{figure}

We benchmark the time required to evaluate photon counting probabilities as a function of circuit size and total detected photon number. The circuit comprises identical SPDC sources based on the source model described in Sec~\ref{sec:gkp-example}. Their signal and idler outputs are sent to separate ideal interferometers described by independent Haar-random unitary matrices (Supplementary Fig.~\ref{fig:gbs}a). Before photon counting, each covariance matrix is reduced using the covariance-Gram basis described in Sec~\ref{sec:spectral-reduction}.

Circuit size is defined as number of rails $M$ including both signals and idlers. For each circuit size and target total detected photon number $N$, we select ten random collision-free patterns and evaluate their frequency-unresolved PNR probabilities three times. Haar matrix is fixed for same $M$. The benchmark reports the mean computation time per pattern probability evaluation using single-precision and double-precision complex arithmetic (\texttt{complex64} and \texttt{complex128}, respectively) (Supplementary Fig.~\ref{fig:gbs}b, c).

\clearpage

\renewcommand{\refname}{Supplementary References}
\phantomsection
\addcontentsline{toc}{section}{Supplementary References}

\bibliographystyle{unsrt}
\bibliography{sn-bibliography}


%
